\documentclass[%
  twocolumn,
 showpacs,
 showkeys,
 preprintnumbers,
 amsmath,amssymb,
 aps,
  pra,
  longbibliography,
 floatfix,
 ]{revtex4-1}

\usepackage{mathptmx}

\usepackage{amssymb,amsthm,amsmath}

\usepackage{listings}
\usepackage{graphicx}

\usepackage[dvipsnames]{xcolor}

\usepackage{tikz}
\usetikzlibrary{calc,decorations.pathreplacing,decorations.markings,positioning,shapes}

\usepackage{hyperref}
\hypersetup{
    colorlinks,
    linkcolor={blue!80!black},
    citecolor={red!75!black},
    urlcolor={blue!80!black}
}

\begin{document}

\title{What makes quantum clicks special?}
\thanks{This is a largely revised and extended version of a paper which has been published somewhat hidden in Section~12.9 of my recent book on (A)Causality~\cite{svozil-pac}.}

\author{Karl Svozil}
\email{svozil@tuwien.ac.at}
\homepage{http://tph.tuwien.ac.at/~svozil}

\affiliation{Institute for Theoretical Physics,
Vienna  University of Technology,
Wiedner Hauptstrasse 8-10/136,
1040 Vienna,  Austria}

\date{\today}

\begin{abstract}
This is an elaboration of the ``extra'' advantage of the performance of quantized physical systems over classical ones, both in terms of single outcomes as well as probabilistic predictions. From a formal point of view, it is based on entities related to (dual) vectors in (dual) Hilbert spaces, as compared to the Boolean algebra of subsets of a set and the additive measures they support.
\end{abstract}


\maketitle

\tableofcontents

\section{Quantum Hocus Pocus}

Time after time, scientists in other areas  as well as theologians, philosophers, and artists, articulate an interest in understanding and comprehending
what is so special about quantum physics.
They~demand to know the gist of quantization and its novel capacities.
I have witnessed that
individual physicists, in order to cope with such inquiries, often respond with at least two extreme strategies, both
amounting to a `why bother?' approach~\cite{bell-a}:

(I) The first strategy is to play the ``magic (joker) card'' and respond by claiming,
``quantum~mechanics is magic''.

Alas, such types of hocus pocus~\cite{svozil-2016-quantum-hokus-pokus} tend to leave the enquirer
in an uneasy state. Because, first of all, it is frustrating to accept incomprehensibility
in physics proper; quasi at the very heart and inner sanctum of contemporary natural science.
Secondly, theologians have a very understandable natural suspicion with regards to any claims of sanctity and consecration,
as this is supposed to be their own bread-and-butter domain. They resent evangelical physicists~\cite{clauser-talkvie}
joining their own realm, and ``throwing parties'' there.

Two subvariants of this ``magic card'' optional response are
(i)~either the acknowledgement~\cite{mermin-1989-shutup,mermin-2004-shutup} that~(Chapter~6 in~\cite{feynman-law})
``nobody understands quantum mechanics''.
So in order not to
``get 'down the drain', into a blind alley from which nobody has
yet escaped''  it is suggested that
one should avoid asking, ``how can it be like that?''--(ii)~or claims that quantum mechanics needs no (further) interpretations~\cite{fuchs-peres} and explanations~\cite{Englert2013}.

(II) The second strategy to cope with quantum mechanics is a formalistic and nominalistic one:
to develop the mathematical formalism, mostly Hilbert spaces,
very often not exceeding a good undergraduate text~\cite{halmos-vs}, and functional analysis.
This is sometimes concealed by the rather hefty discussions of applications involving
solutions of differential equations which at desperate times,
may even employ asymptotic divergent (perturbation) series~\cite{PhysRev.85.631,LeGuillou-Zinn-Justin,PhysRevD.57.1144}.

This latter strategy is frustrating to mathematicians and philosophers alike;
as to the former, it seems that one tries to convince them
that quantum mechanics is either trivial or plagued by inconsistencies and divergences.
The latter group of philosophers would not take formulae they mostly hardly understand as a satisfactory answer:
they suspect that syntax can never substitute~semantics.

Another conceivable claim is a humble one inspired by Popper~\cite{popper} and Lakatos~\cite{lakatosch}:
all our scientific knowledge is preliminary and transitory; we know very little, and what we presume to know changes constantly
as we move on to new ideas and concepts. There is no recognizable ``ontological convergence'' of concepts
which we tend to approximate as we
progress, no ``truth'' we seem to approach. Hence all our theories are embedded in, and part of, the history of thought.

With these caveats or provisions we might state that the current quantum phenomena indicate that we are living
in a ``vector world'' rather than in a world spanned by subsets of sets (that is, power sets), and naive set-theoretical operations among them.
In particular, quantum logic~\cite{birkhoff-36} suggests that the logico-algebraic relations and operations among (quantum) propositions
need to be represented in terms of linear vector space entities (and their duals) endowed with a scalar product.
In certain situations, this is radically different and departing from the ``old'' Boolean (sub)algebra ways.

In what follows we shall investigate those departures from classicality.
They will manifest themselves in various ways and forms, and both in single outcomes as well as probabilistically.
Some forms have no direct operational realization, and all of them must be based on
idealistic constructions involving counterfactual observables.
In extreme cases, the assumption of their (formal or physical) existence would result in a complete contradiction.

\section{General Principles for Object/Observable Construction}
\label{2020-b-opc}

One needs to be aware~\cite{berkeley} that our cognition is not a priori given but could be imagined as a self-sustained
``emergent'' construction of the neuronal activities in our body (mostly located in the brain),
which is subjected and exposed to ``experiences'' by our organs, and by self-reflection.
Even if there would exist an ``ontological anchor out there'' (aka real entities)
our perception would be bound to epistemological constraints, a fact known already since antiquity~(514a--520a in \cite{plato-republic}).
In~particular, physical observables need to be understood not as an objective fact
about some physical reality independent of our perception,
but subject to mental constructions involving conventions and presumptions.
What are the intuitive conditions for some phenomenon to be called ``observable''---and likewise, for a collection of ``stuff'' to be termed ``object''?
In addition, what exactly is it that we ``observe'' when measuring such ``observables?'' on such ``objects''?

What has been discussed so far suggests that any Ansatz, or rather hint or suggestion, of what might qualify as an ``object'' or ``observable''
cannot merely be based on physical quantities alone but should also involve the entire human cognitive apparatus.
Therefore, the notion of object originates in stuff which gets organized (by some cognitive agent)
in spatial-temporal lumps encountered in the environment of an individual or a species.
(To this end evolutionary psychology, and the evolution of cognition in general, is relevant).

Indeed, from an evolutionary point of view
it can be expected that at least in some instances,
a spread might develop between
``good'' and ``bad'' representations/predictions---that is,
what might be considered an appropriate representation of the (potentially dangerous) environment,
and a more pragmatic superstition about it which has a selective advantage.
Because for survival pragmatism might, at least in the short term and for ``ancient'' setting
encountered in a savanna, be more favorable than a careful evaluation of a situation: sometimes
``to run for an escape'' yield san advantage ``to evaluate risks of danger''~\cite{Gigerenzer-2007,Chabris-Simons-2010,kahneman2011thinking,Urbaniok-2020}.

Of course, one could argue that in the long run, careful analysis and reflection on, say, the ``reality/existence of observables/object''
(Kahneman's System 2)
offers more selective advantages than a mere (impulsive) ``blink~\cite{Gladwell-blink}/guts feeling'' (Kahneman's System 1).
This is effectively what happens as science progresses: the concepts and entities/objects/observables involved become increasingly adapted to the phenomena.
Often those evolving concepts are more abstract and formalized than old ones.
This does not necessarily mean that a ``conceptual convergence'' evolves insofar as ``new, progressive theories/representations''
are not necessarily extensions of ``old, degenerative theories/representations''~\cite{lakatosch}.

Therefore, we have to be suspicious of our own perception and its twisted capacity to comprehend the phenomena.
We may even come to the conclusion that pragmatism (Kahneman's {\em System 1}) is ``more effective''
(eg, in terms of Occam's razor) to conceptualize an observable;
and yet, in the long run, this conceptualization might turn out to be degenerative~\cite{lakatosch} and a failure.
Often, overconfidence indicates a lack of competence~\cite{chamorroconfidence}.

Prima facie, objects originally qualified evolutionary either as being:
\begin{itemize}
\item negative; that is, dangerous, such as poisonous snakes, predators, atmospheric phenomena; or
\item positive; that which qualifies as prey/loot/prize with respect to nutrition or joy,
such as eating/drinking/reproducing/breathing.
\end{itemize}

From that perspective, Kant was partially right,
in as much as he suggested that such evolutionary ingrained environmental notions,
structures and patterns appear to be hard-wired into our cognition,
and thus suggest themselves as being ``evident'': by evolutionary selection and substantiation they ``appear natural''.
However, Kant was deceived by not realizing that this sort of ``evidence'' and ``naturalness''
might actually be very deceptive:
the cognitive concepts we inherited from, or share with,
the plant/animal world serve well for survival and present many obvious immediate advantages
-- alas they are only functional with respect to, and serve as relative responses to,
the environmental challenges encountered in the evolutionary past of those species ``inheriting'' them.

This is why we often have not the least inclination to seriously indulge
in the idea that all stuff is an empty vacuum, pierced by extensionless elementary particles,
as modern-day science suggests: we just did not need this idea to survive (so far).
However, for example, we needed the concept of ``snake'' to survive in the desert.
(I actually had this inspiration while contemplating a rock painting of a shaman
taming a yellow snake at Spitzkoppe, Damaraland, Namibia.)

So it could be speculated that early object constructions were not formed by (sub)conscious cognitive processes.
They rather developed as response patterns already at the earliest stages in the evolution of stuff capable of some
(even very restricted, non-universal) forms of ``computation'' and endowed with cognitive capacities:
nerves, brains, adaptive cycles which gain an advantage over
nonadaptive behaviors by being capable of reaction against danger and opportunity.
What we do when performing an object construction---for instance, creating narratives of
what kind of observables are operational, or models of our brain et cetera---
is just a more or less sophisticated extension thereof.
In particular,
\begin{itemize}
\item What qualifies a lump of stuff to be subjected to object/observable construction and become ``an object'' or ``an observable''
is its function with respect to us: otherwise---that is if it does not kill us or we cannot eat it et cetera---we might as well not perceive it as an individual entity separate from the rest of the stuff surrounding us.

\item One might also speculate that every cub or human infant reenacts this structuralization of the environment--which was previously perceived ubiquitous, as a whole and non-separated (cf.~also Piaget) from the cognitive agent--the whole issue of ``external'' versus ``internal'' comes into mind.

\item as a consequence we as scientists have to be aware of these ``hard-wired''
conceptualizations or object constructions we and our species grew up with as ``evident'',
which have served our species well, but which eventually are too rigid and non-adaptive to be useful for the upcoming (deo volente) progressive research programs of Nature.
\end{itemize}

Particular models and instances of object/observable construction can be given in terms of (intertwining) Boolean subalgebras,
resulting in partition logics~\cite{svozil-2018-b} and orthomodular structures such as quantum~logic discussed later.

\section{Context and Greechie Orthogonality Hypergraphs}
\label{2017-b-cagod}

Henceforth a context will be any Boolean (sub-)algebra of compatible propositions that represent simultaneously measurable observables.
The terms context, block, maximal observable, basis, clique, and classical mini-universe will be used synonymously.

In classical physics, there is only a single context, namely the entire set of observables.
There exist models such as partition logics~\cite{dvur-pul-svo,svozil-2001-eua,svozil-2008-ql}
-- realizable by Wright's generalized urn model~\cite{wright} or automaton logic~\cite{schaller-92,svozil-93,schaller-95,schaller-96},
-- which are still quasi-classical but have more than one, possibly intertwined, contexts.
Two contexts are intertwined if they share one or more common elements.
In what follows we shall only consider contexts which if at all, intertwine at a single atomic proposition.

For such configurations Greechie has proposed a kind of  orthogonality hypergraph~\cite{greechie:71,kalmbach-83,svozil-tkadlec}
in which
\begin{enumerate}
\item
entire contexts (Boolean subalgebras, blocks) are drawn as smooth lines, such as straight (unbroken) lines, circles or ellipses;
\item
the atomic propositions of the context are drawn as circles; and
\item
contexts intertwining at a single atomic proposition are represented as non-smoothly connected lines, broken at that proposition.
\end{enumerate}

In Hilbert space realizations,
the straight lines or smooth curves depicting contexts represent orthogonal bases (or, equivalently, maximal observables, Boolean subalgebras or blocks),
and points on these straight lines or smooth curves represent elements of these bases;
that is, two points on the same straight line or smooth curve represent two orthogonal basis elements.
From dimension three onwards, bases may intertwine~\cite{Gleason} by possessing common elements.

\section{General Principles for Probabilities of Objects/Observables}

In this section, a very brief review of probability theory in arbitrary setups will be given.
These~axioms or requirements apply to all systems---classical as well as quantized and even more exotic ones---and therefore are uniform.
In what follows we shall only consider finite configurations.
Every maximal set of mutually exclusive and mutually co-measurable (in quantum mechanical terms: non-complimentary)
observables will be called  context~(cf. Section~\ref{2017-b-cagod} for a detailed discussion).

In what follows we shall assume the following axioms:
\begin{enumerate}
\item[A1:] classical (sub)sets of finite (possibly extended) propositional/observable structures
entail Kolmogorovian-type probabilities.
In particular, they imply that within one and the same context, the corresponding probabilities are
\begin{enumerate}
\item[K1] (non-negativity):  non-negative real numbers;
\item[K2] (unity):  of unit measure; that is, the probability of the occurrence of a complete set of propositions/observables is one;
\item[K3] (additivity):  the probabilities of mutually exclusive events $
\left\{
E_1, \ldots , E_n
\right\}
$ add up; that is, the probability
of occurrence of all of them is the sum of the probabilities of occurrence  of all of them; that is,
$
P\left(E_1\vee \ldots \vee E_n \right) =
P\left(E_1\right) + \cdots + P\left(E_n \right)
$.
\end{enumerate}
\item[A2] (extended unity):
Suppose there are two contexts ${\cal C}_1=\{ {\bf E}_1, \ldots {\bf E}_m\}$
and
${\cal C}_2=\{ {\bf F}_1, \ldots {\bf F}_n\}$.
Then the sum of the conditional probabilities of all the elements of the second context,
relative to any single element of the first context, adds up to one~\cite{svozil-2019-k}.
\end{enumerate}

The latter Axiom A2 deals with situations that are characterized by empirical logics with more than one Boolean sublogics.
These sublogics need not necessarily be ``connected'' or intertwined at one or more elements; they can be isolated.
A2 intuitively states that if one selects one particular outcome of an observable, then the sum of
the relative probabilities of all outcomes of another observable with respect to the previously chosen outcome of the first observable, must be one.

\section{Classical Predictions: Truth Assignments and Probabilities}

In what follows, we shall study observables whose classical and quantum predictions do not coincide and differ in various escalation levels---from ``not very much'' to ``total''.
Such predictions will come in two varieties: (i) stochastic/probabilistic, requiring a lot of individual observations; as~well~as (ii) individual outcome specific.
To be able to make classical predictions we need to develop classical probability theory and logic.
These classical predictions need then to be related so quantized systems with similar observables,
and the respective differences in predictions need to be quantified.

Classical truth assignments will be formalized by two-valued measures whose image is either 0 or~1, associated with the logical values true and false, respectively.
Classical probabilities and expectations will be introduced as the convex combination of ``extreme'' cases--associated with allowed (e.g., consistent)--which will be formalized by two-valued measures.

An upfront caveat:
The alleged physical ``existence'' (ontology) of more than one context
need not be---and, in general, due to quantum complementarity, for quanta---is not operational.
That is, ``most of the alleged ``observables'' in those collections of ``observables''
are not simultaneously measurable on any single individual particle (in any state).
Therefore, as has already been pointed out by Specker~\cite{specker-60},
the following arguments make essential use of counterfactuals.
Such counterfactuals are idealistic constructions of the mind that are identified with observables which could in principle have been measured
yet have not been measured since the experimenter chose to measure another complementary observable.
The earlier discussion in Section~\ref{2020-b-opc}
of the construction of objects and physical observables is particularly pertinent to counterfactuals
because one should not take for granted that all conceivable observables are defined simultaneously.
Indeed, if such an assumption---the simultaneous existence of complementary/counterfactual objects of physical reality/observables is abandoned
the entire chain of argument henceforth developed breaks down.

\subsection{Truth Assignments as Two-Valued Measures, Frame Functions and Admissibility of Probabilities}
\label{2017-b-admissability}

In what follows we shall use notions of ``truth assignments'' on elements of logics which carry different names for related concepts:
\begin{enumerate}
\item
The (quantum) logic community uses the term {\em two-valued state;} or, alternatively, {\em valuation}
\index{two-valued state}
\index{valuation}
for a {\em total} function $v$ on all elements of some logic $L$  mapping
$v: L \rightarrow [0,1]$  such that~(Definition~2.1.1, p.~20 in \cite {pulmannova-91})
\begin{enumerate}
\item
$v (\mathbb{I}) = 1$,
\item
if $\{ a_i, i \in \mathbb{N}\}$ is a sequence of mutually orthogonal elements in $L$---in particular,  this applies to
atoms within the same context (block, Boolean subalgebra)---
then the two-valued state is additive on those elements $a_i$;  that is, additivity holds:
\begin{equation}
v\left( \bigvee_{i \in \mathbb{N}} \right) = \sum_{i \in \mathbb{N}} v(a_i).
\label{2017-ch-pu-qlff}
\end{equation}
\end{enumerate}

\item
Gleason has used the term  frame function~\cite{Gleason} (p.~886)
\index{frame function}
of weight $1$ for a separable Hilbert space $\mathfrak{H}$ as a
total,
real-valued (not necessarily two-valued) function $f$ defined on the (surface of the)
unit sphere of $\mathfrak{H}$  such that if
$\{ a_i, i \in \mathbb{N}\}$ represents an orthonormal basis  of  $\mathfrak{H}$, then  additivity
\begin{equation}
 \sum_{i \in \mathbb{N}} f(a_i) = 1.
\label{2017-ch-pu-glff}
\end{equation}
holds for all orthonormal bases
(contexts, blocks) of the logic based on  $\mathfrak{H}$.

\item
A dichotomic total function $v: L \rightarrow [0,1]$  will be called  strongly admissible
if
\begin{enumerate}
\item[SAD1:]  within every context $C = \{ a_i, i \in \mathbb{N}\}$, a single atom $a_j$ is assigned the value one: \mbox{$v(a_j)=1$};~and
\item[SAD2:]   all other atoms in that context are assigned the value zero: $v(a_i\neq a_j )=0$.
Physically this amounts to only one elementary proposition being true; the rest of them are false.
(One may think of an array of mutually exclusively firing detectors.)
\item[SAD3:]    Non-contextuality, stated explicitly:
The value of any observable, and, in particular, of an atom in which two contexts intertwine, does not depend on the context. It is context-independent.
\end{enumerate}

\item
To cope with value indefiniteness (cf. Section~\ref{2017-b-c-eokst}),
a weaker form of admissibility was proposed~\cite{Abbott:2010uq,2012-incomput-proofsCJ,PhysRevA.89.032109,2015-AnalyticKS}
which is no total function but rather is a {\em partial} function which may remain undefined (indefinite) on some elements of $L$:
A dichotomic partial function  $v: L \rightarrow [0,1]$  will be called
{\em admissible}
\index{admissibility}
if  the following two conditions hold for every context $C$ of $L$:
\begin{enumerate}
\item[WAD1]  if there exists a $a\in C$ with $v(a)=1$, then $v(b)=0$ for all $b\in C\setminus\{a\}$;
\item[WAD2]   if there exists a $a\in C$ with $v(b)=0$ for all $b\in C\setminus\{a\}$, then $v(a)=1$;
\item[WAD3]    the value assignments of all other elements of the logic not covered by, if necessary, successive application of the admissibility rules,
are undefined and thus the atom remains value indefinite.
\end{enumerate}
\end{enumerate}

Unless otherwise mentioned (such as for contextual value assignments or admissibility discussed in Section~\ref{2017-b-c-eokst})
the quantum logical~(I), Gleason type~(II), strong admissibility~(III) notions of two-valued states will be used.
Such two-valued states (probability measures) are interpretable as (pre-existing) truth assignments; they are
sometimes also referred to as a Kochen-Specker value assignment~\cite{Yu-2012}.

\subsection{Boole's Conditions of Possible Experience}
\label{2017-b-eh}

Already George Boole pointed out that
\begin{itemize}
\item[(i)] the classical probabilities of certain events, as well as
\item[(ii)] the classical probabilities of their (joint) occurrence,
formalizable by products of the former ``elementary'' probabilities~(i),
\end{itemize}
are subject to {\em linear}
constraints~\cite{Boole,Boole-62,Frechet1935,Hailperin-1965,Hailperin-86,Ursic1984,Ursic:1986:GFL:3023712.3023752,Ursic1988,Beltrametti-1991,Pykacz-1991,Pulmannova-1992,Beltrametti-1993,Beltrametti-1994,DvurLaen-1994,Beltrametti1995,Beltrametti-1995,Noce-1995,Laenger1995,DvurLaen-1995,DvurLaen-1995b,Beltrametti-1996,Pulmannova-2002}.
A typical problem considered by Boole is this~\cite{Boole-62} (p.~229):
{\em ``Let $p_1, p_2,\ldots , p_n$ represent the probabilities given in the data. As
these will in general not be the probabilities of unconnected events, they will be subject
to other conditions than that of being positive proper fractions,~$\ldots$.
Those other conditions will, as will hereafter be shown, be capable of expression by
equations or inequations reducible to the general form
$a_1 p_1 + a_2p_2 + \cdots + a_n p_n +a \ge 0$,
$a_1, a_2, \ldots , a_n,a$ being numerical constants which differ for the different conditions in
question. These~$\ldots$ may be termed the conditions of possible
experience''.}
\index{conditions of possible experience}

Spool forward a century to
Bell's derivation~\cite{bell} of some bounds on classical joint probabilities
which could be perceived as special cases of Boole's ``conditions of possible
experience''.
Those classical probabilities characterize a setup of classical observables.
However, these classical observables also have a {\em quantum double}~\cite{Arthaud}:
such a corresponding quantized system
has a very similar empirical logic~\cite{Foulis1976}; that is, its structure of observables closely resembles the classical setup.
However, one difference to its classical counterpart is its limited operational capacities;
in particular, complementarity: Because of complementarity this needs to be done by measuring subsets, or {\em contexts
of mutually compatible observables}
(possibly by Einstein-Podolsky-Rosen type~\cite{epr} counterfactual inference)---one at a time; e.g., one after another---
on different distinct subensembles prepared in the same state.
The predictions of the quantized system based on quantum probabilities can be tested,
thereby falsifying the classical Boole-Bell
type predictions based on classical (joint) probabilities.
Please note that as observed by Sakurai~\cite{Sakurai-1994} (pp.~241--243) and  Pitowsky~(Footnote~13 in \cite{pitowsky-89a})
the present form of the ``Bell inequalities'' is due to Wigner~\cite{wigner-70,Fine-82}.

Froissart~\cite{froissart-81,cirelson} suggested a geometric interpretation of Boole's linear ``conditions of possible
experience'' (without explicitly mentioning Boole): In referring to a later paper by Bell~\cite{Bell-71}, Froissart
proposed a general constructive method to produce all ``maximal'' (in the sense of tightest)
constraints on classical probabilities and correlations for certain experimentally realizable quantum mechanical configurations.
This method uses all conceivable types of classical correlated outcomes, represented as matrices (or higher-dimensional objects)
which are the vertices~\cite{froissart-81} (p.~243)  {\em ``of a polyhedron which is their convex hull.
Another way of describing this convex polyhedron is to view it as an intersection
of half-spaces, each one corresponding to a face. The points of the
polyhedron thus satisfy as many inequations as there are faces. Computation
of the face equations is straightforward but tedious''.}
In Froissart's perspective certain ``optimal'' Bell-type inequalities can be interpreted as
defining half-spaces (``below-above'', ``inside-outside'')
which represent the faces of a convex correlation polytope.

Later Pitowsky pointed out the connection to Boole;
in particular that any Bell-type inequality can be interpreted as
Boole's condition of possible experience~\cite{pitowsky-86,pitowsky,pitowsky-89a,Pit-91,Pit-94,2000-poly}.
Pitowsky  does not quote Froissart but mentions~\cite{pitowsky-86} (p.~1556) that he
had been motivated by a (series of) paper(s) by Garg and Mermin~\cite{Garg1984}
(who incidentally did not mention Froissart either)
on   Farkas' Lemma.

A very similar question had also been pursued by Chochet theory~\cite{Bishop-Leeuw-1959},
Vorob'ev~\cite{Vorobev-1962} and Kellerer~\cite{Kellerer-1964,Kellerer-1984},
who inspired Klyachko~\cite{Klyachko-2008},
as neither one of the previous authors are mentioned.
(To~be fair, in the reference section of an unpublished previous paper~\cite{Klyachko-2002} Klyachko
mentions Pitowsky two times; one reference not being cited in the main text).

\subsection{The Convex Polytope Method}
\label{2017-b-chm}

The essence of the convex polytope method is based on the observation that
any classical probability distribution can be written as a convex sum of all of the conceivable ``extreme'' cases.
These ``extreme'' cases can be interpreted as classical truth assignments; or, equivalently,
as two-valued states. A two-valued state is a function on the propositional structure of elementary observables,
assigning any proposition the values ``$0$'' and ``$1$'' if they are (for a particular ``extreme'' case)
``false'' or ``true'', respectively.
``Extreme'' cases are subject to criteria defined later in Section~\ref{2017-b-admissability}.
The first explicit use~\cite{svozil-2001-cesena,svozil-2001-eua,svozil-2008-ql,svozil-2016-s}
(see Pykacz~\cite{Pykacz1989} for early use of two-valued states)
of the polytope method for deriving bounds using two-valued states on logics with intertwined contexts
seems to have been for the pentagon logic, discussed in Section~\ref{2017-b-kcbsi}) and cat's cradle logic
(also called ``K\"afer'', the German word for ``bug'', by Specker), discussed in Section~\ref{2017-b-speckerbug}.

More explicitly, suppose that there be as many, say, $k$, ``weights'' $\lambda_1, \ldots ,\lambda_k$ as there are two-valued states
(or ``extreme'' cases, or truth assignments, if you prefer this denominations).
Then convexity demands that all of these weights are positive and sum up to one; that is,
\begin{equation}
\begin{split}
\lambda_1, \ldots , \lambda_k  \ge 0
\text{, and }
\\
\lambda_1 + \ldots + \lambda_k   = 1
.
\label{2017-b-convexity}
\end{split}
\end{equation}

Suppose  that for any particular $i$th two-valued state
(or the $i$th ``extreme'' case, or the $i$th truth assignment, if you prefer this denomination),
all the, say, $m$, ``relevant'' terms
--
relevance here merely means that we want them to contribute to the linear bounds
denoted by Boole as conditions of possible experience, as discussed in Section~\ref{2017-b-wtmeccp}---are ``lumped'' or combined together
and identified as vector components of a vector $\vert {\bf x}_i\rangle $
in an $m$-dimensional vector space $\mathbb{R}^m$; that is,
\begin{equation}
\begin{split}
\vert {\bf x}_i\rangle
=
\begin{pmatrix}
x_{i_1},
x_{i_2},
\ldots .
x_{i_m}
\end{pmatrix}^\intercal
.
\label{2017-b-vivector}
\end{split}
\end{equation}

Note that any particular convex see Equation~(\ref{2017-b-convexity}) combination
\begin{equation}
\begin{split}
\vert {\bf w} ( \lambda_1 ,\ldots , \lambda_k ) \rangle
=
\lambda_1 \vert  {\bf x}_1 \rangle  + \cdots + \lambda_k \vert  {\bf x}_k \rangle
\label{2017-b-cow}
\end{split}
\end{equation}
of the $k$ weights $\lambda_1, \ldots ,\lambda_k$
yields a valid---that is consistent, subject to the criteria defined in Section~\ref{2017-b-admissability}---
classical probability distribution, characterized by the vector $\vert {\bf w} ( \lambda_1 ,\ldots , \lambda_k ) \rangle  $.
These~$k$ vectors
$\vert  {\bf x}_1 \rangle , \ldots , \vert  {\bf x}_k \rangle$
can be identified with
 vertices  or  extreme points
\index{vertex}
\index{extreme point}
(which cannot be represented as convex combinations of other vertices  or   extreme points),
 associated with the $k$ two-valued states  (or ``extreme'' cases, or truth assignments).
Let
$
V
=
\left\{
\vert  {\bf x}_1 \rangle , \ldots , \vert  {\bf x}_k \rangle
\right\}
$
be the set of all such vertices.

For any such subset $V$ (of  vertices  or   extreme points)
of $\mathbb{R}^m$, the  convex hull
\index{convex hull}
is defined as the smallest convex set in $\mathbb{R}^m$
containing $V$~(Section~2.10, p.~6 in \cite{Fukuda-techrep}).
Based on its vertices a convex ${\cal V}$-polytope can be defined as the subset of $\mathbb{R}^m$ which is the convex hull of a finite set of
vertices or extreme points
$V
=
\left\{
\vert  {\bf x}_1 \rangle , \ldots , \vert  {\bf x}_k \rangle
\right\}$ in $\mathbb{R}^m$:
\begin{equation}
\begin{split}
P=\text{Conv}(V) =
\\=\Big\{
\sum_{i=1}^k   \lambda_i \vert  {\bf x}_i \rangle
\Big|
\lambda_1, \ldots , \lambda_k  \ge 0 ,
\;
\sum_{i=1}^k   \lambda_i   = 1   ,
\;
\vert  {\bf x}_i \rangle \in V
\Big\}
.
\label{2017-b-V-polytope}
\end{split}
\end{equation}

A convex ${\cal H}$-polytope can also be defined as the intersection of a finite set of half-spaces,
that is, the solution set of a finite system of $n$ linear inequalities:
\begin{equation}
\begin{split}
P=P(A,b) =\Big\{
 \vert  {\bf x} \rangle \in \mathbb{R}^m
\Big|
\textsf{\textbf{A}}_i \vert {\bf x} \rangle   \le  \vert {\bf b} \rangle \text{ for } 1 \le i \le  n
\Big\}
,
\label{2017-b-H-polytope}
\end{split}
\end{equation}
with the condition that the set of solutions is bounded, such that there is a constant $c$ such that $\| \vert {\bf x} \rangle  \| \le c$ holds for
all $\vert {\bf x} \rangle  \in P$.   $\textsf{\textbf{A}}_i$ are matrices and $ \vert {\bf b} \rangle$ are  vectors with real  components, respectively.
Due to the Minkoswki-Weyl ``main'' representation theorem~\cite{ziegler,Henk-Ziegler-polytopes,Avis:1997:GCH:280651.280652,mcmullen-71,Schrijver,gruenbaum-2003,Fukuda-techrep}
every ${\cal V}$-polytope has a description  by a finite set of inequalities.
Conversely,
every ${\cal H}$-polytope is the convex hull of a finite set of points.
Therefore the  ${\cal H}$-polytope representation in terms of inequalities  as well as the ${\cal V}$-polytope representation in terms of vertices,
are equivalent, and  the term convex polytope can be used for both and interchangeably.
A $k$-dimensional
convex polytope has  a variety of  faces  which are again convex polytopes of various dimensions between
0 and $k - 1$. In particular, the 0-dimensional faces are called  vertices, the 1-dimensional faces
are called {\em edges}, and the $k - 1$-dimensional faces are called facets.

The solution of the   hull problem, or the
convex hull computation,
is the determination of the convex hull for a given finite set of
$k$ extreme points $V = \{ \vert  {\bf x}_1 \rangle  , \ldots , \vert  {\bf x}_k \rangle \}$ in $\mathbb{R}^m$
(the general hull problem would also tolerate points inside the convex polytope);
in particular, its representation as the intersection of half-spaces defining the facets of this polytope
-- serving as criteria of what lies ``inside'' and ``outside'' of the polytope---or,
more precisely, as a set of solutions to a minimal system of linear inequalities.
As~long as the polytope has a non-empty interior and is full-dimensional
(with respect to the vector space into which it is embedded)
 there are only inequalities;
otherwise, if the polytope lies on a hyperplane one obtains also equations.

For the sake of a familiar example, consider the regular 3-cube,
which is the convex hull of the 8 vertices in  $\mathbb{R}^3$  of
$V= \big\{
\left(0, 0, 0\right)^\intercal$,   $
\left(0, 0, 1\right)^\intercal$,   $
\left(0, 1, 0\right)^\intercal$,   $
\left(1, 0, 0\right)^\intercal$,   $
\left(0, 1, 1\right)^\intercal$,   $
\left(1, 1, 0\right)^\intercal$,   $
\left(1, 0, 1\right)^\intercal$,   $
\left(1, 1, 1\right)^\intercal
\big\}
$.
The cube has 8 vertices, 12 edges, and 6 facets.
The half-spaces defining the  regular 3-cube can be written in terms of the 6 facet inequalities
$0 \le x_1,x_2,x_3 \le 1$.

Finally, the correlation polytope can be defined
as the convex hull of all the vertices or extreme points
$\vert  {\bf x}_1 \rangle , \ldots , \vert  {\bf x}_k \rangle$
in $V$
representing the ($k$ per two-valued state) ``relevant'' terms  evaluated for all the two-valued states
(or ``extreme'' cases, or truth assignments); that is,
\begin{equation}
\begin{split}
\text{Conv}(V) =\Big\{
\vert {\bf w} ( \lambda_1 ,\ldots , \lambda_k ) \rangle
\Big| \\   \Big|
\vert {\bf w} ( \lambda_1 ,\ldots , \lambda_k ) \rangle
=
\lambda_1 \vert  {\bf x}_1 \rangle  + \cdots + \lambda_k \vert  {\bf x}_k \rangle  \;  ,
\\     \;
\lambda_1, \ldots , \lambda_k  \ge 0 ,
\;
\lambda_1 + \ldots + \lambda_k   = 1   ,
\;
\vert  {\bf x}_i \rangle \in V
\Big\}
.
\label{2017-b-correlationpolytope}
\end{split}
\end{equation}

The convex ${\cal H}$-polytope---associated with the convex ${\cal V}$-polytope in~(\ref{2017-b-correlationpolytope})---which is the intersection of a finite number of half-spaces,  can be identified with Boole's conditions of possible
experience.

A similar argument can be put forward for bounds on expectation values, as
the expectations of dichotomic $E   \in \{-1,+1\}$-observables can be considered to be affine transformations
of two-valued states $v  \in \{0,1\}$; that is, $E = 2 v - 1$.
One might even imagine such bounds on arbitrary values of observables, as long as affine transformations are applied.
Joint expectations from products of probabilities transform non-linearly, as,
for instance  $E_{12}= (2v_1-1)(2v_2-1)= 4 v_1v_2 - 2(v_1+v_2)-1$.
Therefore, given some bounds on (joint) expectations, these can be translated into bounds on (joint) probabilities
by substituting $2 v_i - 1$ for expectations $E_i$.
The converse is also true:  bounds on (joint) probabilities can be translated into bounds on (joint)
expectations by $v_i = (E_i +1)/2$.

This method of finding classical bounds must fail if, such as for  Kochen-Specker configurations, there are no or ``too few''
(such that there exist two or more atoms which cannot be distinguished by any two-valued state)
two-valued states.
In this case, one may ease the assumptions; in particular, abandon admissibility, arriving at what has been called
non-contextual inequalities~\cite{cabello:210401}.

\subsubsection{Why Consider Classical Correlation Polytopes when Dealing with Quantized Systems?}
\label{2017-b-wccp}

A caveat seems to be in order from the very beginning:
in what follows correlation polytopes arise from classical (and quasi-classical)
situations.
The considerations are relevant for quantum mechanics only insofar as the quantum probabilities could violate classical bounds;
that is if the quantum tests violate those bounds by ``lying outside'' of the classical correlation polytope.

There exist at least two good reasons to consider (correlation) polytopes for bounds on classical probabilities, correlations, and expectation values:
\begin{enumerate}
\item  they represent a systematic way of enumerating the probability distributions and deriving constraints---
Boole's conditions of possible experience---on them;
\item   one can be sure that these constraints and bounds are  optimal
in the sense that they are guaranteed to yield inequalities which
are the best criteria for classicality.
\end{enumerate}

It is not evident to see why, with the methods by which they have been obtained, Bell's original inequality~\cite{bell-66,Bell-71}
or the Clauser-Horne-Shimony-Holt inequality~\cite{chsh} should be ``optimal'' at the time they were presented.
Their derivation involves estimates which appear  ad hoc, and it is not immediately obvious that bounds based on these estimates  
could not be improved.
The correlation polytope method, on the other hand, offers a conceptually clear framework for a derivation of all classical bounds on
higher-order distributions.

\subsubsection{What Terms May Enter Classical Correlation Polytopes?}
\label{2017-b-wtmeccp}

What can enter as terms in such correlation polytopes?
To quote Pitowsky~\cite{pitowsky-89a} (p.~38),
{\em ``Consider $n$ events $A_1 , A_2, \ldots ,A_n$, in a classical event
space~$\ldots$
Denote
$p_i = \text{probability} (A_i)$,
$p_{ij} = \text{probability} (A_i \cap A_j)$,
and more generally
$p_{{i_1}{i_2}\cdots {i_k}} = \text{probability} \left (   A_{i_1} \cap A_{i_2} \cap \cdots  \cap  A_{i_k} \right)$,
whenever $1 \le i_1 < i_2 < \ldots < i_k \le n$.
We assume no particular relations among the events. Thus $A_1 ,  \ldots ,A_n$ are not
necessarily distinct, they can be dependent or independent, disjoint or non-disjoint
etc''.}

However, although the events $A_1 , \ldots ,A_n$ may be in any  relation to one another, one has to make sure that
the respective probabilities, and, in particular, the extreme cases---the two-valued states interpretable as truth assignments---properly encode the logical or empirical relations among events. In particular, when it comes to an enumeration of cases, consistency must be retained.
For example, suppose one considers the following three
propositions:
$A_1$: ``it rains in Vienna'',
$A_3$: ``it rains in Vienna or it rains in Auckland''.
It cannot be that $A_2$ is less likely than $A_1$;
therefore, the two-valued states interpretable as truth assignments must obey
$p(A_2) \ge p(A_1)$, and in particular, if $A_1$ is true, $A_2$~must be true as well.
(It may happen though that $A_1$ is false while $A_2$ is true.)
Also, mutually exclusive events cannot be true simultaneously.

These admissibility and consistency requirements are considerably softened in the case of non-contextual inequalities~\cite{cabello:210401},
where subclassicality--the requirement that among a complete (maximal) set of mutually exclusive observables only one is true and all others are false
(equivalent to one important criterion for Gleason's frame function~\cite{Gleason})--is abandoned.
To put it pointedly, in such scenarios, the simultaneous existence of inconsistent events such as
$A_1$: ``it rains in Vienna'',
$A_2$: ``it does not rain in Vienna''
are allowed; that is,
$p(\text{``it rains in Vienna''}) = p(\text{``it does not rain in Vienna''}) =1$.
The reason for this rather desperate step is that for Kochen-Specker type configurations,
there are no classical truth assignments satisfying the classical admissibility rules;
therefore the latter is abandoned.
(With the admissibility rules goes the classical Kolmogorovian probability axioms even within classical Boolean subalgebras.)

It is no coincidence that most calculations are limited---or rather limit themselves because there are no formal reasons to go to higher orders--to the joint probabilities or expectations of just two observables:
there is no easy ``workaround'' of quantum complementarity.
The Einstein-Podolsky-Rosen setup~\cite{epr}
offers one for just two complementary contexts at the price of counterfactuals,
but there seems to be no generalization to three or more complementary contexts in sight~\cite{schimpf-svozil}.

\subsubsection{General Framework for Computing Boole's Conditions of Possible Experience}
\label{2017-b-gfcbcpe}

As pointed out earlier, Froissart and  Pitowsky, among others such as  Tsirelson, have
sketched a very precise algorithmic framework for constructively finding all conditions of possible experience.
In particular,
Pitowsky's later method~\cite{pitowsky,pitowsky-89a,Pit-91,Pit-94,2000-poly},
with slight modifications for very general non-distributive propositional structures
such as the pentagon logic~\cite{svozil-2001-eua,svozil-2008-ql,svozil-2016-s}, goes like this:
\begin{enumerate}
\item
define the terms which should enter the bounds;

\item
\begin{enumerate}
\item
if the bounds should be on the probabilities: evaluate all two-valued measures interpretable as truth assignments;

\item
if the bounds should be on the expectations: evaluate all value assignments of the observables;

\item
if (as for non-contextual inequalities) the bounds should be on some pre-defined quantities: evaluate all such value definite pre-assigned quantities;
\end{enumerate}

\item
arrange these terms into vectors whose components are all evaluated for a fixed two-valued state, one state at a time;
one vector per two-valued state (truth assignment), or (for expectations)
per value assignments of the observables, or (for non-contextual inequalities) per value-assignment;

\item
consider the set of all obtained vectors as vertices of a convex polytope;

\item
solve the convex hull problem \index{hull problem} by computing the convex hull, thereby
finding the smallest convex polytope containing all these vertices.
The solution can be represented as
the half-spaces (characterizing the facets of the polytope) formalized by (in)equalities---
(in)equalities which can be identified with  Boole's conditions of possible experience.

\end{enumerate}

Froissart~\cite{froissart-81} and Tsirelson~\cite{cirelson} are not very different;
they arrange joint probabilities for two random variables into matrices instead of ``delineating'' them as vectors;
but this difference is notational only. We shall explicitly apply the method to various configurations next.

The convex hull problem---finding the smallest convex polytope containig all these vertices, given a collection of such vertices---will be evaluated with Fukuda's  cddlib package cddlib-094h~\cite{cdd-pck} (using GMP~\cite{gmplib})
implementing the double description method~\cite{FP96,Avis:1997:GCH:280651.280652,Avis-2002}.
The respective computer codes are listed in the Supplementary Material.

\subsection{Non-Intertwined Contexts: Einstein-Podolsky-Rosen Type ``Explosion'' Setups of Joint Distributions}

The first non-trivial (in the sense that the joint quantum probabilities and joint quantum expectations violate the classical bounds) instance
occurs for four observables in an Einstein-Podolski-Rosen type ``explosion'' setup~\cite{epr},
where $n>1$ observables are measured on both sides, respectively.

Instead of a lengthy derivation of, say the Clauser-Horne-Shimony-Holt case of 2 observers, 2~measurement configurations per observer
the reader is referred to standard computations thereof~\mbox{\cite{pitowsky-89a,Pit-94,pitowsky,svozil-2016-s,svozil-2016-pu-book}}.
At this point, it might be instructive to realize how exactly the approach of  Froissart and Tsirelson blends in~\cite{froissart-81,cirelson}.
The only difference to the Pitowsky method---which enumerates the (two-particle) correlations and expectations as  vector components---is that
Froissart and later  and Tsirelson arrange the two-particle correlations and expectations as matrix components
For instance, Froissart explicitly mentions~\cite{froissart-81} (pp.~242, 243) 10 extremal configurations of the two-particle correlations,
associated with 10 matrices
\begin{equation}
\begin{pmatrix}p_{13}=p_1p_3 &p_{14}=p_1p_4 \\p_{23}=p_2p_3 &p_{24}=p_2p_4 \end{pmatrix}
\end{equation}
containing $0$ s and $1$ s
(the indices ``1, 2'' and ``3, 4'' are associated with the two sides of the Einstein-Podolsky-Rosen ``explosion''-type setup, respectively),
arranged in Pitowsky's case as vector
\begin{equation}
\begin{pmatrix}p_{13}=p_1p_3, p_{14}=p_1p_4, p_{23}=p_2p_3, p_{24}=p_2p_4 \end{pmatrix}.
\end{equation}

For probability correlations the number of different matrices or vectors is 10 (and not 16 as could be expected from the 16 two-valued measures),
since, as enumerated in Table~\ref{2017-CHSH-tvs} some such measures yield identical results on the two-particle correlations; in particular,
$v_1, v_2, v_3, v_4, v_5, v_9, v_{13}$  yield identical matrices (in the Froissart case) or vectors (in the Pitowsky case).
 \begin{table}
\centering
 \caption{\label{2017-CHSH-tvs}  The 16 two-valued states on the 2 particle two observables per particle configuration.}
\begin{ruledtabular}
 \begin{tabular}{ccccccccccccccccccccccc}
\textbf{\#} &\boldmath{$a_1$}&\boldmath{$a_2$}&\boldmath{$a_3$}&\boldmath{$a_4$}&\boldmath{${  a_{13}}$}&\boldmath{${  a_{14}}$}&\boldmath{${  a_{23}}$}&\boldmath{${  a_{24}}$}\\
\hline 
$v_1   $&0  &  0 &   0 &   0  & {  0}&   {  0} &  {  0}  &  {  0}\\
$v_2   $&0  &  0 &   0 &   1  & {  0}&   {  0} &  {   0}  &  {  0}\\
$v_3   $&0  &  0 &   1 &   0  & {  0}&   {  0} &  {   0}  &  {  0}\\
$v_4   $&0  &  0 &   1 &   1  & {  0}&   {  0} &  {   0}  &  {  0}\\
$v_5   $&0  &  1 &   0 &   0  & {  0}&   {  0} &  {   0}  &  {  0}\\
$v_6   $&0  &  1 &   0 &   1  & {  0}&   {  0} &  {  0}  &  {  1}\\
$v_7   $&0  &  1 &   1 &   0  & {  0}&   {  0} &  {  1}  &  {  0}\\
$v_8   $&0  &  1 &   1 &   1  & {  0}&   {  0} &  {  1}  &  {  1}\\
$v_9   $&1  &  0 &   0 &   0  & {  0}&   {  0} &  {   0}  &  {  0}\\
$v_{10}$&1  &  0 &   0 &   1  & {  0}&   {  1} &  {  0}  &  {  0}\\
$v_{11}$&1  &  0 &   1 &   0  & {  1}&   {  0} &  {  0}  &  {  0}\\
$v_{12}$&1  &  0 &   1 &   1  & {  1}&   {  1} &  {  0}  &  {  0}\\
$v_{13}$&1  &  1 &   0 &   0  & {  0}&   {  0} &  {   0}  &  {  0}\\
$v_{14}$&1  &  1 &   0 &   1  & {  0}&   {  1} &  {  0}  &  {  1}\\
$v_{15}$&1  &  1 &   1 &   0  & {  1}&   {  0} &  {  1}  &  {  0}\\
$v_{16}$&1  &  1 &   1 &   1  & {  1}&   {  1} &  {  1}  &  {  1}
\\
\end{tabular}
\end{ruledtabular}
 \end{table}
\vspace{-12pt}
Going beyond the Clauser-Horne-Shimony-Holt case with 2 observers but more measurement configurations per observer is straightforward
but increasingly demanding in terms of computational complexity~\cite{Pit-91}.
The calculation for the facet inequalities for two observers and three measurement configurations per observer
yields  684 inequalities~\cite{2000-poly,sliwa-2003,collins-gisin-2003}.
If one considers (joint) expectations one arrives at novel ones
which are not of the Clauser-Horne-Shimony-Holt type; for instance~(p.~166, Equation~(4) in \cite{sliwa-2003}),
\begin{equation}
\begin{split}
-4   \le    -E_2  +E_3 -E_4  -E_5  +E_{14} -E_{15} +    \\+E_{24}  +E_{25}  +E_{26} -E_{34} -E_{35}  +E_{36},     \\
\text{or}\\
-4 \le   E_1  +E_2   +E_4  +E_5    +E_{14}  +E_{15} + \\+E_{16}  +E_{24}  +E_{25} -E_{26}  +E_{34} -E_{35}.
\end{split}
\label{2017-b-2-3-e-full}
\end{equation}

As already mentioned earlier, these bounds on classical expectations~\cite{sliwa-2003}
translate into bounds on classical probabilities~\cite{2000-poly,collins-gisin-2003} (and  vice versa)
if the affine transformations
$E_i = 2 v_i - 1$ [and conversely $v_i = (E_i +1)/2$] are applied.

Here a word of warning is in order: if one only evaluates the vertices from the joint expectations (and not also the single particle expectations),
one never arrives at the novel inequalities of the type listed in Equation~(\ref{2017-b-2-3-e-full}),
but obtains
90 facet inequalities; among them 72
instances  of the Clauser-Horne-Shimony-Holt inequality form.
They can be combined to yield  (see also Ref.~\cite{sliwa-2003} p.~166, Equation~(4)).
\begin{equation}
\begin{split}
-4 \le   E_{14} + E_{15}    + E_{24}   + E_{26}    + E_{35}  - E_{36}  \le   4.            \\
\end{split}
\label{2017-b-2-3-enovel}
\end{equation}

For the case of $3$~\cite{2000-poly} and more qubits, algebraic methods
different than the hull problem for polytopes were suggested in Refs.~\cite{Werner-2001,Zukowski-02,Pitowsky-2002mbo,schachner-2003}.

\subsection{Truth Assignments and Predictions for Intertwined Contexts}

Let us first contemplate on the question or objection
{ ``why should we be bothered with the classical interpretation and probabilities of multiple contexts?''}
This could already have been asked for isolated contexts discussed earlier, and it becomes more compelling if intertwined contexts are considered.
After all, ``classical'' empirical configurations require a single context---Boole's algebra of observables''---only.
Why consider the simultaneous existence of a multitude of those; and even more so if they are
somehow connected and intertwined by assuming that one and the same observable may occur in different contexts?

A historic answer is this: ``because with the advent of quantum complementarity we were confronted with such observables organized in distinct contexts,
and we had to cope with them.''
In~particular, one could investigate the issue of whether or not such collections of quantum contexts and the observables therein would allow a classical interpretation
relative to the assumptions made.
Therefore, one crucial assumption was the {\em independence of the value of the observable from the particular context in which it appears},
a property often called non-contextuality.
The adverse assumption~\cite{bohr-1949,bell-66} is often referred to as contextuality.

Another motivation for studying intertwined contexts comes from partition logic~\cite{svozil-2001-eua,svozil-2008-ql},
and, in~particular, from generalized urn models and finite automata.
These cases still allow a quasi-classical interpretation although they are not strictly ``classical'' in the sense of a single Boolean algebra
(although they allow a faithful, structure-preserving embedding into a single Boolean algebra).

In the following, we shall present a series of logics
encountered by studying certain finite collections of quantum observables.
The contexts (representable by  maximal observables, Boolean subalgebras, blocks, or orthogonal bases)
formed by those collections of quantum observables
are intertwined; but ``not very much'' so: by assumption and for convenience, any two contexts intertwine in only one element; it does not happen that two contexts
are  pasted~\cite{greechie:71,kalmbach-83,nav:91,pulmannova-91}
along with two or more atoms.
Such intertwines---connecting contexts by pasting them together---can only occur from Hilbert space dimension three onwards,
because contexts in lower-dimensional spaces cannot have the same element unless they are identical.

Any such construction is usually based on a succession of  auxiliary
gadget graphs~\cite{tutte_1954,SZABO2009436,Ramanathan-18}
\index{gadget graph}
stitched together to yield the desired properties.
Therefore, gadgets are formed from gadgets of ever-increasing size and functional performance
(see also Chapter~12 of Ref. \cite{svozil-2016-pu-book}):
\begin{enumerate}
\item 0th order gadget:  a single context (aka clique/block/Boolean (sub)algebra/maximal observable/orthonormal basis);

\item 1st order ``firefly'' gadget: two contexts connected in a single intertwining atom;

\item 2nd order gadget:  two 1st order  firefly  gadgets connected in a single intertwining atom;

\item 3rd order house/pentagon/pentagram gadget:  one firefly and one 2nd order gadget connected in two intertwining atoms to form a cyclic orthogonality diagram (hypergraph);

\item 4th order true-implies-false (TIFS)/01-(maybe better 10)-gadget:  e.g., a Specker bug consisting of two pentagon gadgets connected by an entire context;
as well as extensions thereof to arbitrary angles for terminal (``extreme'') points;

\item 5th order  true-implies-true (TITS)/11-gadget:  e.g.,  Kochen and Specker's $\Gamma_1$, consisting of one 10-gadget and one firefly gadget,
connected at the respective terminal points;

\item 6th order gadget:  e.g.,  Kochen and Specker's $\Gamma_3$,
consisting of a combo of two 11-gadgets, connected  by their common firefly gadgets;

\item 7th order construction:  consisting of one 10- and one 11-gadget, with identical terminal points serving as constructions of Pitowsky's
principle of indeterminacy~\cite{pitowsky:218,2015-AnalyticKS,svozil-2018-whycontexts} and the Kochen-Specker theorem;

\end{enumerate}

In Section~\ref{2017-b-ffl} we shall first study the ``firefly case'' with just two contexts intertwined in one atom;
then, in Section~\ref{2017-b-kcbsi}, proceed to the pentagon configuration with five contexts intertwined cyclically,
then, in Section~\ref{2017-b-speckerbug}, paste two such pentagon logics to form a cat's cradle (or, by another term, Specker's bug) logic;
and finally, in Section~\ref{2017-b-bugscombino}, connect two Specker bugs to arrive at a logic which has a so ``meager''
set of states that it can no longer separate two atoms.
As pointed out already by Kochen and Specker~(p.~70, Theorem~0 in \cite{kochen1}) this is no longer embeddable into some Boolean algebra.
It thus cannot be represented by a partition logic, and thus has neither any generalized urn and finite automata models
nor classical probabilities separating different events.
The case of logics allowing no two-valued states will be covered consecutively.

\subsubsection{Probabilities on the Firefly Gadget}
\label{2017-b-ffl}



History:
Cohen presented~\cite{cohen} (pp. 21--22) a classical realization of the first logic with just two contexts and one intertwining atom: a firefly in a box,
observed from two sides of this box which are divided into two windows; assuming the possibility that sometimes the firefly does not shine at all.
This firefly logic, which is sometimes also denoted by $L_{12}$ because it has 12 elements (in a Hasse diagram) and 5 atoms in
two contexts  defined by
 ${\color{orange}\{a_1,a_2,a_3\}}$ and ${\color{blue}\{a_3,a_4,a_5\}}$.
In shorthand we may arrange the contexts, as well as the atomic propositios/observables containing them, in a set of sets; that is, $\{
{\color{orange}\{a_1,a_2,a_3\}}
,
{\color{blue}\{a_3,a_4,a_5\}}
\}$. The orthogonality hypergraph of the firefly gadget is depicted in Figure~\ref{2020-f1-gd}a.

Classical interpretations:
The five two-valued states on the firefly logic are enumerated in Table~\ref{2017-b-t-fireflytvs}.
 \begin{table}
\centering
 \caption{\label{2017-b-t-fireflytvs}  Two-valued states on the firefly logic.}
\begin{ruledtabular}
 \begin{tabular}{ccccccccccc}
 \textbf{\#}     &\boldmath{$a_1$}&\boldmath{$a_2$}&\boldmath{$a_3$}&\boldmath{$a_4$}&\boldmath{$a_5$}\\
\hline 
$v_1$&$0$&$0$&$0$&$0$&$1$  \\
$v_2$&$0$&$1$&$0$&$1$&$0$  \\
$v_3$&$0$&$1$&$1$&$0$&$0$  \\
$v_4$&$1$&$0$&$1$&$0$&$0$  \\
$v_5$&$1$&$0$&$0$&$1$&$0$
\\
 \end{tabular}
\end{ruledtabular}
 \end{table}

\vspace{-6pt}
As long there are ``sufficiently many'' two-valued states---to be more precise,
as long as there is a separating set of two-valued states~(Theorem~0 in \cite{kochen1})--a canonical partition logic
can be extracted from this set (modulo permutations)
by an ``inverse construction''~\cite{svozil-2001-eua,svozil-2008-ql}:
because of admissibility constraints SAD1-SAD3 (cf. Section~\ref{2017-b-admissability})
every two-valued state has exactly one entry ``1'' per context, and all others ``0'',
by including the index $i$ of the $i$'th measure $v_i$ in a subset of all indices of measures which acquire the value ``1'' on a particular atom
one obtains a subset representation for that atom which is an element of the partition of the index set.
This amounts to enumerating the set of all two-valued states as in Table~\ref{2017-b-t-fireflytvs} and searching its columns for entries ``1'':
in the firefly gadget case~\mbox{\cite{dvur-pul-svo,svozil-2008-ql}}, this~results in a  canonical partition logic  (modulo permutations) of
\begin{equation}
\begin{split}
\{
\{
\{4,5\},\{2,3\},\{1\}
\}
,\\
\{
\{3,4\},\{2,5\},\{1\}
\}
\}.
\end{split}
\end{equation}

This canonical partition, in turn, induces all classical probability distributions, as enumerated in~(Figure~12.4 in \cite{svozil-2016-pu-book}).

Lov\'asz~\cite{lovasz-79,lovasz-89}
defined a faithful orthogonal representation~\cite{Portillo-2015}
of a graph in some finite-dimensional Hilbert space by identifying
vertices with vectors, and adjacency with orthogonality.
No faithful orthogonal representation of the firefly gadget in $\mathbb{R}^3$ is given here, but it is straightforward--just two orthogonal tripods with one identical leg will do (or can be read off from
logics containing more such intertwined fireflies).

\begin{figure*}[htb] 
\begin{center}
\begin{tabular}{ c c c }
\begin{tikzpicture}  [scale=1]

\tikzstyle{every path}=[line width=1pt]

\newdimen\ms
\ms=0.1cm
\tikzstyle{s1}=[color=red,rectangle,inner sep=3.5]
\tikzstyle{c3}=[circle,inner sep={\ms/8},minimum size=5*\ms]
\tikzstyle{c2}=[circle,inner sep={\ms/8},minimum size=3*\ms]
\tikzstyle{c1}=[circle,inner sep={\ms/8},minimum size=2*\ms]


\coordinate (a21) at ({0.6+0},{1.7+0});
\coordinate (a22) at ({0.6+2},{1.7+2});
\coordinate (a23) at ({0.6+1},{1.7+1});
\coordinate (a24) at ({0.6+0},{1.7+2});
\coordinate (a25) at ({0.6+2},{1.7+0});


\draw [color=orange] (a21) -- (a22);
\draw [color=blue] (a24) -- (a25);


\draw (a21) coordinate[c1,fill=orange,label=below:$a_1$];

\draw (a22) coordinate[c1,fill=orange,label=below:$a_2$];

\draw (a23) coordinate[c2,fill=blue,label=above:$a_3$];
\draw (a23) coordinate[c1,fill=orange];

\draw (a24) coordinate[c1,fill=blue,label=below:$a_4$];

\draw (a25) coordinate[c1,fill=blue,label=below:$a_5$];



\coordinate (a1) at (0,2);
\coordinate (a2) at (0,1);
\coordinate (a3) at (0,0);
\coordinate (a4) at (1,0);
\coordinate (a5) at (2,0);


\draw [color=orange] (a1) -- (a3);
\draw [color=blue] (a3) -- (a5);


\draw (a1) coordinate[c1,fill=orange,label=left:$a_1$];

\draw (a2) coordinate[c1,fill=orange,label=left:$a_2$];

\draw (a3) coordinate[c2,fill=blue,label=below:$a_3$];
\draw (a3) coordinate[c1,fill=orange];

\draw (a4) coordinate[c1,fill=blue,label=below:$a_4$];

\draw (a5) coordinate[c1,fill=blue,label=below:$a_5$];

\end{tikzpicture}
&
\begin{tikzpicture}  [scale=1]

\tikzstyle{every path}=[line width=1pt]

\newdimen\ms
\ms=0.1cm
\tikzstyle{s1}=[color=red,rectangle,inner sep=3.5]
\tikzstyle{c3}=[circle,inner sep={\ms/8},minimum size=5*\ms]
\tikzstyle{c2}=[circle,inner sep={\ms/8},minimum size=3*\ms]
\tikzstyle{c1}=[circle,inner sep={\ms/8},minimum size=2*\ms]


\coordinate (a1) at (0,2);
\coordinate (a2) at (0,1);
\coordinate (a3) at (0,0);
\coordinate (a4) at (1,0);
\coordinate (a5) at (2,0);
\coordinate (a6) at (2,1);
\coordinate (a7) at (2,2);


\draw [color=orange] (a1) -- (a3);
\draw [color=blue] (a3) -- (a5);
\draw [color=red] (a5) -- (a7);


\draw (a1) coordinate[c1,fill=orange,label=left:$a_1$];

\draw (a2) coordinate[c1,fill=orange,label=left:$a_2$];

\draw (a3) coordinate[c2,fill=blue,label=below:$a_3$];
\draw (a3) coordinate[c1,fill=orange];

\draw (a4) coordinate[c1,fill=blue,label=below:$a_4$];

\draw (a5) coordinate[c2,fill=red,label=below:$a_5$];
\draw (a5) coordinate[c1,fill=blue];

\draw (a6) coordinate[c1,fill=red,label=right:$a_6$];

\draw (a7) coordinate[c1,fill=red,label=right:$a_7$];

\end{tikzpicture}
&
\begin{tikzpicture}  [scale=1]

\tikzstyle{every path}=[line width=1pt]

\newdimen\ms
\ms=0.1cm
\tikzstyle{s1}=[color=red,rectangle,inner sep=3.5]
\tikzstyle{c3}=[circle,inner sep={\ms/8},minimum size=5*\ms]
\tikzstyle{c2}=[circle,inner sep={\ms/8},minimum size=3*\ms]
\tikzstyle{c1}=[circle,inner sep={\ms/8},minimum size=2*\ms]


\coordinate (a1) at (0,2);
\coordinate (a2) at (0,1);
\coordinate (a3) at (0,0);
\coordinate (a4) at (1,0);
\coordinate (a5) at (2,0);
\coordinate (a6) at (2,1);
\coordinate (a7) at (2,2);
\coordinate (a8) at (1.5,{2+(3.5-2)/2});
\coordinate (a9) at (1,3.5);
\coordinate (a10) at (0.5,{2+(3.5-2)/2});


\draw [color=orange] (a1) -- (a3);
\draw [color=blue] (a3) -- (a5);
\draw [color=red] (a5) -- (a7);
\draw [color=green] (a7) -- (a9);
\draw [color=gray] (a9) -- (a1);


\draw (a1) coordinate[c2,fill=orange,label=left:$a_1$];
\draw (a1) coordinate[c1,fill=gray];

\draw (a2) coordinate[c1,fill=orange,label=left:$a_2$];

\draw (a3) coordinate[c2,fill=blue,label=below:$a_3$];
\draw (a3) coordinate[c1,fill=orange];

\draw (a4) coordinate[c1,fill=blue,label=below:$a_4$];

\draw (a5) coordinate[c2,fill=red,label=below:$a_5$];
\draw (a5) coordinate[c1,fill=blue];

\draw (a6) coordinate[c1,fill=red,label=right:$a_6$];

\draw (a7) coordinate[c2,fill=green,label=right:$a_7$];
\draw (a7) coordinate[c1,fill=red];

\draw (a8) coordinate[c1,fill=green,label=above right:$a_8$];

\draw (a9) coordinate[c2,fill=gray,label=above:$a_9$];
\draw (a9) coordinate[c1,fill=green];

\draw (a10) coordinate[c1,fill=gray,label=above left:$a_{10}$];

\end{tikzpicture}
\\
(a)&(b)&(c)
\\
\begin{tikzpicture}  [scale=1]

\tikzstyle{every path}=[line width=1pt]

\newdimen\ms
\ms=0.1cm
\tikzstyle{s1}=[color=red,rectangle,inner sep=3.5]
\tikzstyle{c3}=[circle,inner sep={\ms/8},minimum size=5*\ms]
\tikzstyle{c2}=[circle,inner sep={\ms/8},minimum size=3*\ms]
\tikzstyle{c1}=[circle,inner sep={\ms/8},minimum size=2*\ms]


\coordinate (a1) at  (1,2);
\coordinate (a2) at (1.5,{(1-(1-0.5)/2)*2});
\coordinate (a3) at (2,1);
\coordinate (a4) at (2,0);
\coordinate (a5) at (2,-1);
\coordinate (a6) at (1.5,{(-0.5-(1-0.5)/2)*2});
\coordinate (a7) at (1,-2);
\coordinate (a8) at (0.5,{(-0.5-(1-0.5)/2)*2});
\coordinate (a9) at (0,-1);
\coordinate (a10) at (0,0);
\coordinate (a11) at (0,1);
\coordinate (a12) at (0.5,{(1-(1-0.5)/2)*2});
\coordinate (a13) at (1,0);


\draw [color=orange] (a1) -- (a3);
\draw [color=blue] (a3) -- (a5);
\draw [color=red] (a5) -- (a7);
\draw [color=green] (a7) -- (a9);
\draw [color=gray] (a9) -- (a11);
\draw [color=magenta] (a11) -- (a1);
\draw [color=cyan] (a10) -- (a4);


\draw (a1) coordinate[c2,fill=orange,label=above:$a_1$];
\draw (a1) coordinate[c1,fill=gray];

\draw (a2) coordinate[c1,fill=orange,label=above right:$a_2$];

\draw (a3) coordinate[c2,fill=blue,label=right:$a_3$];
\draw (a3) coordinate[c1,fill=orange];

\draw (a4) coordinate[c2,fill=cyan,label=right:$a_4$];
\draw (a4) coordinate[c1,fill=blue];

\draw (a5) coordinate[c2,fill=red,label=right:$a_5$];
\draw (a5) coordinate[c1,fill=blue];

\draw (a6) coordinate[c1,fill=red,label=below right:$a_6$];

\draw (a7) coordinate[c2,fill=green,label=below:$a_7$];
\draw (a7) coordinate[c1,fill=red];

\draw (a8) coordinate[c1,fill=green,label=below left:$a_8$];

\draw (a9) coordinate[c2,fill=gray,label=left:$a_9$];
\draw (a9) coordinate[c1,fill=green];

\draw (a10) coordinate[c2,fill=gray,label=left:$a_{10}$];
\draw (a10) coordinate[c1,fill=cyan];

\draw (a11) coordinate[c2,fill=magenta,label=left:$a_{11}$];
\draw (a11) coordinate[c1,fill=gray];

\draw (a12) coordinate[c1,fill=magenta,label=above left:$a_{12}$];

\draw (a13) coordinate[c1,fill=cyan,label=above:$a_{13}$];

\end{tikzpicture}
&
\multicolumn{2}{c}{
\begin{tikzpicture}  [scale=1]

\tikzstyle{every path}=[line width=1pt]

\newdimen\ms
\ms=0.1cm
\tikzstyle{s1}=[color=red,rectangle,inner sep=3.5]
\tikzstyle{c3}=[circle,inner sep={\ms/8},minimum size=4*\ms]
\tikzstyle{c2}=[circle,inner sep={\ms/8},minimum size=3*\ms]
\tikzstyle{c1}=[circle,inner sep={\ms/8},minimum size=2*\ms]


\coordinate (a1) at  (1,2);
\coordinate (a2) at (1.5,{(1-(1-0.5)/2)*2});
\coordinate (a3) at (2,1);
\coordinate (a4) at (2,0);
\coordinate (a5) at (2,-1);
\coordinate (a6) at (1.5,{(-0.5-(1-0.5)/2)*2});
\coordinate (a7) at (1,-2);
\coordinate (a8) at (0.5,{(-0.5-(1-0.5)/2)*2});
\coordinate (a9) at (0,-1);
\coordinate (a10) at (0,0);
\coordinate (a11) at (0,1);
\coordinate (a12) at (0.5,{(1-(1-0.5)/2)*2});
\coordinate (a13) at (1,0);

\coordinate (a14) at (4,0);

\coordinate (a21) at  ({6+1},2);
\coordinate (a27) at ({6+1},-2);


\draw [color=orange] (a1) -- (a3);
\draw [color=blue] (a3) -- (a5);
\draw [color=red] (a5) -- (a7);
\draw [color=green] (a7) -- (a9);
\draw [color=gray] (a9) -- (a11);
\draw [color=magenta] (a11) -- (a1);
\draw [color=cyan] (a10) -- (a4);

\draw [color=lime] (a1) to [out=0,in={90+45}] (a14) to [out={180+90+45},in=180]  (a27);
\draw [color=olive] (a7) to [out=0,in={180+45}] (a14) to [out=45,in=180] (a21);



\draw (a1) coordinate[c3,fill=lime,label=above:$a_1$];
\draw (a1) coordinate[c2,fill=orange];
\draw (a1) coordinate[c1,fill=gray];

\draw (a2) coordinate[c1,fill=orange,label=below left:$a_2$];

\draw (a3) coordinate[c2,fill=blue,label=right:$a_3$];
\draw (a3) coordinate[c1,fill=orange];

\draw (a4) coordinate[c2,fill=cyan,label=right:$a_4$];
\draw (a4) coordinate[c1,fill=blue];

\draw (a5) coordinate[c2,fill=red,label=right:$a_5$];
\draw (a5) coordinate[c1,fill=blue];

\draw (a6) coordinate[c1,fill=red,label=above left:$a_6$];

\draw (a7) coordinate[c3,fill=olive,label=below:$a_7$];
\draw (a7) coordinate[c2,fill=green];
\draw (a7) coordinate[c1,fill=red];

\draw (a8) coordinate[c1,fill=green,label=below left:$a_8$];

\draw (a9) coordinate[c2,fill=gray,label=left:$a_9$];
\draw (a9) coordinate[c1,fill=green];

\draw (a10) coordinate[c2,fill=gray,label=left:$a_{10}$];
\draw (a10) coordinate[c1,fill=cyan];

\draw (a11) coordinate[c2,fill=magenta,label=left:$a_{11}$];
\draw (a11) coordinate[c1,fill=gray];

\draw (a12) coordinate[c1,fill=magenta,label=above left:$a_{12}$];

\draw (a13) coordinate[c1,fill=cyan,label=above:$a_{13}$];

\draw (a14) coordinate[c2,fill=lime,label=above:$c$];
\draw (a14) coordinate[c1,fill=olive];

\draw (a21) coordinate[c1,fill=olive,label=above:$b_1$];

\draw (a27) coordinate[c1,fill=lime,label=below:$b_7$];

%
%
%
%
%
%
%
%
%
%
%
%

\end{tikzpicture}
}
\\
(d)&\multicolumn{2}{c}{(e)}
\\
\multicolumn{3}{c}{
\begin{tikzpicture}  [scale=1]

\tikzstyle{every path}=[line width=1pt]

\newdimen\ms
\ms=0.1cm
\tikzstyle{s1}=[color=red,rectangle,inner sep=3.5]
\tikzstyle{c3}=[circle,inner sep={\ms/8},minimum size=4*\ms]
\tikzstyle{c2}=[circle,inner sep={\ms/8},minimum size=3*\ms]
\tikzstyle{c1}=[circle,inner sep={\ms/8},minimum size=2*\ms]


\coordinate (a1) at  (1,2);
\coordinate (a2) at (1.5,{(1-(1-0.5)/2)*2});
\coordinate (a3) at (2,1);
\coordinate (a4) at (2,0);
\coordinate (a5) at (2,-1);
\coordinate (a6) at (1.5,{(-0.5-(1-0.5)/2)*2});
\coordinate (a7) at (1,-2);
\coordinate (a8) at (0.5,{(-0.5-(1-0.5)/2)*2});
\coordinate (a9) at (0,-1);
\coordinate (a10) at (0,0);
\coordinate (a11) at (0,1);
\coordinate (a12) at (0.5,{(1-(1-0.5)/2)*2});
\coordinate (a13) at (1,0);

\coordinate (a14) at (4,0);

\coordinate (a21) at  ({6+1},2);
\coordinate (a22) at ({6+1.5},{(1-(1-0.5)/2)*2});
\coordinate (a23) at ({6+2},1);
\coordinate (a24) at ({6+2},0);
\coordinate (a25) at ({6+2},-1);
\coordinate (a26) at ({6+1.5},{(-0.5-(1-0.5)/2)*2});
\coordinate (a27) at ({6+1},-2);
\coordinate (a28) at ({6+0.5},{(-0.5-(1-0.5)/2)*2});
\coordinate (a29) at ({6+0},-1);
\coordinate (a210) at ({6+0},0);
\coordinate (a211) at ({6+0},1);
\coordinate (a212) at ({6+0.5},{(1-(1-0.5)/2)*2});
\coordinate (a213) at ({6+1},0);


\draw [color=orange] (a1) -- (a3);
\draw [color=blue] (a3) -- (a5);
\draw [color=red] (a5) -- (a7);
\draw [color=green] (a7) -- (a9);
\draw [color=gray] (a9) -- (a11);
\draw [color=magenta] (a11) -- (a1);
\draw [color=cyan] (a10) -- (a4);

\draw [color=lime] (a1) to [out=0,in={90+45}] (a14) to [out={180+90+45},in=180]  (a27);
\draw [color=olive] (a7) to [out=0,in={180+45}] (a14) to [out=45,in=180] (a21);

\draw [color=orange!40!white] (a21) -- (a23);
\draw [color=blue!40!white] (a23) -- (a25);
\draw [color=red!40!white] (a25) -- (a27);
\draw [color=green!40!white] (a27) -- (a29);
\draw [color=gray!40!white] (a29) -- (a211);
\draw [color=magenta!40!white] (a211) -- (a21);
\draw [color=cyan!40!white] (a210) -- (a24);


\draw (a1) coordinate[c3,fill=lime,label=above:$a_1$];
\draw (a1) coordinate[c2,fill=orange];
\draw (a1) coordinate[c1,fill=gray];

\draw (a2) coordinate[c1,fill=orange,label=below left:$a_2$];

\draw (a3) coordinate[c2,fill=blue,label=right:$a_3$];
\draw (a3) coordinate[c1,fill=orange];

\draw (a4) coordinate[c2,fill=cyan,label=right:$a_4$];
\draw (a4) coordinate[c1,fill=blue];

\draw (a5) coordinate[c2,fill=red,label=right:$a_5$];
\draw (a5) coordinate[c1,fill=blue];

\draw (a6) coordinate[c1,fill=red,label=above left:$a_6$];

\draw (a7) coordinate[c3,fill=olive,label=below:$a_7$];
\draw (a7) coordinate[c2,fill=green];
\draw (a7) coordinate[c1,fill=red];

\draw (a8) coordinate[c1,fill=green,label=below left:$a_8$];

\draw (a9) coordinate[c2,fill=gray,label=left:$a_9$];
\draw (a9) coordinate[c1,fill=green];

\draw (a10) coordinate[c2,fill=gray,label=left:$a_{10}$];
\draw (a10) coordinate[c1,fill=cyan];

\draw (a11) coordinate[c2,fill=magenta,label=left:$a_{11}$];
\draw (a11) coordinate[c1,fill=gray];

\draw (a12) coordinate[c1,fill=magenta,label=above left:$a_{12}$];

\draw (a13) coordinate[c1,fill=cyan,label=above:$a_{13}$];

\draw (a14) coordinate[c2,fill=lime,label=above:$c$];
\draw (a14) coordinate[c1,fill=olive];

\draw (a21) coordinate[c3,fill=olive,label=above:$b_1$];

\draw (a27) coordinate[c3,fill=lime,label=below:$b_7$];

\draw (a21) coordinate[c2,fill=orange!40!white];
\draw (a21) coordinate[c1,fill=gray!40!white];

\draw (a22) coordinate[c1,fill=orange!40!white,label=above right:$b_2$];

\draw (a23) coordinate[c2,fill=blue!40!white,label=right:$b_3$];
\draw (a23) coordinate[c1,fill=orange!40!white];

\draw (a24) coordinate[c2,fill=cyan!40!white,label=right:$b_4$];
\draw (a24) coordinate[c1,fill=blue!40!white];

\draw (a25) coordinate[c2,fill=red!40!white,label=right:$b_5$];
\draw (a25) coordinate[c1,fill=blue!40!white];

\draw (a26) coordinate[c1,fill=red!40!white,label=below right:$b_6$];

\draw (a27) coordinate[c2,fill=green!40!white];
\draw (a27) coordinate[c1,fill=red!40!white];

\draw (a28) coordinate[c1,fill=green!40!white,label=above right:$b_8$];

\draw (a29) coordinate[c2,fill=gray!40!white,label=left:$b_9$];
\draw (a29) coordinate[c1,fill=green!40!white];

\draw (a210) coordinate[c2,fill=gray!40!white,label=left:$b_{10}$];
\draw (a210) coordinate[c1,fill=cyan!40!white];

\draw (a211) coordinate[c2,fill=magenta!40!white,label=left:$b_{11}$];
\draw (a211) coordinate[c1,fill=gray!40!white];

\draw (a212) coordinate[c1,fill=magenta!40!white,label=below right:$b_{12}$];

\draw (a213) coordinate[c1,fill=cyan!40!white,label=above:$b_{13}$];

\end{tikzpicture}
}
\\
\multicolumn{3}{c}{(f)}
\end{tabular}
\end{center}
\caption{\label{2020-f1-gd}
Orthogonality hypergraphs representing historic configurations
of complementary observables arranged in 3-element blocks.
(\textbf{a}) Two renditions of the firefly gadget;
(\textbf{b}) Two firefly gadgets with a common block;
(\textbf{c}) house/pentagon/pentagram gadget;
(\textbf{d}) Specker bug/cat's cradle logic;
(\textbf{e}) extended Specker bug logic with a 1-1/true-implies-true property;
(\textbf{f}) combo of interlinked Specker bugs with a non-separable set of two-valued states (at $a_1$-$b_1$ as well as $a_7$-$b_7$).
}
\end{figure*}

\subsubsection{Probabilities on the House/Pentagon/Pentagram}
\label{2017-b-kcbsi}


Admissibility of two-valued states imposes conditions and restrictions on the two-valued states already for a single context (Boolean subalgebra):
if one atom is assigned the value 1, all other
atoms have to have value assignment(s) $0$.
This is even more so for intertwining contexts.
For the sake of an example, consider two firefly logics pasted along an entire block,
in short
$
\{
{\color{orange}\{a_1,a_2,a_3\}}
$, $
{\color{blue}\{a_3,a_4,a_5\}}
$, $
{\color{red}\{a_5,a_6,a_7\}}
\}
$.
The orthogonality hypergraph is depicted in Figure~\ref{2020-f1-gd}b (see also Figure~12.5 in~\cite{svozil-2016-pu-book}).
For such logic we can state a ``true-and-true implies true'' rule:
if the two-valued measure at the ``outer extremities'' is 1, then it must be 1 at its center atom.

History:
We can pursue this path of ever-increasing restrictions through the construction of pasted; that is, intertwined, contexts.
Let us proceed by stitching/pasting more firefly logics together cyclically in ``closed circles''.
The two simplest such pastings--two firefly logics forming either a triangle or a square Greechie orthogonal hypergraph--have no realization in three dimensional Hilbert space.
The next diagram realizably is obtained by pasting three firefly logics.
It is the house/pentagon/pentagram
(the graphs of the pentagon and the pentagram are related by an isomorphic transformation of the vertices
$a_1 \mapsto a_1$, and
$a_9 \mapsto a_5 \mapsto a_3 \mapsto a_7 \mapsto a_9$)
logic also denoted as  orthomodular house~(p.~46, Figure~4.4 in \cite{kalmbach-83})
and discussed in Ref.~\cite{Beltrametti-1995};
see also Birkhoff's distributivity criterion~(p.~90, Theorem~33 in \cite{beran}),
stating that in particular,  if some lattice contains a pentagon as a sublattice,
then it is not distributive~\cite{birkhoff_1934}.

This cyclic logic is, in short,
$
\{
{\color{orange}\{a_1,a_2,a_3\}}
$, $
{\color{blue}\{a_3,a_4,a_5\}}
$, $
{\color{red}\{a_5,a_6,a_7\}},
{\color{green}\{a_5,a_6,a_7\}},
{\color{gray}\{a_5,a_6,a_7\}}
\}
$.
The~orthogonality hypergraph of the house/pentagon/pentagram gadget is depicted in Figure~\ref{2020-f1-gd}c.

Classical interpretations:
Such house/pentagon/pentagram logics allow ``exotic'' probability measures~\cite{wright:pent}:
as pointed out by Wright~\cite{wright:pent} (p.~268) the pentagon allows 11 ``ordinary'' two-valued states $v_1,\ldots ,v_{11}$,
and one ``exotic'' dispersionless state $v_e$,
which was shown by Wright to have neither a classical nor a quantum interpretation;
all defined on the 10 atoms $a_1, \ldots , a_{10}$.
They are enumerated in Table~\ref{2017-b-t-sotp}.
These 11 ``ordinary'' two-valued states  directly translate into the classical probabilities~(Figure~12.4 in \cite{svozil-2016-pu-book}).
 \begin{table}
\centering
 \caption{\label{2017-b-t-sotp} (Color online) Two-valued states on the pentagon.}
\begin{ruledtabular}
 \begin{tabular}{ccccccccccc}
\textbf{\#}
&
\boldmath{$a_1$}&
\boldmath{$a_2$}&
\boldmath{$a_3$}&
\boldmath{$a_4$}&
\boldmath{$a_5$}&
\boldmath{$a_6$}&
\boldmath{$a_7$}&
\boldmath{$a_8$}&
\boldmath{$a_9$}&
\boldmath{$a_{10}$}
\\
\hline 
$v_1$&1&  0&  0&   1&  0&  1&  0&  1&  0&  0 \\
$v_2$&1&  0&  0&   0&  1&  0&  0&  1&  0&  0\\
$v_3$&1&  0&  0&   1&  0&  0&  1&  0&  0&  0\\
$v_4$&0&  0& 1&    0&  0&  1&  0&  1&  0& 1\\
$v_5$&0&  0& 1&    0&  0&  0&  1&  0&  0& 1\\
$v_6$&0&  0& 1&    0&  0&  1&  0&  0&  1&  0\\
$v_7$&0&  1&  0&   0&  1&  0&  0&  1&  0& 1\\
$v_8$&0&  1&  0&   0&  1&  0&  0&  0&  1&  0\\
$v_9$&0&  1&  0&   1&  0&  0&  1&  0&  0& 1\\
$v_{10}$&0& 1&0&   1&  0&  1&  0&  0&  1&  0\\
$v_{11}$&0& 1&0&   1&  0&  1&  0&  1&  0& 1 \\
\hline 
$v_e$&$\frac{1}{2}$& 0&  $\frac{1}{2}$& 0&  $\frac{1}{2}$& 0&  $\frac{1}{2}$& 0&  $\frac{1}{2}$& 0
 \end{tabular}
 \end{ruledtabular}
 \end{table}

\vspace{-6pt}
The pentagon logic has quasi-classical realizations in terms of partition logics~\cite{dvur-pul-svo,svozil-2001-eua,svozil-2008-ql},
such as generalized urn models~\cite{wright:pent,wright} or automaton logics~\cite{schaller-92,svozil-93,schaller-95,schaller-96}.
The canonical partition logic (up to permutations)  is
\begin{equation}
\begin{split}
\{
\{
\{1,2,3\},\{7,8,9,10,11\},\{4,5,6\}
\}
,\\
\{
\{4,5,6\},\{1,3,9,10,11\},\{2,7,8\}
\}
,\\
\{
\{2,7,8\},\{1,4,6,10,11\},\{3,5,9\}
\}
,\\
\{
\{3,5,9\},\{1,2,4,7,11\},\{6,8,10\}
\}
,\\
\{
\{6,8,10\},\{4,5,7,9,11\},\{1,2,3\}
\}
\}.
\end{split}
\end{equation}

{The classical probabilities can be directly read off from the canonical partition logic: they are thee respective convex combination
of the eleven two-valued states ($0 \le \lambda_i \le 1$ and $\sum_{i=1}^{11}\lambda_i=1$): on $a_1$, $a_2$ and $a_3$ it is, for instance,
$p_1=\lambda_1+\lambda_2+\lambda_3$,
$p_1=\lambda_7+\lambda_8+\lambda_9+\lambda_{10}+\lambda_{11}$,
$p_3= \lambda_4+\lambda_5+\lambda_6$,~respectively.}

An early realization in terms of three-dimensional (quantum) Hilbert space can, for instance, be found in Ref.~\cite{svozil-tkadlec} (pp.~5392, 5393);
other such parametrizations are discussed in Refs.~\cite{Klyachko-2008,Bub-2009,Bub-2010,Badziag-2011}.

{The full hull problem, including all joint expectations of dichotomic $\pm 1$ observables yields 64 inequalities.
The full hull computations for the probabilities $p_1, \ldots , p_{10}$
on all atoms $a_1, \ldots , a_{10}$ reduces to 16 inequalities.
If one considers only the five probabilities on the intertwining atoms,
then the Bub-Stairs inequalit $p_1+p_3+p_5+p_7+p_9 \le 2$ results~\cite{Bub-2009,Bub-2010,Badziag-2011}.
Concentration on the four non-intertwining atoms yields $p_2+p_4+p_6+p_8+p_{10} \ge 1$.
Limiting the hull computation to adjacent pair expectations of dichotomic $\pm 1$ observables
yields the Klyachko-Can-Biniciogolu-Shumovsky inequality~\cite{Klyachko-2008}.}

\subsubsection{Deterministic Predictions and Probabilities on the Specker Bug}
\label{2017-b-s-tif}
\label{2017-b-speckerbug}

Next, we shall study a quasiclassical collection of observables with the property that the preparation of the system in a particular state
entails the prediction that another particular observable must have a particular null/zero outcome.
The respective quantum observables violate that classical constraint by predicting a non-zero probability of the outcome.
Therefore, by observing an outcome one could ``certify'' non-classicality relative to the (idealistic) assumptions
that this particular quasiclassical collection of counterfactual observables ``exists'' and has relevance to the situation.

History:
The pasting of two house/pentagon/pentagram logics with three common contexts results in ever tighter conditions for two-valued measures and thus truth-value assignments:
consider the orthogonality hypergraph of a logic drawn in Figure~\ref{2020-f1-gd}d.
Specker~\cite{Specker-priv} called this the ``K\"afer'' (German for bug) because its graph remotely (for Specker)
resembled the shape of a bug.
In 1965 it was introduced by Kochen and Specker~(Figure~1, p.~182 in \cite{kochen2})
who subsequently used it as a subgraph in the diagrams $\Gamma_1$, $\Gamma_2$ and $\Gamma_3$ demonstrating
the existence of quantum propositional structures with the ``true implies true'' property (cf. Section \ref{2017-b-s-tit}),
the non-existence of any two-valued state (cf. Section~\ref{2017-b-c-lwtvs}),
and the existence of a non-separating set of two-valued states (cf. Section~\ref{2017-b-bugscombino}), respectively~\cite{kochen1}.

Pitowsky called this gadget ``cat's cradle''~\cite{Pitowsky2003395,pitowsky-06}.
See also
Figure~1, p.~123 in~\cite{Greechie1974} (reprinted in Ref.~\cite{Greechie-Suppes1976}),
a subgraph in Figure~21, pp.~126--127 in \cite{redhead},
Figure~B.l. p.~64 in \cite{Belinfante-73},
pp.~588--589 in \cite{stairs83},
Figure~2, p.~446 in \cite{clifton-93},  and
Figure~2.4.6,  p.~39 in \cite{pulmannova-91} for early discussions.

The Specker bug/cat's cradle logic is a pasting of seven intertwined contexts
$\{{\color{orange}\{a_1,a_2,a_3\} }
$, $
{\color{blue}\{a_3,a_4,a_5\} }
$, $
{\color{red}\{a_5,a_6,a_7\} }
$, $
{\color{green}\{a_7,a_8,a_9\} }
$, $
{\color{gray}\{a_{9},a_{10},a_{11}\} }
$, $
{\color{magenta}\{a_{11},a_{12},a_1\} }
$, $
{\color{cyan}\{a_4,a_{13},a_{10}\}}
\}$
from two houses/pentagons/pentagrams
$\{{\color{orange}\{a_1,a_2,a_3\} }
$, $
{\color{blue}\{a_3,a_4,a_5\} }
$, $
{\color{gray}\{a_{9},a_{10},a_{11}\} }
$, $
{\color{magenta}\{a_{11},a_{12},a_1\} }
$, $
{\color{cyan}\{a_4,a_{13},a_{10}\}}
\}$
and
$\{
{\color{blue}\{a_3,a_4,a_5\} }
$, $
{\color{red}\{a_5,a_6,a_7\} }
$, $
{\color{green}\{a_7,a_8,a_9\} }
$, $
{\color{gray}\{a_{9},a_{10},a_{11}\} }
$, $
{\color{cyan}\{a_4,a_{13},a_{10}\}}
\}$ with three common blocks
${\color{blue}\{a_3,a_4,a_5\} }$,
${\color{gray}\{a_{9},a_{10},a_{11}\} }$, and
${\color{cyan}\{a_4,a_{13},a_{10}\}}$.
The orthogonality hypergraph of the Specker bug/cat's cradle gadget is depicted in Figure~\ref{2020-f1-gd}d.

Classical interpretations:
The Specker bug/cat's cradle logic allows 14 two-valued states which are listed in Table~\ref{2001-cesena-t2}.

An early realization in terms of partition logics
can be found in Refs.~\cite{svozil-tkadlec,svozil-2008-ql}.
An explicit faithful orthogonal realization  in $\mathbb{R}^3$
consisting of  13 suitably chosen projections~(p.~206, Figure~1 in \cite{tkadlec-96})
(see also~(Figure~4, p.~5387 in \cite{svozil-tkadlec})).
It is not too difficult to read the canonical partition logic (up to permutations) off
the set of all 14 two-valued states which are tabulated in~Table \ref{2001-cesena-t2}:
\begin{equation}
\begin{split}
\{
\{
\{1,2,3\},\{4,5,6,7,8,9\},\{10,11,12,13,14\}
\}
,\\
\{
\{10,11,12,13,14\},\{2,6,7,8\},\{1,3,4,5,9\}
\}
,\\
\{
\{1,3,4,5,9\},\{2,6,8,11,12,14\},\{7,10,13\}
\}
,\\
\{
\{7,10,13\},\{3,5,8,9,11,14\},\{1,2,4,6,12\}
\}
,\\
\{
\{1,2,4,6,12\},\{3,9,13,14\},\{5,7,8,10,11\}
\}
,\\
\{
\{5,7,8,10,11\},\{4,6,9,12,13,14\},\{1,2,3\}
\}
,\\
\{
\{2,6,7,8\},\{1,4,5,10,11,12\},\{3,9,13,14\}
\}
\}.
\end{split}
\end{equation}

 \begin{table}
\centering
\caption{\label{2001-cesena-t2}  The 14 two-valued states on the Specker bug (cat's cradle) logic.}
\begin{ruledtabular}
 \begin{tabular}{ccccccccccccccccccccccc}
\# &$a_1$&$a_2$&$a_3$&$a_4$&$a_5$&$a_6$&$a_7$&$a_8$&$a_9$&$a_{10}$&$a_{11}$&$a_{12}$&$a_{13}$\\
\hline 
$v_1 $&1&0&0&0&1&0&0&0&1&0&0&0&1         \\
$v_2 $&1&0&0&1&0&1&0&0&1&0&0&0&0          \\
$v_3 $&1&0&0&0&1&0&0&1&0&1&0&0&0         \\
$v_4 $&0&1&0&0&1&0&0&0&1&0&0&1&1         \\
$v_5 $&0&1&0&0&1&0&0&1&0&0&1&0&1         \\
$v_6 $&0&1&0&1&0&1&0&0&1&0&0&1&0          \\
$v_7 $&0&1&0&1&0&0&1&0&0&0&1&0&0          \\
$v_8 $&0&1&0&1&0&1&0&1&0&0&1&0&0          \\
$v_9 $&0&1&0&0&1&0&0&1&0&1&0&1&0         \\
$v_{10}$&0&0&1&0&0&0&1&0&0&0&1&0&1         \\
$v_{11}$&0&0&1&0&0&1&0&1&0&0&1&0&1         \\
$v_{12}$&0&0&1&0&0&1&0&0&1&0&0&1&1         \\
$v_{13}$&0&0&1&0&0&0&1&0&0&1&0&1&0         \\
$v_{14}$&0&0&1&0&0&1&0&1&0&1&0&1&0
 \end{tabular}
\end{ruledtabular}
 \end{table}

\vspace{-6pt}
Deterministic prediction:
As  pointed out by Greechie~(Figure~1, p.~122--123 in~\cite{Greechie1974}  and reprinted in Ref.~\cite{Greechie-Suppes1976}),
Pt{\'{a}}k and Pulmannov{\'{a}}~(Figure~2.4.6, p.~39 in \cite{pulmannova-91})
as well as
Pitowsky~\cite{Pitowsky2003395,pitowsky-06}
the reduction of some probabilities of atoms at intertwined contexts yields~(p.~285, Equation~(11.2) in \cite{svozil-2016-s})
\begin{equation}
p_1+p_7=\frac{3}{2}- \frac{1}{2}\left(p_{12}+p_{13}+p_2+p_6+p_8\right)\le \frac{3}{2}
.
\label{2015-s-e2}
\end{equation}

A tighter approximation comes from the explicit parameterization of the classical probabilities on the atoms $a_1$ and $a_7$,
derivable from all the mutually disjoined two-valued states which do not vanish on those atoms~(Figure~12.9 in \cite{svozil-2016-pu-book}):
$p_1= \lambda_1 + \lambda_2 + \lambda_3$, and
$p_7= \lambda_7 + \lambda_{10} + \lambda_{13}$.
Because of additivity the 14 positive weights $\lambda_{1},\ldots ,\lambda_{14} \ge 0$
must add up to 1; that is, $\sum_{i=1}^{14} \lambda_{i}=1$.
Therefore,
\begin{equation}
\begin{split}
p_{1} + p_{7}
= \lambda_1 + \lambda_2 + \lambda_3 + \lambda_7 + \lambda_{10} + \lambda_{13}
\le \sum_{i=1}^{14} \lambda_{i}=1
.
\end{split}
\label{2017-b-spbp17}
\end{equation}

For two-valued measures this yields the ``1-0'' or ``true implies false'' rule~\cite{svozil-2006-omni}:
if $a_1$ is true, then $a_7$ must be false.
This property
has been exploited in Kochen and Specker's graph~$\Gamma_1$ in \cite{kochen1}
implementing a ``1-1'' or ``true implies true'' rule, as well as
to construct both a $\Gamma_3$-logic with a non-separating,
as well as $\Gamma_2$ which does not support a two-valued state.
The former ``1-1'' or ``true implies true''  case will be discussed in the next section.

Probabilistic prediction:
The hull problem yields 23 facet inequalities; one of them relating $p_1$  to $p_7$:
$p_1 + p_2   + p_7  + p_6 \ge  1  + p_4$,
which is satisfied, since, by subadditivity,
$p_1 + p_2 = 1 - p_3$,
$p_7 + p_6 = 1 - p_5$,
and
$p_4 = 1 - p_5 - p_3$.
A restricted hull calculation for the joint expectations on the six edges of the orthogonality hypergraph yields 18 inequalities; among them
\begin{equation}
E_{13}  + E_{57} + E_{9,11} \le   E_{35} + E_{79} +   E_{11,1}
.
\label{2017-b-sp-hcrc}
\end{equation}

Configurations in arbitrary dimensions greater than two can be found in Ref.~\cite{2018-minimalYIYS}, where also another
interesting ``true-implies-triple false'' structure by Yu and Oh~(Figure~2 in \cite{Yu-2012}) is reviewed.
Geometric constraints do not allow faithful orthogonal representations of the Specker bug logic which
``perform better'' than with probability  $\frac{1}{9}$~\cite{Belinfante-73,redhead,cabello-1994,Cabello-1996-diss,svozil-tkadlec}, so that quantum systems prepared in a pure state corresponding to $a_1$
are measured in another pure state $a_7$ with a probability higher than $\frac{1}{9}$.
More recently ``true implies false'' gadgets~\cite{2015-AnalyticKS,svozil-2018-whycontexts,Ramanathan-18}
allow ``tunable'' relative angles between preparation and measurement states,
so that the relative quantum advantage
can be made higher.
In particular, Ref.~\cite{Ramanathan-18}
contains a collection of 34 observables
in 21 intertwined contexts
 {\color{orange} $\{ a_{ 3}, a_{ 7}, a_{ 6}  \}$},
 {\color{blue}  $\{ a_{ 3}, a_{14}, a_{15} \}$},
 {\color{green}  $\{ a_{ 3}, a_{10}, a_{11} \}$},
 {\color{magenta}  $\{ a_{ 8}, a_{ 9}, a_{20} \}$},
 {\color{black}  $\{ a_{ 4}, a_{18}, a_{21} \}$},
 {\color{brown}  $\{ a_{16}, a_{17}, a_{20} \}$},
 {\color{gray}  $\{ a_{12}, a_{13}, a_{20} \}$},
 {\color{violet}  $\{ a_{ 2}, a_{ 5}, a_{19} \}$},
 {\color{TealBlue}  $\{ a_{ 7}, a_{24}, a_{ 8} \}$},
 {\color{Apricot}  $\{ a_{ 5}, a_{ 6}, a_{23} \}$},
 {\color{MidnightBlue}  $\{ a_{ 3}, c, a_{ 1}      \}$},
 {\color{Mulberry}  $\{ a_{10}, a_{34}, a_{ 9} \}$},
 {\color{BrickRed}  $\{ a_{ 2}, a_{29}, a_{ 1} \}$},
 {\color{Emerald}  $\{ a_{19}, a_{30}, a_{22} \}$},
 {\color{YellowGreen}  $\{ a_{ 1}, a_{32}, a_{ 4} \}$},
 {\color{Tan}  $\{ a_{22}, a_{33}, a_{21} \}$},
 {\color{RawSienna}  $\{ a_{20}, c, a_{22}      \}$},
 {\color{SpringGreen}  $\{ a_{14}, a_{25}, a_{13} \}$},
 {\color{Salmon}  $\{ a_{15}, a_{26}, a_{16} \}$},
 {\color{Fuchsia}  $\{ a_{11}, a_{27}, a_{12} \}$},
 {\color{CornflowerBlue}  $\{ a_{17}, a_{31}, a_{18} \}$}.
Its orthogonality hypergraph is drawn in Figure~\ref{2019-c-HH10}.
The set of 89 2-valued states induce a
``true-implies-false'' property on the two pairs $a_1$-$a_{20}$ and $a_1$-$a_{22}$, respectively.
\begin{figure}[htb] 
\centering
\begin{tikzpicture}  [scale=0.5, rotate= 0]

\newdimen\E
\E=0.05 cm
\tikzstyle{every path}=[line width=1 pt]

\tikzstyle{c4}=[circle,inner sep=4pt,minimum size=7pt]
\tikzstyle{c3}=[circle,inner sep=3pt,minimum size=5pt]
\tikzstyle{c2}=[circle,inner sep=2pt,minimum size=3pt]
\tikzstyle{c1}=[circle,inner sep=1.5pt,minimum size=2pt]
\tikzstyle{c0}=[circle,inner sep=0.5pt,minimum size=1pt]


\path
( {4 *2.68838},{4 * 2.05247})   coordinate(1)
( {4 *3.41896},{4 *    0.966553})   coordinate(4)
( {4 *3.13324},{4 * 3.17128})   coordinate(5)
( {4 *2.04217},{4 * 3.18563})   coordinate(6)
( {4 *1.37051},{4 * 1.95877})   coordinate(8)
( {4 *0.8},{4 * 2.80416})   coordinate(11)
( {4 *0.00},{4 * 1.05577})   coordinate(12)
( {4 *0.40372},{4 * 1.55514})   coordinate(14)
( {4 *1.26923},{4 *    0.189747})   coordinate(16)
( {4 *1.83338},{4 * 0.})   coordinate(17)
( {4 *3.11765},{4 * 0.0226601})   coordinate(18)
( {4 *3.45895},{4 *    2.21445})   coordinate(19)
( {4 *1.82887},{4 * 0.922707})   coordinate(20)
( {4 *3.58461},{4 * 0.431072})   coordinate(21)
( {4 *2.79057},{4 *    1.13936})   coordinate(22)
;

\draw [line width=2mm,color=gray!30] (20) --    (22);
\draw [color=RawSienna]           (20) --    (22) coordinate[c2,fill=RawSienna,draw=RawSienna,pos=0.5,label=270:{\color{black}$c$}]    (31);
\draw [line width=2mm,color=gray!30] (1) --    (31);
\draw [color=MidnightBlue]           (1) --    (31) coordinate[c2,fill=MidnightBlue,draw=MidnightBlue,pos=0.5,label=0:{\color{black}$a_3$}]    (3);
\draw [color=orange]              (3) --  (6) coordinate[c2,fill=orange,draw=orange,pos=0.6,label=0:{\color{black}$a_{7}$}]    (7);
\draw [color=blue]              (3) -- (14) coordinate[c2,fill=blue,draw=blue,pos=0.7,label=95:{\color{black}$a_{15}$}]   (15);
\draw [color=green]               (3) --  (11)  coordinate[c2,fill=green,draw=green,pos=0.2]  (10);
\draw [color=magenta]            (8)  --  (20)  coordinate[c2,fill=magenta,draw=magenta,pos=0.6]   (9);
\draw [color=black]               (4) -- (18) coordinate[c2,fill=black,draw=black,pos=0.5,label=0:{\color{black}$a_{21}$}]   (21);
\draw [color=brown]             (17) --  (20) coordinate[c2,fill=brown,draw=brown,pos=0.5,label=0:{\color{black}$a_{16}$}] (16);
\draw [color=gray]                (12) --  (20)   coordinate[c2,fill=gray,draw=gray,pos=0.5,label=270:{\color{black}$a_{13}$}]  (13);
\draw [color=violet]              (5) --  (19) coordinate[c2,fill=violet,draw=violet,pos=0.5,label=0:{\color{black}$a_{2}$}]   (2);
\draw [color=Apricot]             (5) -- (6)  ;
\draw [color=TealBlue]            (7) --    (8);
\draw [color=MidnightBlue]        (3) --   (1);
\draw [color=Mulberry]            (10) --    (9);
\draw [color=BrickRed]            (2) --    (1);
\draw [color=Emerald]             (19) --    (22);
\draw [color=YellowGreen]          (1) --    (4);
\draw [color=Tan]                 (22) --    (21);
\draw [color=SpringGreen]         (14) --    (13);
\draw [color=Salmon]              (15) --    (16);
\draw [color=Fuchsia]             (11) --    (12);
\draw [color=CornflowerBlue]        (17) --    (18);

 \draw (10) coordinate[c1,fill=Mulberry,label=90:{\color{black}$a_{10}$}];
 \draw (13) coordinate[c1,fill=SpringGreen];
 \draw (15) coordinate[c1,fill=Salmon];
 \draw (9) coordinate[c1,fill=Mulberry,label=260:{\color{black}$a_{9}$}];
 \draw (7) coordinate[c1,fill=TealBlue];
 \draw (2) coordinate[c1,fill=BrickRed];
 \draw (21) coordinate[c1,fill=Tan];
 \draw (16) coordinate[c1,fill=Salmon];

 \draw (11) coordinate[c2,fill=green,draw=green,label=180:{\color{black}$a_{11}$}];
 \draw (11) coordinate[c1,fill=Fuchsia];

 \draw (12) coordinate[c2,fill=gray,draw=gray,label=180:{\color{black}$a_{12}$}];
 \draw (12) coordinate[c1,fill=Fuchsia];

 \draw (14) coordinate[c2,fill=blue,draw=blue];
 \draw (14) coordinate[c1,fill=SpringGreen,label=270:{\color{black}$a_{14}$}];

 \draw (3) coordinate[c4,fill=orange,draw=orange];
 \draw (3) coordinate[c3,fill=blue,draw=blue];
 \draw (3) coordinate[c2,fill=green,draw=green];
 \draw (3) coordinate[c1,fill=MidnightBlue];

 \draw (20) coordinate[c4,fill=magenta,draw=magenta];
 \draw (20) coordinate[c3,fill=brown,draw=brown];
 \draw (20) coordinate[c2,fill=gray,draw=gray,label=80:{\color{black}$a_{20}$}];
 \draw (20) coordinate[c1,fill=RawSienna];

 \draw (6) coordinate[c2,fill=orange,draw=orange,label=90:{\color{black}$a_{6}$}];
 \draw (6) coordinate[c1,fill=Apricot];

 \draw (5) coordinate[c2,fill=violet,draw=violet,label=90:{\color{black}$a_{5}$}];
 \draw (5) coordinate[c1,fill=Apricot];

 \draw (19) coordinate[c2,fill=violet,draw=violet,label=0:{\color{black}$a_{19}$}];
 \draw (19) coordinate[c1,fill=Emerald];

 \draw (4) coordinate[c2,fill=black,draw=black,label=0:{\color{black}$a_{4}$}];
 \draw (4) coordinate[c1,fill=YellowGreen];

 \draw (18) coordinate[c2,fill=black,draw=black,label=270:{\color{black}$a_{18}$}];
 \draw (18) coordinate[c1,fill=CornflowerBlue];

 \draw (17) coordinate[c2,fill=brown,draw=brown,label=270:{\color{black}$a_{17}$}];
 \draw (17) coordinate[c1,fill=CornflowerBlue];

 \draw (8) coordinate[c2,fill=magenta,draw=magenta];
 \draw (8) coordinate[c1,fill=Mulberry,label=0:{\color{black}$a_{8}$}];

 \draw (1) coordinate[c3,fill=MidnightBlue,draw=MidnightBlue,label=0:{\color{black}$a_{1}$}];
 \draw (1) coordinate[c2,fill=violet,draw=YellowGreen];
 \draw (1) coordinate[c1,fill=BrickRed];

 \draw (22) coordinate[c3,fill=RawSienna,draw=RawSienna];
 \draw (22) coordinate[c2,fill=Emerald,draw=Emerald,label=0:{\color{black}$a_{22}$}];
 \draw (22) coordinate[c1,fill=Tan];

 \draw (31) coordinate[c1,fill=MidnightBlue];

\end{tikzpicture}
\caption{\label{2019-c-HH10}
Orthogonality hypergraph~\cite{svozil-2020-c} (some observables which are not essential to the argument are not drawn)
from proof of
Theorem~3 in Ref.~\cite{Ramanathan-18}.
The advantage of this true-implies-false gadget
is a straightforward parametric faithful orthogonal representation allowing angles
$0< \angle a_1,a_{22} \le \frac{\pi}{4}$ radians ($45^\circ$)
of, say, the terminal points $a_1$ and $a_{22}$.
The corresponding logic including the completed set of
34~vertices in
21~blocks is set representable by partition logics because the supported 89~two-valued
states are (color) separable.
It is not too difficult to prove (by~contradiction) that say
if both $a_1$ as well as $a_{22}$
are assumed to be $1$, then
$a_2$,
$a_3$,
$a_4$,
as well as
$a_{19}$,
$a_{20}$
and $a_{21}$
should be $0$.
Therefore,
$a_5$ and
$a_{18}$
would need to be true.
As a result,
$a_{6}$ and
$a_{17}$ would need to be false.
Hence,
$a_{7}$ as well as
$a_{16}$ would be $1$,
rendering
$a_{8}$ and
$a_{15}$ to be $0$.
This would imply
$a_{9}$ as well as
$a_{14}$ to be $1$,
which in turn would demand
$a_{10}$ and
$a_{13}$ to be false.
Therefore,
$a_{11}$ and
$a_{12}$ would have to be $1$, which contradicts the admissibility rules  WAD1{\&}WAD2
for value assignments.
It is also a true-inplies-true gadget for the terminal points
$a_1$-$a_{20}$.
}
\end{figure}

\subsubsection{Deterministic Predictions of Kochen-Specker's $\Gamma_1$ ``True Implies True'' Logic}
\label{2017-b-s-tit}

Here we shall study a quasiclassical collection of observables with the property that the preparation of the system in a particular state
entails the prediction that the outcome of another particular observable must occur with certainty; that is, this outcome must always occur.
The~respective quantum observables violate that classical constraint by predicting a  probability of outcome strictly smaller than one.
Therefore, by observing the absence of an outcome one could ``certify'' non-classicality relative to the (idealistic) assumptions
that this particular  quasiclassical collection of counterfactual observables ``exists'' and has relevance to the situation.

Scheme:
As depicted in Figure~\ref{2020-f-tits-scheme} a scheme involving three cyclically arranged three element contexts can serve as a
``true implies true'' gadget. This triangular scheme has to be adopted by substituting a Specker bug/cat's cradle in order to fulfill
faithful orthogonal representability.
\begin{figure}[htb] 
\centering
\begin{tabular}{ c c c }
\begin{tikzpicture}  [scale=1]

\tikzstyle{every path}=[line width=1pt]

\newdimen\ms
\ms=0.1cm
\tikzstyle{sb}=[color=red,regular polygon,regular polygon sides=6,inner sep=8]
\tikzstyle{s1}=[color=red,rectangle,inner sep=3.5]
\tikzstyle{c5}=[circle,inner sep={\ms/8},minimum size=6*\ms]
\tikzstyle{c4}=[circle,inner sep={\ms/8},minimum size=5*\ms]
\tikzstyle{c3}=[circle,inner sep={\ms/8},minimum size=4*\ms]
\tikzstyle{c2}=[circle,inner sep={\ms/8},minimum size=3*\ms]
\tikzstyle{c1}=[circle,inner sep={\ms/8},minimum size=2*\ms]


\coordinate (a1) at  (1,2);
\coordinate (a2) at (1.5,1);
\coordinate (a3) at (2,0);
\coordinate (a4) at (1,0);
\coordinate (a5) at (0,0);
\coordinate (a6) at (0.5,1);


\draw [color=green] (a1) -- (a3);
\draw [color=blue] (a3) -- (a5);
\draw [color=red] (a5) -- (a1);


\draw (a1) coordinate[c2,fill=red,label=above:$a_1$];
\draw (a1) coordinate[c1,fill=green];

\draw (a2) coordinate[c1,fill=green,label=above right:$a_2$];

\draw (a3) coordinate[c2,fill=green,label=below:$a_3$];
\draw (a3) coordinate[c1,fill=blue];

\draw (a4) coordinate[c1,fill=blue,label=below:$a_4$];

\draw (a5) coordinate[c2,fill=blue,label=below:$a_5$];
\draw (a5) coordinate[c1,fill=red];

\draw (a6) coordinate[c1,fill=red,label=above left:$a_6$];

\end{tikzpicture}
& $\qquad$ &
\begin{tikzpicture}  [scale=1]

\tikzstyle{every path}=[line width=1pt]

\newdimen\ms
\ms=0.1cm
\tikzstyle{sb}=[color=red,regular polygon,regular polygon sides=6,inner sep=8]
\tikzstyle{s1}=[color=red,rectangle,inner sep=3.5]
\tikzstyle{c5}=[circle,inner sep={\ms/8},minimum size=6*\ms]
\tikzstyle{c4}=[circle,inner sep={\ms/8},minimum size=5*\ms]
\tikzstyle{c3}=[circle,inner sep={\ms/8},minimum size=4*\ms]
\tikzstyle{c2}=[circle,inner sep={\ms/8},minimum size=3*\ms]
\tikzstyle{c1}=[circle,inner sep={\ms/8},minimum size=2*\ms]


\coordinate (a1) at  (1,2);
\coordinate (a2) at (1.5,1);
\coordinate (a3) at (2,0);
\coordinate (a4) at (1,0);
\coordinate (a5) at (0,0);
\coordinate (a6) at (0.5,1);


\draw [color=green] (a1) -- (a3);
\draw [color=blue] (a3) -- (a5);
\draw [color=red] (a5) -- (a1);


\draw (a1) coordinate[c2,fill=red,label=above:$a_1$];
\draw (a1) coordinate[c1,fill=green];

\draw (a2) coordinate[c1,fill=green,label=above right:$a_2$];

\draw (a3) coordinate[c2,fill=green,label=below:$a_3$];
\draw (a3) coordinate[c1,fill=blue];

\draw (a4) coordinate[c1,fill=blue,label=below:$a_4$];

\draw (a5) coordinate[c2,fill=blue,label=below:$a_5$];
\draw (a5) coordinate[c1,fill=red];

\draw (a6) coordinate[sb,fill=red];

\end{tikzpicture}
\\
(a)&&(b)
\end{tabular}
\caption{\label{2020-f-tits-scheme}
Orthogonality hypergraph (\textbf{a}) of the scheme of a true implies true gadget realizable by a partition logic but not
as a faithful orthogonal representation in 3-dimensional Hilbert space
(in~three-dimensional Hilbert space all three ``inner'' vertices $a_2$, $a_4$ and $a_6$ would ``collapse''
into any of the three ``outer'' vertices $a_1$, $a_3$ or $a_5$): suppose $a_1=1$; then, according to the admissability
axioms WAD1{\&}WAD2, $a_3=a_5=0$
and thus $a_4=1$.
(\textbf{b}) To achieve a faithful orthogonal representation one could modify this triangular scheme by following
Kochen and Specker~([$\Gamma_1$, p.~68 in \cite{kochen1}): they substituted one of the three contexts, say $\{a_5,a_6,a_1\}$, by a Specker bug
which also provides a ``true implies false'' property at its terminal vertices $a_1$ and $a_5$, and thereby acts just as admissibility rule~WAD1.
}
\end{figure}

History:
A small extension of the Specker bug logic by two contexts extending from $a_1$ and $a_7$,
both intertwining at a point $c$ renders a logic which facilitates that
whenever $a_1$ is true, so must be an atom $b_1$, which is element in the context $\{a_7,c,b_1\}$.
The orthogonality hypergraph of the extended Specker bug (Kochen-Specker's $\Gamma_1$~(\cite{kochen1} (p.~68)))
gadget is depicted in Figure~\ref{2020-f1-gd}e.

Other ``true implies true'' logics were introduced by Belinfante~(Figure~C.l. p.~67 in \cite{Belinfante-73}),
Pitowsky~\cite{Pitowsky-1982-subs} (p.~394),
Clifton~\cite{clifton-93,Johansen-1994,Vermaas-1994},
as well as Cabello and G. Garc{\'{i}}a-Alcaine~(Lemma~1 in \cite{Cabello-1996-bks-fd}).
More recently ``true implies true'' gadgets
allowing arbitrarily small relative angles between preparation and measurement states,
so that the relative quantum advantage
can be made arbitrarily high~\cite{2015-AnalyticKS,svozil-2018-whycontexts}.
Configurations in arbitrary dimensions greater than two can be found in Ref.~\cite{2018-minimalYIYS}.

Classical interpretations:
The reduction of some probabilities of atoms at intertwined contexts yields
($q_1, q_7$ are the probabilities on $b_1, b_7$, respectively),
additionally to Equation~(\ref{2015-s-e2}),
\mbox{$
p_1 - p_7 = q_1 - q_7
$.}
This implies that for all the 112 two-valued states, if $p_1=1$, then [from Equation~(\ref{2015-s-e2})] $p_7=0$,
and $q_1=1$ as well as $q_7 = 1 - q_1 = 0$.

\subsection{Beyond Classical Embedability}

The following examples of observables are situated in-between classical (structure-preserving) faithful embeddability into
Boolean algebras on the one hand, and a total absence of any classical valuations on the other hand.
Kochen-Specker showed that this kind of classical embeddability of a (quantum) logic  can be characterized by a
separability criterion~(Theorem~0 in \cite{kochen1}) related to its set of two-valued states:
a logic has a separable set of two-valued states if any arbitrarily chosen pair $a_i$, $a_j$, $i\neq j$,
of different atoms/elementary observable propositions of that logic  can be ``separated''
by at least one of those states, say $v$ such that ``$v$ discriminates between $a_i$ and $a_j$'';
that is, $v(a_i)  \neq v(a_j)$.

\subsubsection{Deterministic Predictions on a Combo of Two Interlinked Specker Bugs}
\label{2017-b-bugscombino}

The following collection of observables allows a faithful orthogonal representation (in three- and higher-dimensional Hilbert spaces) and thus
a quantum interpretation. However, it does not allow a set of 24 two-valued states, enumerated in Table~\ref{2020-b-t-combo}
which empirically separates any observable from any other:
some observables always have the same classical value, and therefore (unlike quantum observables) cannot be differentiated by any classical means.

\setlength{\tabcolsep}{0.33em}

\begin{table*}
 \begin{center}
 \caption{\label{2020-b-t-combo}  The 24 two-valued states on the interconnected Specker combo $\Gamma_3$~\cite[p.~70]{kochen1}.}
\begin{ruledtabular}
 \begin{tabular}{ccccccccccccccccccccccccccccccccccccccccc}
\# &$a_1$&$a_2$&$a_3$&$a_4$&$ a_{5} $&$  a_{6} $&$ a_{7}$&$ a_{8} $&$ a_{9} $&$ a_{10} $&$  a_{11} $&$ a_{12} $&$ a_{13} $&$ b_{1} $&$  b_{2} $&$ b_{3} $&$ b_{4} $&$ b_{5} $&$ b_{6} $&$ b_{7} $&$ b_{8} $&$ b_{9} $&$  b_{10} $&$ b_{11} $&$ b_{12} $&$ b_{13}$&$ c$\\
\hline
$v_1   $&1 & 0 &  0 &  1 &  0 &  1 &  0 &  0 &  1 & 0 & 0 & 0 & 0 & 1 & 0 & 0 & 0 & 1 & 0 & 0 & 0 & 1 & 0 & 0 & 0 & 1 & 0 \\
$v_2   $&1 & 0 &  0 &  0 &  1 &  0 &  0 &  1 &  0 & 1 & 0 & 0 & 0 & 1 & 0 & 0 & 0 & 1 & 0 & 0 & 0 & 1 & 0 & 0 & 0 & 1 & 0 \\
$v_3   $&1 & 0 &  0 &  0 &  1 &  0 &  0 &  0 &  1 & 0 & 0 & 0 & 1 & 1 & 0 & 0 & 1 & 0 & 1 & 0 & 0 & 1 & 0 & 0 & 0 & 0 & 0 \\
$v_4   $&1 & 0 &  0 &  0 &  1 &  0 &  0 &  0 &  1 & 0 & 0 & 0 & 1 & 1 & 0 & 0 & 0 & 1 & 0 & 0 & 1 & 0 & 1 & 0 & 0 & 0 & 0 \\
$v_5   $&0 & 1 &  0 &  1 &  0 &  1 &  0 &  1 &  0 & 0 & 1 & 0 & 0 & 0 & 1 & 0 & 1 & 0 & 1 & 0 & 1 & 0 & 0 & 1 & 0 & 0 & 1 \\
$v_6   $&0 & 1 &  0 &  1 &  0 &  1 &  0 &  1 &  0 & 0 & 1 & 0 & 0 & 0 & 1 & 0 & 1 & 0 & 1 & 0 & 0 & 1 & 0 & 0 & 1 & 0 & 1 \\
$v_7   $&0 & 1 &  0 &  1 &  0 &  1 &  0 &  1 &  0 & 0 & 1 & 0 & 0 & 0 & 1 & 0 & 0 & 1 & 0 & 0 & 1 & 0 & 1 & 0 & 1 & 0 & 1 \\
$v_8   $&0 & 1 &  0 &  1 &  0 &  1 &  0 &  1 &  0 & 0 & 1 & 0 & 0 & 0 & 0 & 1 & 0 & 0 & 1 & 0 & 1 & 0 & 1 & 0 & 1 & 0 & 1 \\
$v_9   $&0 & 1 &  0 &  1 &  0 &  1 &  0 &  0 &  1 & 0 & 0 & 1 & 0 & 0 & 1 & 0 & 1 & 0 & 1 & 0 & 1 & 0 & 0 & 1 & 0 & 0 & 1 \\
$v_{10}$&0 & 1 &  0 &  1 &  0 &  1 &  0 &  0 &  1 & 0 & 0 & 1 & 0 & 0 & 1 & 0 & 1 & 0 & 1 & 0 & 0 & 1 & 0 & 0 & 1 & 0 & 1 \\
$v_{11}$&0 & 1 &  0 &  1 &  0 &  1 &  0 &  0 &  1 & 0 & 0 & 1 & 0 & 0 & 1 & 0 & 0 & 1 & 0 & 0 & 1 & 0 & 1 & 0 & 1 & 0 & 1 \\
$v_{12}$&0 & 1 &  0 &  1 &  0 &  1 &  0 &  0 &  1 & 0 & 0 & 1 & 0 & 0 & 0 & 1 & 0 & 0 & 1 & 0 & 1 & 0 & 1 & 0 & 1 & 0 & 1 \\
$v_{13}$&0 & 1 &  0 &  1 &  0 &  0 &  1 &  0 &  0 & 0 & 1 & 0 & 0 & 0 & 0 & 1 & 0 & 0 & 0 & 1 & 0 & 0 & 0 & 1 & 0 & 1 & 0 \\
$v_{14}$&0 & 1 &  0 &  0 &  1 &  0 &  0 &  1 &  0 & 1 & 0 & 1 & 0 & 0 & 1 & 0 & 1 & 0 & 1 & 0 & 1 & 0 & 0 & 1 & 0 & 0 & 1 \\
$v_{15}$&0 & 1 &  0 &  0 &  1 &  0 &  0 &  1 &  0 & 1 & 0 & 1 & 0 & 0 & 1 & 0 & 1 & 0 & 1 & 0 & 0 & 1 & 0 & 0 & 1 & 0 & 1 \\
$v_{16}$&0 & 1 &  0 &  0 &  1 &  0 &  0 &  1 &  0 & 1 & 0 & 1 & 0 & 0 & 1 & 0 & 0 & 1 & 0 & 0 & 1 & 0 & 1 & 0 & 1 & 0 & 1 \\
$v_{17}$&0 & 1 &  0 &  0 &  1 &  0 &  0 &  1 &  0 & 1 & 0 & 1 & 0 & 0 & 0 & 1 & 0 & 0 & 1 & 0 & 1 & 0 & 1 & 0 & 1 & 0 & 1 \\
$v_{18}$&0 & 0 &  1 &  0 &  0 &  1 &  0 &  1 &  0 & 1 & 0 & 1 & 0 & 0 & 1 & 0 & 1 & 0 & 1 & 0 & 1 & 0 & 0 & 1 & 0 & 0 & 1 \\
$v_{19}$&0 & 0 &  1 &  0 &  0 &  1 &  0 &  1 &  0 & 1 & 0 & 1 & 0 & 0 & 1 & 0 & 1 & 0 & 1 & 0 & 0 & 1 & 0 & 0 & 1 & 0 & 1 \\
$v_{20}$&0 & 0 &  1 &  0 &  0 &  1 &  0 &  1 &  0 & 1 & 0 & 1 & 0 & 0 & 1 & 0 & 0 & 1 & 0 & 0 & 1 & 0 & 1 & 0 & 1 & 0 & 1 \\
$v_{21}$&0 & 0 &  1 &  0 &  0 &  1 &  0 &  1 &  0 & 1 & 0 & 1 & 0 & 0 & 0 & 1 & 0 & 0 & 1 & 0 & 1 & 0 & 1 & 0 & 1 & 0 & 1 \\
$v_{22}$&0 & 0 &  1 &  0 &  0 &  0 &  1 &  0 &  0 & 1 & 0 & 1 & 0 & 0 & 0 & 1 & 0 & 0 & 0 & 1 & 0 & 0 & 0 & 1 & 0 & 1 & 0 \\
$v_{23}$&0 & 0 &  1 &  0 &  0 &  0 &  1 &  0 &  0 & 0 & 1 & 0 & 1 & 0 & 1 & 0 & 1 & 0 & 0 & 1 & 0 & 0 & 0 & 1 & 0 & 0 & 0 \\
$v_{24}$&0 & 0 &  1 &  0 &  0 &  0 &  1 &  0 &  0 & 0 & 1 & 0 & 1 & 0 & 0 & 1 & 0 & 0 & 0 & 1 & 0 & 0 & 1 & 0 & 1 & 0 & 0
\end{tabular}
\end{ruledtabular}
 \end{center}
 \end{table*}
Scheme:
A ``bowtie'' scheme depicted in Figure~\ref{2020-f-Gamma3-scheme}
serves as a straightforward implementation of a logic with an non-separating set of two-valued states
\begin{figure}[htb] 
\begin{tabular}{ c c c }
\begin{tikzpicture}  [scale=1]

\tikzstyle{every path}=[line width=1pt]

\newdimen\ms
\ms=0.1cm
\tikzstyle{sb}=[color=red,regular polygon,regular polygon sides=6,inner sep=8]
\tikzstyle{s1}=[color=red,rectangle,inner sep=3.5]
\tikzstyle{c5}=[circle,inner sep={\ms/8},minimum size=6*\ms]
\tikzstyle{c4}=[circle,inner sep={\ms/8},minimum size=5*\ms]
\tikzstyle{c3}=[circle,inner sep={\ms/8},minimum size=4*\ms]
\tikzstyle{c2}=[circle,inner sep={\ms/8},minimum size=3*\ms]
\tikzstyle{c1}=[circle,inner sep={\ms/8},minimum size=2*\ms]


\coordinate (a1) at  (0,2);
\coordinate (a2) at (0,1);
\coordinate (a3) at (0,0);
\coordinate (a4) at (1,1);
\coordinate (a5) at (2,2);
\coordinate (a6) at (2,1);
\coordinate (a7) at (2,0);


\draw [color=red] (a1) -- (a3);
\draw [color=red!40!white] (a5) -- (a7);
\draw [color=lime] (a1) -- (a7);
\draw [color=olive] (a3) -- (a5);


\draw (a1) coordinate[c2,fill=lime,label=left:$a_1$];
\draw (a1) coordinate[c1,fill=red];

\draw (a2) coordinate[c1,fill=red,label=left:$a_2$];

\draw (a3) coordinate[c2,fill=red,label=left:$a_3$];
\draw (a3) coordinate[c1,fill=olive];

\draw (a4) coordinate[c2,fill=olive,label=above:$c$];
\draw (a4) coordinate[c1,fill=lime];

\draw (a5) coordinate[c2,fill=red!40!white,label=right:$b_1$];
\draw (a5) coordinate[c1,fill=olive];

\draw (a6) coordinate[c1,fill=red!40!white,label=right:$a_6$];

\draw (a7) coordinate[c2,fill=red!40!white,label=right:$b_3$];
\draw (a7) coordinate[c1,fill=lime];

\end{tikzpicture}
& $\qquad$ &
\begin{tikzpicture}  [scale=1]

\tikzstyle{every path}=[line width=1pt]

\newdimen\ms
\ms=0.1cm
\tikzstyle{sb}=[color=red,regular polygon,regular polygon sides=6,inner sep=8]
\tikzstyle{s1}=[color=red,rectangle,inner sep=3.5]
\tikzstyle{c5}=[circle,inner sep={\ms/8},minimum size=6*\ms]
\tikzstyle{c4}=[circle,inner sep={\ms/8},minimum size=5*\ms]
\tikzstyle{c3}=[circle,inner sep={\ms/8},minimum size=4*\ms]
\tikzstyle{c2}=[circle,inner sep={\ms/8},minimum size=3*\ms]
\tikzstyle{c1}=[circle,inner sep={\ms/8},minimum size=2*\ms]


\coordinate (a1) at  (0,2);
\coordinate (a2) at (0,1);
\coordinate (a3) at (0,0);
\coordinate (a4) at (1,1);
\coordinate (a5) at (2,2);
\coordinate (a6) at (2,1);
\coordinate (a7) at (2,0);


\draw [color=red] (a1) -- (a3);
\draw [color=red!40!white] (a5) -- (a7);
\draw [color=lime] (a1) -- (a7);
\draw [color=olive] (a3) -- (a5);


\draw (a1) coordinate[c2,fill=lime,label=left:$a_1$];
\draw (a1) coordinate[c1,fill=red];

\draw (a2) coordinate[sb,fill=red];

\draw (a3) coordinate[c2,fill=red,label=left:$a_3$];
\draw (a3) coordinate[c1,fill=olive];

\draw (a4) coordinate[c2,fill=olive,label=above:$c$];
\draw (a4) coordinate[c1,fill=lime];

\draw (a5) coordinate[c2,fill=red!40!white,label=right:$b_1$];
\draw (a5) coordinate[c1,fill=olive];

\draw (a6) coordinate[sb,fill=red!40!white];

\draw (a7) coordinate[c2,fill=red!40!white,label=right:$b_3$];
\draw (a7) coordinate[c1,fill=lime];

\end{tikzpicture}
\\
(a)&&(b)
\end{tabular}
\centering
\caption{\label{2020-f-Gamma3-scheme}
Orthogonality hypergraph (\textbf{a}) of the scheme of a logic with a nonseparable set of two-valued states  not
as a faithful orthogonal representation in 3-dimensional Hilbert space: suppose $a_1=1$; then, according to the admissability
axioms WAD1{\&}WAD2, $a_3=b_3=c=0$
and thus $b_1=1$ (and, by symmetry,  vice versa: $B_1=1$ implies $a_1=1$).
(\textbf{b}) To achieve a faithful orthogonal representation one could modify this bowtie scheme by following
Kochen and Specker~($\Gamma_3$, p.~70 in~\cite{kochen1}): they substituted two of the four contexts $\{a_1,a_2,a_3\}$ and $\{b_1,b_2,b_3\}$
by Specker bugs
which also provide a ``true implies false'' property at their terminal vertices $a_1$ and $a_3$, as well as $b_1$ and $b_3$, respectively,
and thereby act just as admissibility rule WAD1.
}
\end{figure}

History:
As we are heading toward  logics with less and less ``rich'' set of two-valued states we are approaching
a logic   depicted in Figure~\ref{2020-f1-gd}(f)
which is a combination of two Specker bug logics linked by two external contexts.
It is the $\Gamma_3$-configuration of Kochen-Specker~\cite{kochen1} (p.~70)
with a set of two-valued states which is no longer separating:
In this case, one obtains the ``one-one'' and ``zero-zero rules''~\cite{svozil-2006-omni},
stating that  $a_1$ occurs if and only if $b_1$ occurs
(likewise, $a_7$ occurs if and only if $b_7$ occurs):
Suppose $v$ is a two-valued state on the $\Gamma_3$-configuration of Kochen-Specker.
Whenever $v(a_1)=1$, then $v(c)=0$ because it is in the same context $\{a_1,c,b_7\}$ as $a_1$.
Furthermore, because of Equation~(\ref{2015-s-e2}), whenever $v(a_1)=1$, then $v(a_7)=0$.
Because $b_1$ is in the same context $\{a_7,c,b_1\}$ as $a_7$ and $c$, because of admissibility, $v(b_1)=1$.
Conversely, by symmetry, whenever $v(b_1)=1$, so must be $v(a_1)=1$.
Therefore it can never happen that either one of the two atoms $a_1$ and $b_1$ have different dichotomic values.
The same is true for the pair of atoms  $a_7$ and $b_7$.

If oone ties together two Specker bug logics  (at their ``true implies false'' extremities)
one  obtains non-separability;
just extending one to the
Kochen-Specker $\Gamma_1$ logic~\cite{kochen1} (p.~68)
discussed earlier to obtain ``true implies true'' would be insufficient.
Because in this case a consistent two-valued state exists for which $v(b_1)=v(b_7)=1$ and $v(a_1)=v(a_7)=0$,
thereby separating $a_1$ from $b_1$, and {\it vice versa}.
A second Specker bug logic is neded to elimitate this case; in particular, $v(b_1)=v(b_7)=1$.
The~orthogonality hypergraph of this extended combo of two Specker bug (Kochen-Specker's $\Gamma_3$)
logic is depicted in Figure~\ref{2020-f1-gd}f.

Besides the  quantum mechanical realization of this logic in terms of propositions which are projection operators
corresponding to vectors in three-dimensional Hilbert space suggested by Kochen and Specker~\cite{kochen1},
Tkadlec has given~(p.~206, Figure~1 in \cite{tkadlec-96}) an explicit collection of such vectors
(see also the proof of Proposition~7.2 in Ref.~\cite{svozil-tkadlec} (p.~5392)).

\setlength{\tabcolsep}{0.33em}

Seizing the classical remainder of this logic results in the ``folding'' or ``merging''
of the non-separable observables $a_1$-$b_1$ as well as $a_7$-$b_7$ by identifying those pairs, respectively.
As~a result also the two contexts $\{ a_1,c,b_7\}$ as well as $\{a_7,c,b_1\}$ merge and fold into each other,
leaving a structure depicted in Figure~\ref{2020-f-clr-SpeckerBugCombo}.
Formally, such a structure is obtained in two steps: (i) by enumerating all two-valued states on the original
(non-classical) logic; followed by (ii) an inverse reconstruction of the classical
partition logic resulting from the set of two-valued states obtained in step~(i).
\begin{figure}[htb] 
\centering
\begin{tikzpicture}  [scale=0.9]

\tikzstyle{every path}=[line width=1.1pt]

\newdimen\ms
\ms=0.1cm
\tikzstyle{s1}=[color=red,rectangle,inner sep=3.5]
\tikzstyle{c5}=[circle,inner sep={\ms/8},minimum size=6*\ms]
\tikzstyle{c4}=[circle,inner sep={\ms/8},minimum size=5*\ms]
\tikzstyle{c3}=[circle,inner sep={\ms/8},minimum size=4*\ms]
\tikzstyle{c2}=[circle,inner sep={\ms/8},minimum size=3*\ms]
\tikzstyle{c1}=[circle,inner sep={\ms/8},minimum size=2*\ms]


\coordinate (a1) at  (4,3);
\coordinate (a2) at (3,2);
\coordinate (a3) at (2,1);
\coordinate (a4) at (2,0);
\coordinate (a5) at (2,-1);
\coordinate (a6) at (3,-2);
\coordinate (a7) at (4,-3);
\coordinate (a8) at (2,-2);
\coordinate (a9) at (0,-1);
\coordinate (a10) at (0,0);
\coordinate (a11) at (0,1);
\coordinate (a12) at (2,2);
\coordinate (a13) at (1,0);

\coordinate (a14) at (4,0);

\coordinate (a21) at  ({6+1},2);
\coordinate (a22) at (6,2);
\coordinate (a23) at ({6+2},1);
\coordinate (a24) at ({6+2},0);
\coordinate (a25) at ({6+2},-1);
\coordinate (a26) at (6,-2);
\coordinate (a27) at ({6+1},-2);
\coordinate (a28) at (5,-2);
\coordinate (a29) at ({6+0},-1);
\coordinate (a210) at ({6+0},0);
\coordinate (a211) at ({6+0},1);
\coordinate (a212) at (5,2);
\coordinate (a213) at ({6+1},0);


\draw [color=orange] (a1) -- (a3);
\draw [color=blue] (a3) -- (a5);
\draw [color=red] (a5) -- (a7);
\draw [color=green] (a7) -- (a9);
\draw [color=gray] (a9) -- (a11);
\draw [color=magenta] (a11) -- (a1);
\draw [color=cyan] (a10) -- (a4);

\draw [color=lime] (a1) -- (a14) -- (a7);

\draw [color=orange!40!white] (a1) -- (a23);
\draw [color=blue!40!white] (a23) -- (a25);
\draw [color=red!40!white] (a25) -- (a7);
\draw [color=green!40!white] (a7) -- (a29);
\draw [color=gray!40!white] (a29) -- (a211);
\draw [color=magenta!40!white] (a211) -- (a1);
\draw [color=cyan!40!white] (a210) -- (a24);


\draw (a1) coordinate[c5,fill=lime,label=above:$a_1$];
\draw (a1) coordinate[c4,fill=orange];
\draw (a1) coordinate[c3,fill=gray];
\draw (a1) coordinate[c2,fill=orange!40!white];
\draw (a1) coordinate[c1,fill=gray!40!white];

\draw (a2) coordinate[c1,fill=orange,label=below right:$a_2$];

\draw (a3) coordinate[c2,fill=blue,label=right:$a_3$];
\draw (a3) coordinate[c1,fill=orange];

\draw (a4) coordinate[c2,fill=cyan,label=right:$a_4$];
\draw (a4) coordinate[c1,fill=blue];

\draw (a5) coordinate[c2,fill=red,label=right:$a_5$];
\draw (a5) coordinate[c1,fill=blue];

\draw (a6) coordinate[c1,fill=red,label=above right:$a_6$];

\draw (a7) coordinate[c5,fill=olive,label=below:$a_7$];
\draw (a7) coordinate[c4,fill=green];
\draw (a7) coordinate[c3,fill=red];
\draw (a7) coordinate[c2,fill=green!40!white];
\draw (a7) coordinate[c1,fill=red!40!white];

\draw (a8) coordinate[c1,fill=green,label=below left:$a_8$];

\draw (a9) coordinate[c2,fill=gray,label=left:$a_9$];
\draw (a9) coordinate[c1,fill=green];

\draw (a10) coordinate[c2,fill=gray,label=left:$a_{10}$];
\draw (a10) coordinate[c1,fill=cyan];

\draw (a11) coordinate[c2,fill=magenta,label=left:$a_{11}$];
\draw (a11) coordinate[c1,fill=gray];

\draw (a12) coordinate[c1,fill=magenta,label=above left:$a_{12}$];

\draw (a13) coordinate[c1,fill=cyan,label=above:$a_{13}$];

\draw (a14) coordinate[c1,fill=lime,label=right:$c$];


\draw (a22) coordinate[c1,fill=orange!40!white,label=above right:$b_2$];

\draw (a23) coordinate[c2,fill=blue!40!white,label=right:$b_3$];
\draw (a23) coordinate[c1,fill=orange!40!white];

\draw (a24) coordinate[c2,fill=cyan!40!white,label=right:$b_4$];
\draw (a24) coordinate[c1,fill=blue!40!white];

\draw (a25) coordinate[c2,fill=red!40!white,label=right:$b_5$];
\draw (a25) coordinate[c1,fill=blue!40!white];

\draw (a26) coordinate[c1,fill=red!40!white,label=below right:$b_6$];

\draw (a28) coordinate[c1,fill=green!40!white,label=above left:$b_8$];

\draw (a29) coordinate[c2,fill=gray!40!white,label=left:$b_9$];
\draw (a29) coordinate[c1,fill=green!40!white];

\draw (a210) coordinate[c2,fill=gray!40!white,label=left:$b_{10}$];
\draw (a210) coordinate[c1,fill=cyan!40!white];

\draw (a211) coordinate[c2,fill=magenta!40!white,label=left:$b_{11}$];
\draw (a211) coordinate[c1,fill=gray!40!white];

\draw (a212) coordinate[c1,fill=magenta!40!white,label=below left:$b_{12}$];

\draw (a213) coordinate[c1,fill=cyan!40!white,label=above:$b_{13}$];

\end{tikzpicture}
\caption{\label{2020-f-clr-SpeckerBugCombo}
Orthogonality hypergraph of the classical remainder of
a combo of interlinked Specker bugs with a non-separable set of two-valued states
depicted in Figure~\ref{2020-f1-gd}f.
}
\end{figure}

Embeddability: As every algebra embeddable in a Boolean algebra must have a separating set of two-valued states~(Theorem~0 in \cite{kochen1}),
this logic is no longer ``classical''
in the sense of ``homomorphically (structure-preserving) imbeddable''.
Nevertheless, there may still exist ``plenty'' of two-valued states--indeed,
of them. It is just that these states can no longer differentiate
between the pairs of atoms $(a_1,b_1)$ as well as $(a_7,b_7)$.
Partition logics and their generalized urn or finite automata models fail to reproduce
two linked Specker bug logics resulting in a Kochen-Specker $\Gamma_3$ logic even at this stage.
Of course, the situation will become more dramatic with the non-existence of any kind of two-valued state
(interpretable as truth assignment) on certain logics associated with quantum~propositions.

Chromatic inseparability: The ``true implies true'' rule is associated with
chromatic separability;
in particular, with the impossibility to separate two atoms $a_7$ and $b_7$
with less than four colors.
A proof is presented in~(Figure~12.4 in \cite{svozil-2016-pu-book}).
That chromatic separability on the unit sphere requires 4 colors is implicit in Refs.~\cite{godsil-zaks,havlicek-2000}.

\subsubsection{Deterministic Predictions on Observables with a Nun-Unital Set Of Two-Valued States}
\label{2020-b-nonunital}

In our ``escalation of non-classicality'', we ``scratch the current borderline'' between non-classical observables allowing
a faithful orthogonal (and thus quantum mechanical) representation with a very ``meager'' set of two-valued states:
this is ``as bad as it might get'' before the complete absence of all two-valued states interpretable as classical truth assignments
(subject to admissibility rules)
Indeed, these states are so scarce that they do not acquire the vale ``1'' in at least one (but usually many) of its atomic
propositions: every classical representation requires these observables not to occur at all times; regardless of the preparation state.
These sets of states are called unital~\cite{wright:pent,kalmbach-86,tkadlec-96}.

Schemes and historic examples:
Figure~\ref{2020-f-nun-unital-schemes}a--d present hypergraphs of logics with a non-unital set of two-valued states.
Whereas the propositional structures depicted in~Figure~\ref{2020-f-nun-unital-schemes}a,c have no faithful orthogonal representation in three-dimensional Hilbert space
(they contain disallowed ``triangles), logics schematically depicted in~Figure~\ref{2020-f-nun-unital-schemes}b,d may have
a faithful orthogonal representation if the true-implies-false gadget they contain are suitable.
Based on a (non-intuitive) configuration (invented for the rendition of a classical tautology which is no quantum tautology)
by Sch{\"u}tte~\cite{Schuette} and mentioned in a dissertation by Clavadetscher-Seeberger~\cite{clavadetscher}
a concrete example of a logic with a non-unital set of two-valued states which has a faithful orthogonal representation in three-dimensional Hilbert space
was given by Tkadlec~(Figure~2, p.~207 in \cite{tkadlec-96}).
It consists of 36 atoms in 26 intertwined contexts; in short
 $\{\{ a_{ 1}, a_{ 2}, a_{ 3} \}$,
 $\{ a_{ 1}, a_{ 4}, a_{ 5} \}$,
 $\{ a_{ 4}, a_{ 6}, a_{ 7} \}$,
 $\{ a_{ 8}, a_{ 9}, a_{31} \}$,
 $\{ a_{ 7}, a_{ 8}, a_{23} \}$,
 $\{ a_{17}, a_{ 9}, a_{ 2} \}$,
 $\{ a_{ 7}, a_{33}, a_{10} \}$,
 $\{ a_{28}, a_{17}, a_{18} \}$,
 $\{ a_{ 5}, a_{18}, a_{19} \}$,
 $\{ a_{19}, a_{20}, a_{31} \}$,
 $\{ a_{19}, a_{21}, a_{23} \}$,
 $\{ a_{ 4}, a_{14}, a_{15} \}$,
 $\{ a_{15}, a_{20}, a_{26} \}$,
 $\{ a_{14}, a_{17}, a_{35} \}$,
 $\{ a_{13}, a_{15}, a_{16} \}$,
 $\{ a_{16}, a_{22}, a_{36} \}$,
 $\{ a_{21}, a_{22}, a_{27} \}$,
 $\{ a_{ 3}, a_{13}, a_{23} \}$,
 $\{ a_{ 2}, a_{22}, a_{24} \}$,
 $\{ a_{11}, a_{12}, a_{13} \}$,
 $\{ a_{10}, a_{11}, a_{34} \}$,
 $\{ a_{24}, a_{25}, a_{32} \}$,
 $\{ a_{ 5}, a_{11}, a_{25} \}$,
 $\{ a_{ 9}, a_{12}, a_{29} \}$,
 $\{ a_{ 6}, a_{24}, a_{30} \}$,
 $\{ a_{ 2}, a_{10}, a_{20} \}\}$
allowing only six two-valued states enumerated in Table~\ref{2020-b-t-tkadlec}.

\begin{figure*}
\centering
\scalebox{0.95}[0.95]{
\begin{tabular}{ c c c c c c c c }
\begin{tikzpicture}  [scale=1]

\tikzstyle{every path}=[line width=1pt]

\newdimen\ms
\ms=0.1cm
\tikzstyle{sb}=[color=red,regular polygon,regular polygon sides=6,inner sep=8]
\tikzstyle{s1}=[color=red,rectangle,inner sep=3.5]
\tikzstyle{c5}=[circle,inner sep={\ms/8},minimum size=6*\ms]
\tikzstyle{c4}=[circle,inner sep={\ms/8},minimum size=5*\ms]
\tikzstyle{c3}=[circle,inner sep={\ms/8},minimum size=4*\ms]
\tikzstyle{c2}=[circle,inner sep={\ms/8},minimum size=3*\ms]
\tikzstyle{c1}=[circle,inner sep={\ms/8},minimum size=2*\ms]


\coordinate (a1) at (1,2);
\coordinate (a2) at (1.5,1);
\coordinate (a3) at (2,0);
\coordinate (a4) at (1,0);
\coordinate (a5) at (0,0);
\coordinate (a6) at (0.5,1);
\coordinate (a7) at (1,1);


\draw [color=red] (a1) -- (a3);
\draw [color=olive] (a3) -- (a5);
\draw [color=green] (a5) -- (a1);
\draw [color=blue] (a1) -- (a4);


\draw (a1) coordinate[c3,fill=blue,label=above:$a_1$];
\draw (a1) coordinate[c2,fill=green];
\draw (a1) coordinate[c1,fill=red];

\draw (a2) coordinate[c1,fill=red,label=above right:$a_2$];

\draw (a3) coordinate[c2,fill=red,label=below:$a_3$];
\draw (a3) coordinate[c1,fill=olive];

\draw (a4) coordinate[c2,fill=olive,label=below:$a_4$];
\draw (a4) coordinate[c1,fill=blue];

\draw (a5) coordinate[c2,fill=green,label=below:$a_5$];
\draw (a5) coordinate[c1,fill=olive];

\draw (a6) coordinate[c1,fill=green,label=above left:$a_6$];

\draw (a7) coordinate[c1,fill=blue,label={[label distance=.005cm]275:$a_7$}];

\end{tikzpicture}
&
\begin{tikzpicture}  [scale=1]

\tikzstyle{every path}=[line width=1pt]

\newdimen\ms
\ms=0.1cm
\tikzstyle{di}=[star,star points=7]
\tikzstyle{sb}=[regular polygon,regular polygon sides=6,inner sep=3]
\tikzstyle{s1}=[color=red,rectangle,inner sep=3.5]
\tikzstyle{c5}=[circle,inner sep={\ms/8},minimum size=6*\ms]
\tikzstyle{c4}=[circle,inner sep={\ms/8},minimum size=5*\ms]
\tikzstyle{c3}=[circle,inner sep={\ms/8},minimum size=4*\ms]
\tikzstyle{c2}=[circle,inner sep={\ms/8},minimum size=3*\ms]
\tikzstyle{c1}=[circle,inner sep={\ms/8},minimum size=2*\ms]


\coordinate (a1) at (1,2);
\coordinate (a2) at (1.5,1);
\coordinate (a3) at (2,0);
\coordinate (a4) at (1,0);
\coordinate (a5) at (0,0);
\coordinate (a6) at (0.5,1);
\coordinate (a7) at (1,1);


\draw [color=red] (a1) -- (a3);
\draw [color=olive] (a3) -- (a5);
\draw [color=green] (a5) -- (a1);
\draw [color=blue] (a1) -- (a4);


\draw (a1) coordinate[c3,fill=blue,label=above:$a_1$];
\draw (a1) coordinate[c2,fill=green];
\draw (a1) coordinate[c1,fill=red];

\draw (a2) coordinate[di,fill=red];

\draw (a3) coordinate[c2,fill=red,label=below:$a_3$];
\draw (a3) coordinate[c1,fill=olive];

\draw (a4) coordinate[c2,fill=olive,label=below:$a_4$];
\draw (a4) coordinate[c1,fill=blue];

\draw (a5) coordinate[c2,fill=green,label=below:$a_5$];
\draw (a5) coordinate[c1,fill=olive];

\draw (a6) coordinate[di,fill=green];

\draw (a7) coordinate[c1,fill=blue,label={[label distance=.005cm]275:$a_7$}];

\end{tikzpicture}
&
\begin{tikzpicture}  [scale=1]

\tikzstyle{every path}=[line width=1pt]

\newdimen\ms
\ms=0.1cm
\tikzstyle{sb}=[regular polygon,regular polygon sides=6,inner sep=8]
\tikzstyle{s1}=[,rectangle,inner sep=3.5]
\tikzstyle{c5}=[circle,inner sep={\ms/8},minimum size=6*\ms]
\tikzstyle{c4}=[circle,inner sep={\ms/8},minimum size=5*\ms]
\tikzstyle{c3}=[circle,inner sep={\ms/8},minimum size=4*\ms]
\tikzstyle{c2}=[circle,inner sep={\ms/8},minimum size=3*\ms]
\tikzstyle{c1}=[circle,inner sep={\ms/8},minimum size=2*\ms]


\coordinate (a1) at (1,4);
\coordinate (a2) at (1.5,3.5);
\coordinate (a3) at (2,3);
\coordinate (a4) at (1.5,2.5);
\coordinate (a5) at (1,2);
\coordinate (a6) at (1.5,1);
\coordinate (a7) at (2,0);
\coordinate (a8) at (1,0);
\coordinate (a9) at (0,0);
\coordinate (a10) at (0.5,1);
\coordinate (a11) at (0.5,2.5);
\coordinate (a12) at (0,3);
\coordinate (a13) at (0.5,3.5);
\coordinate (a14) at (1,3);
\coordinate (a15) at (1.5,1.8);
\coordinate (a16) at (0.5,1.8);


\draw [color=red] (a1) -- (a3);
\draw [color=olive] (a3) -- (a5);
\draw [color=green] (a5) -- (a7);
\draw [color=lime] (a7) -- (a9);
\draw [color=orange] (a9) -- (a5);
\draw [color=blue] (a5) -- (a12);
\draw [color=brown] (a12) -- (a1);
\draw [color=Apricot] (a1) -- (a5);
\draw [color=Mulberry] (a10) -- (a11);
\draw [color=SpringGreen] (a4) -- (a6);


\draw (a1) coordinate[c3,fill=brown,label=above:$a_1$];
\draw (a1) coordinate[c2,fill=SpringGreen];
\draw (a1) coordinate[c1,fill=red];

\draw (a2) coordinate[c1,fill=red,label=above right:$a_2$];

\draw (a3) coordinate[c2,fill=red,label=right:$a_3$];
\draw (a3) coordinate[c1,fill=olive];

\draw (a4) coordinate[c2,fill=olive,label=below right:$a_4$];
\draw (a4) coordinate[c1,fill=SpringGreen];

\draw (a5) coordinate[c5,fill=olive,label=below:$a_5$];
\draw (a5) coordinate[c4,fill=green];
\draw (a5) coordinate[c3,fill=orange];
\draw (a5) coordinate[c2,fill=blue];
\draw (a5) coordinate[c1,fill=Apricot];

\draw (a6) coordinate[c2,fill=green,label=above right:$a_6$];
\draw (a6) coordinate[c1,fill=SpringGreen];

\draw (a7) coordinate[c2,fill=green,label=below:$a_7$];
\draw (a7) coordinate[c1,fill=lime];

\draw (a8) coordinate[c1,fill=lime,label=below:$a_8$];

\draw (a9) coordinate[c2,fill=orange,label=below:$a_9$];
\draw (a9) coordinate[c1,fill=lime];

\draw (a10) coordinate[c2,fill=Mulberry,label=above left:$a_{10}$];
\draw (a10) coordinate[c1,fill=orange];

\draw (a15) coordinate[c1,fill=SpringGreen,label=right:$a_{15}$];

\draw (a16) coordinate[c1,fill=Mulberry,label=left:$a_{16}$];

\draw (a11) coordinate[c2,fill=Mulberry,label=below left:$a_{11}$];
\draw (a11) coordinate[c1,fill=blue];

\draw (a12) coordinate[c2,fill=brown,label=left:$a_{12}$];
\draw (a12) coordinate[c1,fill=blue];

\draw (a13) coordinate[c1,fill=brown,label=above left:$a_{12}$];

\draw (a14) coordinate[c1,fill=Apricot,label=right:$a_{14}$];

\end{tikzpicture}
&
\begin{tikzpicture}  [scale=1]

\tikzstyle{every path}=[line width=1pt]

\newdimen\ms
\ms=0.1cm
\tikzstyle{di}=[star,star points=7]
\tikzstyle{sb}=[regular polygon,regular polygon sides=6,inner sep=3]
\tikzstyle{s1}=[rectangle,inner sep=3.5]
\tikzstyle{c5}=[circle,inner sep={\ms/8},minimum size=6*\ms]
\tikzstyle{c4}=[circle,inner sep={\ms/8},minimum size=5*\ms]
\tikzstyle{c3}=[circle,inner sep={\ms/8},minimum size=4*\ms]
\tikzstyle{c2}=[circle,inner sep={\ms/8},minimum size=3*\ms]
\tikzstyle{c1}=[circle,inner sep={\ms/8},minimum size=2*\ms]


\coordinate (a1) at (1,4);
\coordinate (a2) at (1.5,3.5);
\coordinate (a3) at (2,3);
\coordinate (a4) at (1.5,2.5);
\coordinate (a5) at (1,2);
\coordinate (a6) at (1.5,1);
\coordinate (a7) at (2,0);
\coordinate (a8) at (1,0);
\coordinate (a9) at (0,0);
\coordinate (a10) at (0.5,1);
\coordinate (a11) at (0.5,2.5);
\coordinate (a12) at (0,3);
\coordinate (a13) at (0.5,3.5);
\coordinate (a14) at (1,3);
\coordinate (a15) at (1.5,1.8);
\coordinate (a16) at (0.5,1.8);


\draw [color=red] (a1) -- (a3);
\draw [color=olive] (a3) -- (a5);
\draw [color=green] (a5) -- (a7);
\draw [color=lime] (a7) -- (a9);
\draw [color=orange] (a9) -- (a5);
\draw [color=blue] (a5) -- (a12);
\draw [color=brown] (a12) -- (a1);
\draw [color=Apricot] (a1) -- (a5);
\draw [color=Mulberry] (a10) -- (a11);
\draw [color=SpringGreen] (a4) -- (a6);


\draw (a1) coordinate[c3,fill=brown,label=above:$a_1$];
\draw (a1) coordinate[c2,fill=SpringGreen];
\draw (a1) coordinate[c1,fill=red];

\draw (a2) coordinate[di,fill=red,label=above right:$a_2$];

\draw (a3) coordinate[c2,fill=red,label=right:$a_3$];
\draw (a3) coordinate[c1,fill=olive];

\draw (a4) coordinate[c2,fill=olive,label=below right:$a_4$];
\draw (a4) coordinate[c1,fill=SpringGreen];

\draw (a5) coordinate[c5,fill=olive,label=below:$a_5$];
\draw (a5) coordinate[c4,fill=green];
\draw (a5) coordinate[c3,fill=orange];
\draw (a5) coordinate[c2,fill=blue];
\draw (a5) coordinate[c1,fill=Apricot];

\draw (a6) coordinate[c2,fill=green,label=above right:$a_6$];
\draw (a6) coordinate[c1,fill=SpringGreen];

\draw (a7) coordinate[c2,fill=green,label=below:$a_7$];
\draw (a7) coordinate[c1,fill=lime];

\draw (a8) coordinate[di,fill=lime,label=below:$a_8$];

\draw (a9) coordinate[c2,fill=orange,label=below:$a_9$];
\draw (a9) coordinate[c1,fill=lime];

\draw (a10) coordinate[c2,fill=Mulberry,label=above left:$a_{10}$];
\draw (a10) coordinate[c1,fill=orange];

\draw (a15) coordinate[di,fill=SpringGreen,label=right:$a_{15}$];

\draw (a16) coordinate[di,fill=Mulberry,label=left:$a_{16}$];

\draw (a11) coordinate[c2,fill=Mulberry,label=below left:$a_{11}$];
\draw (a11) coordinate[c1,fill=blue];

\draw (a12) coordinate[c2,fill=brown,label=left:$a_{12}$];
\draw (a12) coordinate[c1,fill=blue];

\draw (a13) coordinate[di,fill=brown,label=above left:$a_{12}$];

\draw (a14) coordinate[c1,fill=Apricot,label=right:$a_{14}$];

\end{tikzpicture}
\\
(a)&(b)&(c)&(d)
\end{tabular}}
\caption{\label{2020-f-nun-unital-schemes}
Orthogonality hypergraph schemes of observables with non-unital sets of two-valued states with impossible
or unknown faithful orthogonal representations
(\textbf{a}) simplest scheme without a faithful orthogonal representation: if
$v(a_1)=1$ then
$v(a_3)=v(a_4)=v(a_5)=0$ which contradicts admissibility rule WAD1;
(\textbf{b}) the same as in (\textbf{a}) but with $a_2$ and $a_6$ substituted by a ``true-implies-false gadget''
(for example, the Specker bug gadget is taken in~Figures~2.4 and~6.3 of Ref.~\cite{svozil-tkadlec});
(\textbf{c}) another scheme for a logic with a non-unital set of two-valued states:  if
 $v(a_1)=1$ then
$v(a_3)=v(a_5)=v(a_{12})=0$;
therefore
$v(a_4)=v(a_{11})=0$; and therefore
$v(a_6)=v(a_{10})=0$
and
$v(a_7)=v(a_9)=1$
which contradicts admissibility rule WAD2;
(\textbf{d})  the same as in (\textbf{a}) but with $a_2$ and $a_6$ substituted by a ``true-implies-false gadget''
(for example, the Specker bug gadget is taken in~Figure~7.3 of Ref.~\cite{svozil-tkadlec})).
}   
\end{figure*}

\begin{table*}
\centering
 \caption{\label{2020-b-t-tkadlec}  The 6 two-valued states on the non-unital Tkadlec logic~(Figure~2, p.~20 in \cite{tkadlec-96}).}
\begin{ruledtabular}
\scalebox{0.95}[0.95]{
 \begin{tabular}{ccccccccccccccccccccccccccccccccccccccccc}
\textbf{\#}
&\boldmath{$a_1$}&\boldmath{$a_2$}&\boldmath{$a_3$}&\multicolumn{32}{c}{\boldmath{$ \ldots \qquad \ldots \qquad \ldots\qquad \ldots\qquad \ldots \qquad \ldots \qquad \ldots \qquad \ldots \qquad \ldots$}}&\boldmath{$a_{36}  $}
\\
\hline 
$v_1$&0& 1& 0& 1& 0& 0& 0& 1& 0& 0& 0& 0& 1& 0& 0& 0& 0& 0& 1& 0& 0& 0& 0& 0& 1& 1& 1& 1& 1& 1& 0& 0& 1& 1& 1& 1\\
$v_2$&0& 1& 0& 1& 0& 0& 0& 0& 0& 0& 1& 0& 0& 0& 0& 1& 0& 1& 0& 0& 0& 0& 1& 0& 0& 1& 1& 0& 1& 1& 1& 1& 1& 0& 1& 0\\
$v_3$&0& 1& 0& 1& 0& 0& 0& 0& 0& 0& 0& 1& 0& 0& 0& 1& 0& 1& 0& 0& 0& 0& 1& 0& 1& 1& 1& 0& 0& 1& 1& 0& 1& 1& 1& 0\\
$v_4$&0& 1& 0& 0& 1& 1& 0& 0& 0& 0& 0& 1& 0& 1& 0& 1& 0& 0& 0& 0& 0& 0& 1& 0& 0& 1& 1& 1& 0& 0& 1& 1& 1& 1& 0& 0\\
$v_5$&0& 1& 0& 0& 1& 1& 0& 0& 0& 0& 0& 1& 0& 0& 1& 0& 0& 0& 0& 0& 0& 0& 1& 0& 0& 0& 1& 1& 0& 0& 1& 1& 1& 1& 1& 1\\
$v_6$&0& 1& 0& 0& 1& 0& 1& 0& 0& 0& 0& 0& 1& 1& 0& 0& 0& 0& 0& 0& 1& 0& 0& 0& 0& 1& 0& 1& 1& 1& 1& 1& 0& 1& 0& 1
\end{tabular}}
\end{ruledtabular}
 \end{table*}
\vspace{-6pt}
A  closer inspection of observable propositions whosecolumn entries are all ``0''
reveals that all those classical cases require no less than eight such
observable propositions
$a_{1}$, $a_{3}$, $a_{9}$, $a_{10}$, $a_{17}$, $a_{20}$, $a_{22}$, $a_{24}$
to be false.
On the other hand, all classical two-valued states require the observable $a_2$ to ``always~happen''.

On the other hand, there exists a faithful orthogonal (quantum) representation~(Figure~2, p.~207 in \cite{tkadlec-96})
of the Tkadlec logic
for which not all of those atoms are mutually collinear or orthogonal.
That is, regardless of the state prepared, some of the respective quantum outcomes are sometimes ``seen''.
With regards to the quantum observable corresponding to $a_2$ one might choose a quantum state perpendicular to
the unit vector representing $a_2$,  and thereby arrive at a complete contradiction to the classical prediction.

\subsubsection{Direct Probabilistic Criteria against Value Definiteness from Constraints on Two-Valued Measures}
\label{2017-b-ss-pc}

The ``1-1'' or  ``true implies true'' rule can be taken as an operational criterion:
Suppose that one prepares a system to be in a pure state
corresponding to $a_1$, such that the preparation ensures that $v(a_1)=1$.
If the system is then measured along $b_1$, and the proposition that
the system is in state $b_1$  is found  to be {\em not} true, meaning that $v(b_1)\neq 1$ (the respective detector does not click),
then  one has established that the system is not performing classically,
because classically the set of two-valued states requires non-separability; that is, $v(a_1)=v(b_1)=1$.
With the Tkadlec directions~(p.~206, Figure~1 in \cite{tkadlec-96})
(see also~(Figure~4, p.~5387 in \cite{svozil-tkadlec})),
$\vert {\bf a}_1\rangle = (1/\sqrt{3}) \left(    1,\sqrt{2},0     \right)^\intercal$ and
$\vert {\bf b}_1\rangle = (1/\sqrt{3})\left(     \sqrt{2},1,0      \right)^\intercal$
so that the probability to find a quantized system prepared along $\vert {\bf a}_1\rangle$
and measured along $\vert {\bf b}_1\rangle$ is
$p_{a_1}(b_1) = \vert \langle {\bf b}_1 \vert {\bf a}_1 \rangle \vert^2=  8/9  $,
and that a violation of classicality should occur with ``optimal''~\mbox{\cite{cabello-1994,Cabello-1996-diss,svozil-tkadlec}} (for any fathful orthogonal representation)
probability $1/9$.
Of course, any other classical prediction, such as the ``1-0'' or ``true implies false'' rule,
or more general  classical predictions such as of Equation~(\ref{2015-s-e2})
can also be taken as empirical criteria for non-classicality~(Section~11.3.2. in \cite{svozil-2016-s})).

Indeed, already Stairs~\cite{stairs83} (pp.~588--589) has argued along similar lines for the Specker bug
``true implies false'' logic (a translation into our nomenclature
is:
$m1(1) \equiv a_1$,
$m2(1) \equiv a_3$,
$m2(2) \equiv a_5$,
$m2(3) \equiv a_4$,
$m3(1) \equiv a_{11}$,
$m3(2) \equiv a_9$,
$m3(3) \equiv a_{10}$,
$m4(1) \equiv a_7$).
Independently Clifton (there is a note added in proof to Stairs~\cite{stairs83} (pp.~588--589))
presents asimilar argument, based on (i) another ``true implies true'' logic~(Sections~II and III, Figure~1 in \cite{clifton-93,Johansen-1994,Vermaas-1994})
inspired by Bell~(Figure~C.l. p.~67 in \cite{Belinfante-73}) (cf. also Pitowsky~\cite{Pitowsky-1982-subs} (p.~394)),
as well as (ii) on the Specker bug logic~(Section~IV, Figure~2 in \cite{clifton-93}).
More recently Hardy~\cite{Hardy-92,Hardy-93,hardy-97}
as well as Cabello and
Garc{\'{i}}a-Alcaine and
others~\cite{Cabello-1995-ppks,cabello-96,cabello-97-nhvp,Badziag-2011,Cabello-2013-HP,Cabello-2013-Hardylike} discussed such scenarios.
These criteria for non-classicality are benchmarks aside from the Boole-Bell type polytope method,
and also different from the full Kochen-Specker theorem.

Any such criteria can be directly applied also to logics with a non-unital set of two-valued states.
In such cases, the situation can be even more stringent because any classical prediction ``locks'' the observables into non-occurrence.
However, when interpreting those configurations of observables quantum mechanically
--
that is, as a faithful orthogonal representation
--
this can never be uniformly achieved:
because, if there is only one observable forced into being ``silent''
then one can always choose a preparation state which is non-orthogonal to the one predicted to be ``silent''.
If there are more ``silent'' and complementary (non-collinear and non-orthogonal)
observables such as in Tkadlec's logic~(Figure~2, p.~207 in \cite{tkadlec-96})
any attempt to accommodate the quantum predictions with the classical ``silent'' ones are doomed from the very beginning:
one just needs to be waiting for the first click in a detector corresponding to one of the classically ``silent'' observables
to be able to claim the non-classicality of quantum mechanics.

Very similar issues relating to chromatic separability and embeddability as have been put forward in the case of non-separating sets of two-valued states can be
made for cases on non-unital sets of two-valued states.

\subsection{Finite Propositional Structures Admitting Neither Truth Assignments nor Predictions}
\label{2017-b-c-lwtvs}



\vspace{-6pt}
\subsubsection{Scarcity of Two-Valued States}

When it comes to the absence of a global two-valued state on quantum logics corresponding to Hilbert
spaces of dimension three and higher--whenever contexts or blocks can be intertwined or pasted~\cite{nav:91} to form chains--Kochen and Specker~\cite{kochen1} pursued a very concrete, ``constructive''
(in the sense of finitary mathematical objects but not in the sense of physical operationalizability~\cite{bridgman})
strategy: they presented finite logics realizable by vectors (from the origin to the unit sphere) spanning one-dimensional subspaces, equivalent
to observable propositions, which allowed for lesser \& lesser two-valued state properties.

History:
For non-homomorphic imbedability~(Theorem~0 in \cite{kochen1}) it is already sufficient to present finite collections of observables with a non-separating
or non-unital set of two-valued states.
Concrete examples have already been exposed by considering the Specker bug combo $\Gamma_3$~\cite{kochen1} (p.~70)
discussed in Section~\ref{2017-b-bugscombino},
and the Tkadlec non-unital logic based on the Sch\"utte rays
discussed in Section~\ref{2020-b-nonunital},~respectively.

Kochen and Specker went further and presented a  proof by contradiction
of the non-existence of two-valued states on a finite number of propositions,
based on their  $\Gamma_1$ ``true implies true'' logic~\cite{kochen1} (p.~68) discussed in Section~\ref{2017-b-s-tit},
and iterating them until they reached a complete contradiction in their $\Gamma_2$ logic~\cite{kochen1} (p.~69).
As was pointed out earlier, their representation as points of the sphere is a little bit ``buggy'' (as could be expected from the formation of so many bugs):
as Tkadlec has observed, Kochen-Specker diagram $\Gamma_2$ it is not a one-to-one representation of the logic, because of some different points
at the diagram represent the same element of corresponding orthomodular
poset [cf. Ref.~\cite{svozil-tkadlec} (p.~5390), and Ref.~\cite{tkadlec-01} (p.~156)].

The early 1990s saw an ongoing flurry of papers recasting the Kochen-Specker proof with ever smaller numbers of,
or more symmetric, configurations of
observables
(see Refs.~\mbox{\cite{peres-91,penrose-ks,Peres:1996fk,Kernaghan-1994,mermin-93,bub,svozil-tkadlec,tkadlec-96,cabello-96,Cabello-1996ega,CalHerSvo-tatra,tkadlec-00,tkadlec-01,pavicic:inria-00070615,Smith-2004,Pavicic-2005,Arends2011,Waegell-2011,1751-8121-44-50-505303,Planat2012,Yu-2012,Cabello-2014-PhysRevA.89.042101,Pavii2019,Pavi_i__2019}} for an incomplete list).
The ``most compact'' proofs (in terms of the number of vectors and their associated observables)
should contain no less than 22 vectors in three-dimensional space~\cite{Uijlen2016},
and  no less than 18 vectors for dimension four~\cite{Pavicic-2005} and higher~\cite{Arends2011,Xu-Chen-Guehnw-2020}.
In four dimensions the most compact explicit realization has been suggested by Cabello, Estebaranz and Garc{\'{i}}a-Alcaine~\cite{cabello-96,cabello-99,Pavicic-2005}.
It consists of 9 contexts, with each of the 18 atoms tightly intertwined in two contexts.
The challenge in such (mostly automated) computations is twofold: to generate (and exclude a sufficient number) of ``candidate graphs'';
and subsequently to find a faithful orthogonal representation~\cite{lovasz-79,lovasz-89,Portillo-2015,Pavii2019,Pavi_i__2019}.

\subsubsection{Chromatic Number of the Sphere}

Graph coloring allows another view on value (in)definiteness.
The chromatic number of a graph
\index{chromatic number}
is defined as the least number of colors needed in any total coloring of a graph;
with the constraint that two adjacent vertices have distinct colors.

Suppose that we are interested in the chromatic number of graphs associated with
both (i) the real and (ii) the rational three-dimensional unit sphere.

More generally, we can consider $n$-dimensional unit spheres with the same adjacency property defined by orthogonality.
An orthonormal basis will be called context (block, maximal observable, Boolean subalgebra),
or, in this particular area, a $n$-clique.
Note that for any such graphs involving $n$-cliques the chromatic number of this graph is at least be $n$
(because the chromatic number of a single $n$-clique or context is $n$).

Therefore vertices of the graph are identified with points on the three-dimensional unit sphere;
with adjacency  defined by orthogonality; that is,
two vertices of the graph are adjacent if and only if the unit vectors from the origin
to the respective two points are orthogonal.

The connection to quantum logic is this: any context
(block, maximal observable, Boolean subalgebra, orthonormal basis)
can be represented by a triple of points on the sphere
such that any two unit vectors from the origin
to two distinct points of that triple of points are orthogonal.
Thus graph adjacency in logical terms indicates
``belonging to some common context (block, maximal observable, Boolean subalgebra, orthonormal basis)''.

In three dimensions, if the chromatic number of  graphs is four or higher,
there does not globally exist any consistent
coloring obeying the rule that adjacent vertices (orthogonal vectors)
must have different colors: if one allows only three different colors,
then somewhere in that graph of a chromatic number higher than three, adjacent vertices must have the same colors
(or else the chromatic number would be three or lower).

By a similar argument, non-separability of two-valued states
--
such as encountered in Section~\ref{2017-b-bugscombino}
with the $\Gamma_3$-configuration of Kochen-Specker~\cite{kochen1} (p.~70)--translates into non-separability by colorings
with colors less or equal to the number of atoms in a block
[cf. Figure~\ref{2020-f1-gd}f].

Godsil and Zaks~\cite{godsil-zaks,havlicek-2000}
proved the following results relevant to this discussion:
\begin{enumerate}
\item[(i)]
the chromatic number of the graph based on points of real-valued unit 2-sphere $S^2$
is four~(Lemma~1.1 in \cite{godsil-zaks}).
\item[(ii)]
The chromatic number of rational points on the unit 2-sphere
$S^2\cap \mathbb{Q}^3$
is three~(Lemma~1.2 in~\cite{godsil-zaks}).
\end{enumerate}

We shall concentrate on (i) and discuss (ii) later.
As was pointed out by Godsil in an email conversation from 13 March 2016~\cite{godsil-pc},
{\em ``the fact that the chromatic number of the unit sphere in $\mathbb{R}^3$
is four is a consequence of  Gleason's theorem,
from which the Kochen-Specker theorem follows by compactness.
Gleason's result implies that there is no subset of the sphere that contains exactly one point from each orthonormal basis''.}

Indeed, any coloring can be mapped onto a two-valued state by identifying
a single color with ``$1$'' and all other colors with ``$0$''.
By reduction, all propositions on two-valued states translate into statements about graph coloring.
In particular, if the chromatic number of any logical structure representable as a graph consisting of $n$-atomic
contexts
(blocks, maximal observables with $n$ outcomes, Boolean subalgebras $2^n$, orthonormal bases with $n$ elements)--for instance, as orthogonality hypergraph of quantum logics--is larger than $n$,
then there cannot be any globally consistent two-valued state (truth-value assignment) obeying adjacency (aka admissibility).
Likewise, if no two-valued states  on a logic
which is a pasting of $n$-atomic contexts
exist, then, by reduction, no global consistent coloring with $n$ different colors exists.
Therefore, the Kochen-Specker theorem proves that the chromatic number of the graph corresponding to
the unit sphere with adjacency defined as orthogonality must be higher than three.

For an inverse construction, one may conjecture that all colorings of a particular
Orthogonal hypergraph are determined by the set of unital two-valued states (if it exists)
(modulo permutations of colors).
This can be made plausible by the following construction:
Choose one context or block and choose the first atom thereof.
Then take one of the two-valued states which acquire the value ``1'' on this atom, and associate the first color with this state.
Accordingly, all atoms whose value is ``1'' in this aforementioned two-valued state become
colored with this particular color as well in all the other (complementary) contexts or blocks.
Then take the second atom of the original block and repeat this procedure with a second color, and so on until the last atom is reached.
All other colorings can be obtained by all variations of two-valued states, respectively.


Based on Godsil and Zaks finding that the chromatic number of rational points on the  unit sphere
$S^2\cap \mathbb{Q}^3$
is three~(Lemma~1.2 in \cite{godsil-zaks})---thereby constructing a two-valued measure on the rational unit sphere
by identifying one color with ``$1$'' and the two remaining colors with ``$0$''---there exist ``exotic'' options to circumvent Kochen-Specker type constructions which were quite aggressively (Cabello has referred to this
as the second contextuality war~\cite{Cabello-talk-Vajo-2017})
marketed by allegedly ``nullifying''~\cite{meyer:99} the respective theorems
under the umbrella of ``finite precision measurements''~\cite{kent:99,clifton:99,mermin-99iks,Breuer-02a,Breuer-02b,Barrett-2004}:
the support of vectors spanning the one-dimensional subspaces associated with atomic propositions could be ``diluted'' yet dense,
so much so that the intertwines of contexts (blocks, maximal observables, Boolean subalgebras, orthonormal bases) break up;
and the contexts themselves become ``free and isolated''.
Under such circumstances the logics decay into horizontal sums;
and the orthogonality hypergraphs are just disconnected stacks of previously intertwined contexts.
As can be expected, proofs of Gleason- or Kochen-Specker-type theorems do no longer exist,
as the necessary intertwines are missing.


The ``nullification'' claim and subsequent ones triggered a lot of papers, some cited in ~\cite{Barrett-2004};
mostly critical---of course, not of the results of Godsil and Zaks's finding;
 how could they?---but with respect to their physical applicability.
Peres even wrote a parody by arguing that ``finite precision measurement nullifies Euclid's postulates''~\cite{peres-2003-fpnep},
so that ``nullification'' of the Kochen-Specker theorem might have to be our least concern.

\subsubsection{Exploring Value Indefiniteness}
\label{2017-b-c-eokst}

``Extensions'' of the Kochen-Specker theorem investigate situations
in which a system is prepared in a state
$\vert {\bf x} \rangle \langle {\bf x} \vert$  along direction
$\vert {\bf x} \rangle$
and measured along a non-orthogonal, non-collinear projection
$\vert {\bf y} \rangle \langle {\bf y} \vert$  along direction
$\vert {\bf y} \rangle$.
Those extensions yield what may be called~\cite{pitowsky:218,hru-pit-2003} {\em indeterminacy}:
Pitowsky's  {\em logical indeterminacy principle}~(Theorem~6, p.~226 in \cite{pitowsky:218})
states that given two linearly independent
non-orthogonal unit vectors
$\vert {\bf x} \rangle$
and
$\vert {\bf y} \rangle$
in $\mathbb{R}^3$,
there is a finite set of unit vectors
$\Gamma ( \vert {\bf x} \rangle , \vert {\bf y} \rangle )$ containing
$\vert {\bf x} \rangle$
and
$\vert {\bf y} \rangle$
for which the following statements hold:  There is no two-valued state $v$ on $\Gamma ( \vert {\bf x} \rangle , \vert {\bf y} \rangle )$ which
satisfies
\begin{enumerate}
\item[(i)]
either $v ( \vert {\bf x} \rangle ) = v ( \vert {\bf y} \rangle ) =1$,
\item[(ii)]
or $v ( \vert {\bf x} \rangle ) = 1$ and $ v ( \vert {\bf y} \rangle ) =0$,
\item[(iii)]
or $v ( \vert {\bf x} \rangle ) = 0$ and $ v ( \vert {\bf y} \rangle ) =1$.
\end{enumerate}

Stated differently~(Theorem~2, p.~183 in \cite{hru-pit-2003}),
let $\vert {\bf x} \rangle$
and
$\vert {\bf y} \rangle$
be two non-orthogonal rays in a Hilbert space $\mathfrak{H}$ of finite
dimension $\ge 3$. Then there is a finite set of rays $\Gamma ( \vert {\bf x} \rangle , \vert {\bf y} \rangle )$ containing
$\vert {\bf x} \rangle$
and
$\vert {\bf y} \rangle$
such that a two-valued
state $v$ on $\Gamma ( \vert {\bf x} \rangle , \vert {\bf y} \rangle )$
satisfies $v ( \vert {\bf x} \rangle ),( \vert {\bf y} \rangle ) \in \{0,1\}$
only if $v ( \vert {\bf x} \rangle ) = v ( \vert {\bf y} \rangle ) =0$.
That~is,
if a system of three mutually exclusive outcomes (such as the spin of a spin-$1$ particle in a particular direction)
is prepared in a definite state $\vert {\bf x} \rangle$  corresponding to $v(\vert {\bf x} \rangle )=1$,
then the state $ v ( \vert {\bf y} \rangle ) $ along  some direction $\vert {\bf y} \rangle $ which is neither collinear nor orthogonal
to  $\vert {\bf x} \rangle$
cannot be (pre-)determined,
because, by an argument {\it via} some set of intertwined rays  $\Gamma ( \vert {\bf x} \rangle , \vert {\bf y} \rangle )$,
both cases would lead to a complete contradiction.

The proofs of the logical indeterminacy principle
presented by  Pitowsky and Hrushovski~\cite{pitowsky:218,hru-pit-2003}
is global in the sense that any ray in the
set of intertwining rays $\Gamma ( \vert {\bf x} \rangle , \vert {\bf y} \rangle )$
in-between  $\vert {\bf x} \rangle$
and
$\vert {\bf y} \rangle$---and thus not necessarily the ``beginning and end points''
$\vert {\bf x} \rangle$
and
$\vert {\bf y} \rangle$--may not have a pre-existing value.
(If you are an omni-realist, substitute ``pre-existing'' by ``non-contextual'':
that is, any ray in the
set of intertwining rays $\Gamma ( \vert {\bf x} \rangle , \vert {\bf y} \rangle )$
may violate the admissibility rules and, in particular, non-contextuality.)
Therefore, one might argue that the cases
(i) as well as (ii); that is,
$v ( \vert {\bf x} \rangle ) = v ( \vert {\bf y} \rangle ) =1$.
as well as
$v ( \vert {\bf x} \rangle ) = 1$ and $ v ( \vert {\bf y} \rangle ) =0$
might still be predefined, whereas at least one ray in $\Gamma ( \vert {\bf x} \rangle , \vert {\bf y} \rangle )$ cannot be pre-defined.
(If you are an omni-realist, substitute ``pre-defined'' with ``non-contextual'').

This possibility was excluded in a series of papers~\cite{Abbott:2010uq,2012-incomput-proofsCJ,PhysRevA.89.032109,2015-AnalyticKS}
localizing value indefiniteness.
Therefore, the strong admissibility rules coinciding with two-valued states which are total function on a logic,
were generalized or extended (if you prefer ``weakened'')
to allow value indefiniteness.
Essentially, by allowing the two-valued state to be a partial function on the logic,
which need not be defined any longer on all of its elements,
admissibility was defined by the rules WAD1-WAD3 of Section~\ref{2017-b-admissability},
as well as counterfactually, in all contexts including $\vert {\bf x} \rangle$ and in mutually orthogonal rays which are orthogonal to $\vert {\bf x} \rangle$,
such as the star-shaped Greechie orthogonal hypergraph configuration~(Figure~5 in \cite{PhysRevA.89.032109}) (see also Figure~1 of Ref.~\cite{svozil-2013-omelette}).

In such a formalism, and relative to the assumptions---in particular,
by the admissibility rules WAD1-WAD3 allowing for value indefiniteness,
sets of intertwined rays  $\Gamma ( \vert {\bf x} \rangle , \vert {\bf y} \rangle )$
can be constructed
which render value indefiniteness of property $\vert {\bf y} \rangle  \langle {\bf y} \vert $
if the system is prepared in state $\vert {\bf x} \rangle$ (and thus $ v( \vert {\bf x} \rangle )=1$).
More specifically,
finite sets of intertwined rays  $\Gamma ( \vert {\bf x} \rangle , \vert {\bf y} \rangle )$
can be found which demonstrate that in accordance with the ``weak'' admissibility rules WAD1-WAD3
of Section~\ref{2017-b-admissability},
in~Hilbert spaces of dimension greater than two,
in accordance with complementarity,
any proposition which is complementary with respect to the state prepared
must be value indefinite ~\cite{Abbott:2010uq,2012-incomput-proofsCJ,PhysRevA.89.032109,2015-AnalyticKS}.

\subsubsection{How Can You Measure a Contradiction?}

Clifton replied with this (rhetorical) question
after I had asked him if he could imagine any possibility to somehow ``operationalize'' the Kochen-Specker theorem.
Indeed, the Kochen-Specker theorem---in particular, not only non-separability but the total absence of any two-valued state---was resilient to attempts to somehow ``measure'' it:
first, as alluded by Clifton, its proof is by contraction---any assumption or attempt to consistently (by admissibility)
construct a two-valued state on certain finite subsets of quantum logic provably fails.

Second, the very absence of any two-valued state on such logics reveals the futility of any attempt to somehow define classical probabilities
or classical predictions;
let alone the derivation of any Boole's conditions of physical experience---both rely on, or are,  the hull spanned by the vertices derivable from two-valued states (if the latter existed) and the respective correlations.
Therefore, in essence, on logics corresponding to Kochen-Specker configurations, such as
the $\Gamma_2$-configuration of Kochen-Specker~\cite{kochen1} (p.~69), or
the Cabello, Estebaranz, and Garc{\'{i}}a-Alcaine logic~\cite{cabello-96,cabello-99}
which (subject to admissibility) have no two-valued states,
classical probability theory breaks down entirely---that is, in the most fundamental way;
by not allowing any two-valued state.

It is amazing how many papers exist which claim to ``experimentally verify'' the Kochen-Specker theorem.
However, without exception, those experiments either prove some kind of Bell-Boole of inequality
on single-particles (to be fair this is referred to as ``proving contextuality'';
such as, for instance, Refs.~\cite{Hasegawa-2003,hasegawa:230401,cabelloFilipp-2008,Bartosik-09,kirch-09});
or show that the quantum predictions yield complete contradictions if  one ``forces'' or assumes the counterfactual
co-existence of
observables in different contexts (and measured in separate, distinct experiments carried out in different subensembles; e.g.,
Refs.~\cite{ghz,cabello-99,Si-Zu-Wein-Ze-2000,panbdwz};
again~these lists of references are incomplete.)

Of course, what one could still do is measuring all contexts, or subsets of compatible observables
(possibly by Einstein-Podolsky-Rosen type~\cite{epr} counterfactual inference)---one at a time---on different subensembles
prepared in the same state by Einstein-Podolsky-Rosen type~\cite{epr} experiments,
and comparing the complete sets of results
with classical predictions~\cite{ghz}.

\subsubsection{Non-Contextual Inequalities}

If one is willing to drop admissibility altogether while at the same time maintaining non-contextuality--- thereby only assuming
that the hidden variable theories assign
values to all the observables~(Section~4, p.~375 in \cite{Bengtsson-2012}),
thereby  only assuming non-contextuality~\cite{cabello:210401},
one arrives at {\em non-contextual inequalities}~\cite{cabello-2013-ncyclea}.
Of course, these value assignments need to be much more general as the admissibility requirements on
two-valued states; allowing all $2^n$ (instead of just $n$ combinations) of contexts with $n$ atoms;
such as $1-1-1- \ldots -1$, or $0-0-\ldots -0$.

\section{Quantum Predictions: Probabilities and Expectations}

Since from Hilbert space dimension higher than two, there do not exist any two-valued states,
the (quasi-)classical Boolean strategy to find (or define) probabilities {\it via}
the convex sum of two-valued states brakes down entirely.
Therefore,
the quantum probabilities have to be ``derived'' or postulated from
entirely new concepts, based on quantities---such as vectors or projection operators---in linear vector spaces equipped with a scalar product~\cite{Dirac621,dirac,Jordan1927,vonNeumann:1927:WAQ,v-neumann-49,v-neumann-55}.
One implicit property and guiding principle of these ``new type of probabilities''
was that among those observables which are simultaneously co-measurable (that is, whose
projection operators commute), the classical Kolmogorovian-type probability theory should hold.

Historically, what is often referred to as
Born rule
\index{Born rule}
for calculating probabilities,
was a statistical re-interpretation of Schr\"odinger's
wave function~(Footnote 1, Anmerkung bei der Korrektur, \cite{born-26-1} (p.~865)),
as outlined by Dirac~\cite{Dirac621,dirac}
(a digression: a small piece~\cite{dirac-81} on ``the futility of war''
by the late Dirac is highly recommended; I had the honour listening to the talk personally), Jordan~\cite{Jordan1927},
von Neumann~\cite{vonNeumann:1927:WAQ,v-neumann-49,v-neumann-55}, and
L\"uders~\cite{Luders-1950,Luders-1950e,Busch2009}.

Rather than stating it as an axiom of quantum mechanics,
Gleason~\cite{Gleason}
derived the Born rule from elementary assumptions; in particular from subclassicality: within contexts---that is,
among mutually commuting and thus simultaneously co-measurable observables---the
quantum probabilities should reduce to the classical, Kolmogorovian, form.
In particular, the probabilities of propositions corresponding to observables which are (i) mutually exclusive
(in geometric terms: correspond to orthogonal vectors/projectors)
as well as (ii) simultaneously co-measurable observables
are (i) non-negative, (ii) normalized, and (iii) finite additive; that is, probabilities
(of atoms within contexts or blocks)
add up~(Section~1 in \cite{sep-probability-interpret}).

As already mentioned earlier, Gleason's paper made a high impact on those in the community capable
of comprehending it~\cite{ZirlSchl-65,kamber65,bell-66,kochen1,c-k-m,r:dvur-93,pitowsky:218,rich-bridge}.
Nevertheless, it might not be unreasonable to state that while a proof of the Kochen-Specker theorem is straightforward,
Gleason's results are less attainable.
However, in what follows we shall be less concerned with either necessity nor with mixed states,
but shall rather concentrate on sufficiency and pure states.
(This will also rid us of the limitations to Hilbert spaces of dimensions higher than two.)

Independently, and presumably motivated from Lov{\'a}sz's faithful orthogonal representation of
graphs by vectors in some Hilbert space~\cite{lovasz-79,lovasz-89,Portillo-2015},
Gr{\"o}tschel, Lov{\'a}sz and  Schrijver (Theorem~3.2, p.~338 in~\cite{GroetschelLovaszSchrijver1986} and \S~9.3, p.~285-303 in~\cite{Grtschel1993})
proposed a (probability) weight~\cite{Knuth1994}
on a given graph which essentially rephrases the Born rule in a graph theoretical setting~\cite{Cabello-2014-gtatqc,svozil-2018-b,Xu-Chen-Guehnw-2020}.

Recall that pure states~\cite{Dirac621,dirac}
as well as elementary yes-no propositions~\cite{v-neumann-49,v-neumann-55,birkhoff-36} can both
be represented by (normalized) vectors in some Hilbert space.
If one prepares a pure state corresponding to a unit vector
$\vert {\bf x} \rangle$ (associated with the one-dimensional projection operator
$\textsf{\textbf{E}}_{\bf x}=\vert {\bf x} \rangle \langle {\bf x} \vert $)
and measures an elementary yes-no proposition, representable by a one-dimensional projection operator
$\textsf{\textbf{E}}_{\bf y}=\vert {\bf y} \rangle \langle {\bf y} \vert $
(associated with the vector
$\vert {\bf y} \rangle$),
then Gleason notes~\cite{Gleason} (p.~885) in the second paragraph that (in Dirac notation),
{\em  ``it is easy to see that such a [[probability]] measure $\mu$
can be obtained by selecting a vector $\vert {\bf y} \rangle$
and, for each closed subspace $A$, taking $\mu ({A})$ as the square of the norm of the
projection 
of $\vert {\bf y} \rangle$  on ${A}$''.}

Since in Euclidean space, the
projection $\textsf{\textbf{E}}_{\bf y}$
of $\vert {\bf y} \rangle$  on $\mathfrak{A} = \text{span} (\vert {\bf x} \rangle)$
is the dot product  (both vectors $\vert {\bf x} \rangle , \vert {\bf y} \rangle$
are supposed to be normalized)
$
\vert {\bf x} \rangle  \langle {\bf x} \vert {\bf y} \rangle  =
\vert {\bf x} \rangle  \cos \angle (\vert {\bf x} \rangle , \vert {\bf y} \rangle )
$,
Gleason's observation amounts to the well-known quantum mechanical cosine square probability law
referring to the probability to find a system prepared a in state in another, observed, state.
(Once this is settled, all self-adjoint observables follow by linearity and the spectral theorem.)

In this line of thought, ``measurement'' contexts (orthonormal bases)
allow ``views'' on  ``prepared'' contexts (orthonormal bases)
by the respective projections.


\subsection{Gleason-Type Continuity}
\label{2017-b-c-lwtvs-gleason}

Gleason's theorem~\cite{Gleason} was a response to Mackey's
problem to {\em ``determine all measures on the closed subspaces of a Hilbert space''} contained in a review~\cite{ma-57} of
Birkhoff and von Neumann's centennial paper~\cite{birkhoff-36} on the logic of quantum mechanics.
Starting from von Neumann's formalization of quantum mechanics~\cite{v-neumann-49,v-neumann-55},
the quantum mechanical probabilities and expectations
(aka the Born rule)
are essentially derived from (sub)additivity
among the quantum context; that is, from subclassicality:
within any context (Boolean subalgebra, block, maximal observable, orthonormal base)
the quantum probabilities sum up to $1$.

Gleason's finding caused ripples in the community,
at least of those who cared and coped with
it~\cite{ZirlSchl-65,kamber65,bell-66,kochen1,c-k-m,r:dvur-93,pitowsky:218,rich-bridge}.
(I recall arguing with Van Lambalgen around 1983, who could not believe that anyone in the larger quantum community
had not heard of Gleason's theorem.
As we approached an elevator at Vienna University of Technology's Freihaus building we realized there was also one very prominent
member of the Vienna experimental community entering the cabin.
I suggested to stage an example by asking; and {\em voila}$\ldots$)

With the possible exception of Specker---who did not explicitly refer to the Gleason's theorem
in independently announcing that two-valued states on quantum logics cannot exist~\cite{specker-60}---he preferred to discuss scholastic philosophy;
at that time the Swiss may have inhabited their biotope of quantum logical thinking---Gleason's theorem directly implies the absence of two-valued states.
Indeed, at least for finite dimensions~\cite{Alda,Alda2},
as Zierler and Schlessinger~\cite{ZirlSchl-65} (even before publication of Bell's review~\cite{bell-66}) noted,
``it should also be mentioned that, in fact, the non-existence of two-valued states is an elementary
geometric fact contained quite explicitly in~(Paragraph~2.8 in \cite{Gleason})''.

Now, Gleason's Paragraph~2.8 contains the following main (necessity) theorem~\cite{Gleason} (p.~888):
{\em ``Every non-negative frame function on the unit sphere $S$ in ${\Bbb R}^3$
is regular''.}
Whereby~\cite{Gleason} (p.~886)
{\em ``a frame function $f$ [[satisfying additivity]]
is regular if and only if there exists a self-adjoint
operator $\textsf{\textbf{T}}$ defined on [[the separable Hilbert space]] $\mathfrak{H}$ such that
$f( \vert x \rangle ) = \langle \textsf{\textbf{T}}x \vert x\rangle$ for all unit vectors $ \vert x \rangle $''.}
(Of course, Gleason did not use the Dirac notation.)

In what follows we shall consider Hilbert spaces of dimension $n=3$ and higher.
Suppose that the quantum system is prepared to be in a
pure state associated with the unit vector $\vert x \rangle$,
or the projection operator $\vert x \rangle \langle x \vert$.

As all self-adjoint operators have a spectral decomposition~\cite{halmos-vs} ({\S}~79),
and the scalar product is (anti)linear in its arguments,
let us, instead of $\textsf{\textbf{T}}$, only consider one-dimensional orthogonal projection operators
$\textsf{\textbf{E}}_i^2=\textsf{\textbf{E}}_i = \vert y_i \rangle \langle y_i \vert$
(formed by the unit vector $ \vert y_i \rangle $ which are elements of an orthonormal basis
$\{  \vert y_1 \rangle , \ldots ,  \vert y_n \rangle \}$)
occurring in the spectral sum of
$\textsf{\textbf{T}}=\sum_{i=1}^{n\ge 3} \lambda_i \textsf{\textbf{E}}_i$,
with
${\Bbb I}_n =\sum_{i=1}^{n\ge 3} \textsf{\textbf{E}}_i$.

Thus if $\textsf{\textbf{T}}$ is restricted to some one-dimensional projection operator
$\textsf{\textbf{E}} = \vert y \rangle \langle y \vert$ along $\vert y \rangle $,
then Gleason's main theorem
states that any frame function
reduces to the absolute square of the scalar product;
and in real Hilbert space to the square of the angle between those vectors spanning the linear subspaces corresponding to the two projectors involved;
that is (note that $\textsf{\textbf{E}}$ is self-adjoint),
$f_y( \vert x \rangle ) =
\langle \textsf{\textbf{E}}x \vert x\rangle  =
\langle x \vert \textsf{\textbf{E}} x\rangle  =
\langle x \vert y \rangle \langle y \vert x\rangle  =
\vert \langle x \vert y \rangle
\vert^2 = \cos^2 \angle (x,y)$.

Hence, unless a configuration of contexts is of the star-shaped orthogonality hypergraph form
as depicted in Figure~5 of Ref.~\cite{PhysRevA.89.032109} (see also Figure~1 of Ref.~\cite{svozil-2013-omelette}),--meaning that they all share one common atom; and,
in terms of geometry, meaning that all orthonormal bases share a common vector--and the two-valued state has value $1$ on its center,
there is no way that any two contexts could have a two-valued assignment;
even if one context has one: it is just not possible by the continuous, $\cos^2$-form
of the quantum probabilities.
That is (at least in this author's believe) the watered-down version of the remark of Zierler and Schlessinger~(p.~259, Example~3.2 in \cite{ZirlSchl-65}).

\subsection{Comparison of Classical and Quantum form of Correlations}

In what follows, quantum configurations corresponding to the logic presented in the earlier sections will be considered.
All of them have quantum realizations in terms of vectors spanning one-dimensional subspaces
corresponding to the respective one-dimensional projection operators.

It is stated without a detailed derivation (see Appendix B of Ref.~\cite{svozil-pac}) that,
whereas on the singlet state the classical correlation function~\cite{peres222}
$
 {2 \over \pi}\theta - 1
$
is linear,
the quantum correlations of two   are of the ``stronger'' cosine form
$
 -\cos (\theta )
$.
A stronger-than-quantum correlation would be a sign function
$
 \text{sgn} (\theta-\pi /2 )
$~\cite{svozil-krenn}.

When translated into the most fundamental empirical level---to two clicks in $2\times 2 =4$ respective detectors, a single click on each side---the resulting differences
\begin{equation}
\begin{split}
\Delta E = E_{\text{c},2,2}(\theta ) - E_{\text{q},2j+1,2}(\theta ) \\
= -1 + {2 \over \pi}\theta + \cos \theta
= {2 \over \pi}\theta + \sum_{k=1}^\infty \frac{(-1)^k \theta^{2k}}{(2k)!}
\label{2017-b-ch-diffev}
\end{split}
\end{equation}
signify a critical difference
with regards to the occurrence of joint events:
both classical and quantum systems perform the same at the three points
$\theta \in \{0, \frac{\pi}{2},\pi\}$.
In the region
$0 < \theta <\frac{\pi}{2}$,
$\Delta E $ is strictly positive, indicating that quantum mechanical systems ``outperform''
classical ones with regard to the production of {\em unequal pairs} ``$+-$''  and ``$-+$'',
as compared to equal pairs  ``$++$''  and ``$--$''.
This gets largest at $\theta_{\text{max}}=\text{arcsin}({2}/{\pi}) \approx 0.69$;
at which point the differences amount to 38\%
of all such pairs, as compared to the classical correlations.
Conversely,
in the region
$ \frac{\pi}{2} < \theta <\pi $,
$\Delta E $ is strictly negative, indicating that quantum mechanical systems ``outperform''
classical ones with regard to the production of {\em  equal pairs} ``$++$''  and ``$--$'',
as compared to unequal pairs   ``$+-$''  and ``$-+$''.
This gets largest at $\theta_{\text{min}}= \pi -\text{arcsin}({2}/{\pi}) \approx 2.45$.
Stronger-than-quantum correlations~\cite{popescu-97,popescu-2014} could be of a sign functional form
$
E_{\text{s},2,2}(\theta )= \text{sgn} (\theta-\pi /2 )
$~\cite{svozil-krenn}.

In correlation experiments, these differences are the reason for violations of Boole's
(classical) conditions of possible experience.
Therefore, it appears not entirely unreasonable to speculate that
the non-classical behavior already is expressed and reflected at the level of these two-particle correlations,
and not in need for any violations of the resulting inequalities.

\subsection{Min-Max Principle}

Violation of  Boole's
(classical) conditions of possible experience
by the quantum probabilities, correlations, and expectations
are indications of some sort of non-classicality;
and are often interpreted as certification of
quantum physics, and quantum physical features~\cite{belrand2010,Um-2013}.
Therefore it is important to know the extent of such violations; as well as the experimental configurations
(if they exist~\cite{specker57})
for which such violations reach a maximum.

The basis of the min-max method are two observations~\cite{filipp-svo-04-qpoly-prl}:
\begin{enumerate}
\item
Boole's bounds  are {\em linear}---indeed linearity is, according to Pitowsky\cite{Pit-94},
the main finding of Boole with regards to {\em conditions of possible (nowadays classical physical) experience}~\cite{Boole,Boole-62}---in the terms entering those bounds, such as probabilities and $n$th order correlations or expectations.
\item
All such terms, in particular, probabilities and $n$th order correlations or expectations,
have a quantum realization as self-adjoint transformations. As coherent superpositions (linear sums and differences) of self-adjoint transformations are again self-adjoint transformations (and thus normal operators), they are subject to the spectral theorem. Therefore, effectively, all those terms are ``bundled together'' to give a single ``comprehensive'' (for Boole's conditions of possible experience) observable.
\item
The spectral theorem, when applied to self-adjoint transformations obtained from substituting the quantum terms for the classical terms, yields an eigensystem consisting of all (pure or non-pure) states, as well as the associated eigenvalues which,
according to the quantum mechanical axioms,  serve as the measurement outcomes corresponding to the combined, bundled, ``comprehensive'', observables. (In the usual Einstein-Podolsky-Rosen ``explosion type'' setup these quantities will be highly non-local.) The important observation is that this ``comprehensive'' (for Boole's conditions of possible experience) observable encodes or includes all possible one-by-one measurements on each one of the single terms alone,
at least insofar as they pertain to Boole's conditions.
\item
By taking the minimal and the maximal eigenvalue in the spectral sum of this comprehensive observable one, therefore, obtains the minimal and the maximal measurement outcomes ``reachable'' by quantization.
\end{enumerate}

Therefore, Boole's conditions of possible
experience are taken as given and for granted, and the computational intractability of their hull problem~\cite{Pit-91}
is of no immediate concern, because nothing needs to be said of actually finding  those conditions of possible experience, whose
calculation may grow exponentially with the number of vertices.
Note also that there might be a possible confusion of the term ``min-max principle''~\cite{halmos-vs} ({\S}~90) with
the term ``maximal operator'~\cite{halmos-vs} ({\S}~84).
Finally, this is no attempt to compute general quantum ranges,
as for instance discussed by Pitowsky~\cite{pitowsky-86,pit:range-2001,Pitowsky-08-ge}
and Tsirelson~\cite{cirelson:80,cirelson:87,cirelson}.

Indeed, functional analysis provides a technique to compute (maximal) violations of Boole-Bell type inequalities~\cite{filipp-svo-04-qpoly,filipp-svo-05}:
the
min-max principle
\index{min-max principle}
also known as
{\em Courant-Fischer-Weyl min-max principle} for self-adjoint transformations
(cf. Ref.~\cite{halmos-vs} ({\S}~90),  Ref.~\cite{reed-sim4} (p.~75ff),
and  (Section~4.4, p.~142ff in Ref.~\cite{Teschl-schr})),
or rather an elementary consequence thereof:
by the spectral theorem any bounded self-adjoint linear operator $\textsf{\textbf{T}}$ has a spectral decomposition
$\textsf{\textbf{T}}=\sum_{i=1}^{n} \lambda_i \textsf{\textbf{E}}_i$, in terms of the sum of products
of bounded eigenvalues times the associated orthogonal projection operators.
Suppose for the sake of demonstration that the spectrum is non-degenerate.
Then we can (re)order the spectral sum so that $\lambda_1 \ge \lambda_2 \ge \ldots \ge \lambda_n$
(in case the eigenvalues are also negative, take their absolute value for the sort),
and consider  the greatest eigenvalue.

In quantum mechanics,  the maximal eigenvalue of a self-adjoint linear operator can be identified
with the maximal value of an observation.
Therefore, the spectral theorem supplies even the state associated with this maximal eigenvalue $\lambda_1$: it is the
eigenvector (linear subspace)  $\vert {\bf e}_1 \rangle $ associated with the orthogonal projector
 $\textsf{\textbf{E}}_i = \vert {\bf e}_1 \rangle \langle  {\bf e}_1 \vert $ occurring in the (re)ordered
spectral sum  of $\textsf{\textbf{T}}$.

With this in mind, computation of maximal violations of all the Boole-Bell type inequalities associated with
 Boole's  (classical) conditions of possible experience
is straightforward:
\begin{enumerate}

\item   take all terms containing probabilities, correlations or expectations
and the constant real-valued coefficients which are their
multiplicative factors;
thereby excluding single constant numerical values $O(1)$
(which could be written on ``the other'' side of the inequality; resulting if what might look like
``$T(p_1,\ldots,p_n, p_{1,2},\ldots,p_{123}, \ldots) \le O(1)$'' (usually, these inequalities,
for reasons of operationalizability, as discussed earlier, do not include highter than 2rd order correlations),
and thereby define a function $T$;
\item
in the transition ``quantization'' step $T \rightarrow \textsf{\textbf{T}}$
substitute all classical probabilities and correlations or expectations with the respective quantum
self-adjoint operators, such as for two spin-$\frac{1}{2}$ particles,
$p_1  \rightarrow  q_1
=
{\frac{1}{2}}\left[\mathbb{I}_2 \pm {\sigma}( \theta_1,\varphi_1)\right]\otimes  \mathbb{I}_2$,
$p_2  \rightarrow  q_2
=
{\frac{1}{2}}\left[\mathbb{I}_2 \pm {\sigma}( \theta_2,\varphi_2)\right]\otimes  \mathbb{I}_2$,
$p_{12} \rightarrow  q_{12}   =
{\frac{1}{2}}\left[\mathbb{I}_2 \pm {\bf \sigma}( \theta_1,\varphi_1)\right]
\otimes
{\frac{1}{2}}\left[\mathbb{I}_2 \pm {\bf \sigma}(  \theta_2,\varphi_2)\right]
$,
$
E_{\text{c}} \rightarrow
\textsf{\textbf{E}}_{\text{q}} =  p_{12++}+ p_{12--}  -p_{12+-} -p_{12-+}
$,
as demanded by the inequality.
Please note that since the coefficients in $\textsf{\textbf{T}}$ are all real-valued, and
because $(A+B)^\dagger =A^\dagger +B^\dagger = (A+B)$ for arbitrary self-adjoint transformations $A,B$,
the real-valued weighted sum $\textsf{\textbf{T}}$ of self-adjoint transformations is again self-adjoint.

\item
Finally, compute the eigensystem of $\textsf{\textbf{T}}$; in particular the
largest eigenvalue  $\lambda_{\text{max}}$ and the associated projector
which, in the non-degenerate case, is the dyadic product of the ``maximal state''
$\vert {\bf e}_{\text{max}} \rangle$, or
$\textsf{\textbf{E}}_{\text{max}} = \vert {\bf e}_{\text{max}} \rangle \langle  {\bf e}_{\text{max}} \vert $.

\item
In a last step, maximize $\lambda_{\text{max}}$
(and find the associated eigenvector $\vert {\bf e}_{\text{max}} \rangle$)
with respect to variations of the parameters incurred in step (ii).
\end{enumerate}

The min-max method yields a feasible, constructive method to explore the quantum bounds on
Boole's (classical) conditions of possible experience.
Its application to other situations is feasible.
A generalization to higher-dimensional cases appears tedious but with the help of automated
formula manipulation straightforward.

\subsubsection{Expectations from Quantum Bounds}

The quantum expectation can be directly computed from spin state operators.
For spin-$\frac{1}{2}$ particles, the relevant operator, normalized to eigenvalues  $\pm 1$, is
\begin{equation}
\begin{split}
\textsf{\textbf{T}} (\theta_1 ,\varphi_1;\theta_2 ,\varphi_2)
=
\left[2 \textsf{\textbf{S}}_\frac{1}{2} (\theta_1 ,\varphi_1)\right]
\otimes
\left[2 \textsf{\textbf{S}}_\frac{1}{2} (\theta_2 ,\varphi_2)\right]
.
\end{split}
\label{e-2017-b-whatever-e6a}
\end{equation}

The eigenvalues are $-1,-1,1,1$ and $0$; with eigenvectors for   $\varphi_1=\varphi_2=\frac{\pi}{2}$,
\begin{equation}
\begin{split}
\left( -e^{-i (\theta_1+\theta_2)},0,0,1\right)^\intercal  ,
\left( 0,-e^{-i (\theta_1-\theta_2)},1,0\right)^\intercal  ,
\\
\left( e^{-i (\theta_1+\theta_2)},0,0,1\right)^\intercal   ,
\left( 0,e^{-i (\theta_1-\theta_2)},1,0 \right)^\intercal  ,
\end{split}
\label{e-2017-b-whatever-e6b}
\end{equation}
respectively.

If the states are restricted to Bell basis states
\index{Bell basis}
$
\vert \Psi^\mp \rangle = \frac{1}{\sqrt{2}}\left(\vert 0   1 \rangle \mp \vert 1   0 \rangle  \right)
$
and
$\vert \Phi^\mp \rangle = \frac{1}{\sqrt{2}}\left(\vert 0   0 \rangle \mp \vert 1   1 \rangle  \right)
$
and the respective projection operators are
$\textsf{\textbf{E}}_{\Psi^\mp}$ and
$\textsf{\textbf{E}}_{\Phi^\mp}$,
then the correlations, reduced to the projected operators
$\textsf{\textbf{E}}_{\Psi^\mp} \textsf{\textbf{E}} \textsf{\textbf{E}}_{\Psi^\mp} $ and
$\textsf{\textbf{E}}_{\Phi^\mp} \textsf{\textbf{E}} \textsf{\textbf{E}}_{\Phi^\mp} $
on those states,
yield extrema at
$-\cos (\theta_1-\theta_2)$ for $\textsf{\textbf{E}}_{\Psi^-}$,
$\cos (\theta_1-\theta_2)$ for $\textsf{\textbf{E}}_{\Psi^+}$,
$-\cos (\theta_1+\theta_2)$ for $\textsf{\textbf{E}}_{\Phi^-}$, and
$\cos (\theta_1+\theta_2)$ for $\textsf{\textbf{E}}_{\Phi^+}$.

\subsubsection{Quantum Bounds on the House/Pentagon/Pentagram Logic}

In a similar way  two-particle correlations of a spin-one system can be defined by
\begin{equation}
\begin{split}
\textsf{\textbf{A}} (\theta_1 ,\varphi_1;\theta_2 ,\varphi_2)
=
\textsf{\textbf{S}}_1 (\theta_1 ,\varphi_1) \otimes
\textsf{\textbf{S}}_1 (\theta_2 ,\varphi_2)
.
\end{split}
\label{e-2017-b-whatever-e6}
\end{equation}

Plugging in these correlations into the Klyachko-Can-Biniciogolu-Shumovsky inequality~\cite{Klyachko-2008}
yields  the Klyachko-Can-Biniciogolu-Shumovsky operator
\begin{equation}
\begin{split}
\textsf{\textbf{KCBS}}( \theta_1, \ldots , \theta_5 ,\varphi_1, \ldots ,\varphi_5)=
\textsf{\textbf{A}}( \theta_1,\varphi_1, \theta_3,\varphi_3) +
\textsf{\textbf{A}} ( \theta_3,\varphi_3, \theta_5,\varphi_5)  \\
+ \textsf{\textbf{A}} ( \theta_5,\varphi_5, \theta_7,\varphi_7 ) +
\textsf{\textbf{A}} ( \theta_7,\varphi_7, \theta_9,\varphi_9)  +
\textsf{\textbf{A}} ( \theta_9,\varphi_9, \theta_1,\varphi_1)
.
\end{split}
\label{e-2017-b-whatever-e7}
\end{equation}

Taking the special values of Tkadlec~\cite{tkadlec-priv-1995},
which, is spherical coordinates, are
$a_{1}~=~\left(   1 , \frac{\pi }{2} , 0  \right)^\intercal$,
$a_{2}~=~\left(   1 , \frac{\pi }{2} , \frac{\pi }{2}  \right)^\intercal$,
$a_{3} = \left(   1 , 0 , \frac{\pi }{2}  \right)^\intercal$,
$a_{4} = \left(   \sqrt{2} , \frac{\pi }{2} , -\frac{\pi }{4}  \right)^\intercal$,
$a_{5} = \left(   \sqrt{2} , \frac{\pi }{2} , \frac{\pi }{4}  \right)^\intercal$,
$a_{6}~=~\left(   \sqrt{6} , \tan ^{-1}\left(\frac{1}{\sqrt{2}}\right) , -\frac{\pi }{4}  \right)^\intercal$,
$a_{7} = \left(   \sqrt{3} , \tan ^{-1}\left(\sqrt{2}\right) , \frac{3 \pi }{4}  \right)^\intercal$,
$a_{8} = \left(   \sqrt{6} , \tan ^{-1}\left(\sqrt{5}\right) , \tan ^{-1}\left(\frac{1}{2}\right)  \right)^\intercal$,
$a_{9} = \left(   \sqrt{2} , \frac{3 \pi }{4} , \frac{\pi }{2}  \right)^\intercal$,
$a_{10} = \left(  \sqrt{2} , \frac{\pi }{4} , \frac{\pi }{2}  \right)^\intercal$,
 yields  eigenvalues of $\textsf{\textbf{KCBS}}$ in
\begin{equation}
\begin{split}
\big\{-2.49546, 2.2288, -1.93988, 1.93988, -1.33721,
\\
1.33721, -0.285881, 0.285881, 0.266666\big\}
\end{split}
\label{e-2017-b-whatever-e7kcbs}
\end{equation}
all violating the Klyachko-Can-Biniciogolu-Shumovsky inequality~\cite{Klyachko-2008}.

\subsubsection{Quantum Bounds on the Cabello, Estebaranz and Garc{\'{i}}a-Alcaine Logic}

As a final exercise we shall compute the quantum bounds on the Cabello, Estebaranz and Garc{\'{i}}a-Alcaine logic~\cite{cabello-96,cabello-99}
which can be used in a parity proof of the Kochen-Specker theorem in 4 dimensions,
as well as  the dichotomic observables~(Equation~(2) in \cite{cabello:210401})
$\textsf{\textbf{A}}_i = 2 \vert {\bf a}_i \rangle \langle {\bf a}_i \vert - \mathbb{I}_4$ is used.
The observables are then ``bundled'' into the respective contexts to which they belong; and the context summed
according to the non-contextual inequalities from the Hull computation and
introduced by Cabello~(Equation~(1) in \cite{cabello:210401}).
As a result (we use Cabello's notation and not ours),
\begin{equation}
\begin{aligned}
\textsf{\textbf{T}}=
   &-   \textsf{\textbf{A}}_{12} \otimes  \textsf{\textbf{A}}_{16} \otimes  \textsf{\textbf{A}}_{17} \otimes   \textsf{\textbf{A}}_{18}
 \\
   &-   \textsf{\textbf{A}}_{34} \otimes  \textsf{\textbf{A}}_{45} \otimes  \textsf{\textbf{A}}_{47} \otimes   \textsf{\textbf{A}}_{48}
 \\
   &-   \textsf{\textbf{A}}_{17} \otimes  \textsf{\textbf{A}}_{37} \otimes  \textsf{\textbf{A}}_{47} \otimes   \textsf{\textbf{A}}_{67}
 \\
   &-   \textsf{\textbf{A}}_{12} \otimes  \textsf{\textbf{A}}_{23} \otimes  \textsf{\textbf{A}}_{28} \otimes   \textsf{\textbf{A}}_{29}
 \\
   &-   \textsf{\textbf{A}}_{45} \otimes  \textsf{\textbf{A}}_{56} \otimes  \textsf{\textbf{A}}_{58} \otimes   \textsf{\textbf{A}}_{59}
 \\
   &-   \textsf{\textbf{A}}_{18} \otimes  \textsf{\textbf{A}}_{28} \otimes  \textsf{\textbf{A}}_{48} \otimes   \textsf{\textbf{A}}_{58}
 \\
   &-   \textsf{\textbf{A}}_{23} \otimes  \textsf{\textbf{A}}_{34} \otimes  \textsf{\textbf{A}}_{37} \otimes   \textsf{\textbf{A}}_{39}
 \\
   &-   \textsf{\textbf{A}}_{16} \otimes  \textsf{\textbf{A}}_{56} \otimes  \textsf{\textbf{A}}_{67} \otimes   \textsf{\textbf{A}}_{69}
 \\
   &-   \textsf{\textbf{A}}_{29} \otimes  \textsf{\textbf{A}}_{39} \otimes  \textsf{\textbf{A}}_{59} \otimes   \textsf{\textbf{A}}_{69}
\end{aligned}
\label{e-2017-b-whatever-e7cabnc}
\end{equation}

The resulting $4^4=256$ eigenvalues of $\textsf{\textbf{T}}$
have numerical approximations as ordered numbers
$ -6.94177 \le  -6.67604\le   \ldots \le  5.78503\le  6.023$,
neither of which violates the non-contextual inequality enumerated in (Equation~(1) in Ref.~\cite{cabello:210401}).

\section{Epistemologic Deceptions}

When reading the book of Nature, she tries to tell us something very sublime yet simple;
but what exactly is it?
I have the feeling that often discussants approach this particular book not with
evenly suspended attention~\cite{Freud-1912,Freud-1912-e} but with strong
-- even ideologic~\cite{clauser-talkvie} or evangelical~\cite{zeil-05_nature_ofQuantum}
-- (pre)dispositions.
This might be one of the reasons why Specker called this area ``haunted''~\cite{Specker-priv-Oct2000}.
With these provisos we shall enter the discussion.

Already in 1935---possibly
based to the Born rule for computing quantum probabilities which
differ from classical probabilities on a global scale involving complementary observables,
and yet coincide within contexts---
Schr\"odinger pointed out (cf. also Pitowsky~(footnote~2, p.~96 in \cite{Pit-94}))
that~\cite{schrodinger-en-10.2307/986572}  (p. 327)
{\em ``at no moment does there exist an ensemble of classical states of the model
that squares with the totality of quantum mechanical statements of this moment''.
This seems to be the gist of what can be learned from the quantum probabilities:
they cannot be accommodated entirely within a classical framework.''}

What can be positively said?
There is operational access to a single  context (block, maximal observable, orthonormal basis, Boolean subalgebra);
and for all that operationally matters, all observables forming that context can be simultaneously value definite.
(It could formally be argued that an entire star of contexts intertwined in a ``true'' proposition
must be valued definite.)
A single context represents the maximal information encodable into a quantum system.
This can be done by state preparation.

Beyond this single context,
one can have views on that single state in which the quantized system was prepared.
However, these views come at a price: value indefiniteness.
(Value indefiniteness is often expressed as contextuality, but this view is distractive,
as it suggests some existing entity which is changing
its value; depending on how---that is, along which context---it is measured~\cite{Svozil-2018-p}.)
It might also come as no surprise that by
artificially forcing classical relations build on subset relations on Hilbert space entities
yields indeterminacies: the standard relation and set theoretical operations on subsets may simply not an adequate framework
to investigate vector-worlds.

Moreover, there are grave issues with regards to interpreting whether or not certain detector clicks
indicate non-classical performance:
the clicks are observed all right; yet what do they mean?
(This aspect relates to a discussion~\cite{Kimble-aposterioriQT,Bouwm-aposterioriQTReply}
about whether or not a particular
quantum teleportation experiment~\cite{Bouwmeester1997} is achieved only as a postdiction.)
Depending on the type of gadget or configuration of observable chosen
the same detector click may simultaneously support very different---indeed even mutually contradicting and exclusive---propositions and conclusions~\cite{svozil-2018-whycontexts,svozil-2020-c}.
This is aggravated by the fact that there are no rules selecting one gadget or cloud of observables over other gadgets or~clouds.

This situation might not be taken as a metaphysical conundrum, but perceived rather Socratically:
it should come as no surprise that intrinsic~\cite{svozil-94}, emdedded~\cite{toffoli:79}
observers have no access to
all the information they subjectively desire, but only to a limited amount of properties
their system---be it a virtual or a physical universe---is capable of expressing.
Therefore there is no omniscience in the wider sense of ``all that observers want to know''
but rather ``all that is operational''.

Indeed, we may have been deceived into believing that all observable we believe to epistemically exist
are ontic.
Anything beyond this narrow ``local omniscience covering a single context''
in which the quantized system was prepared
appears to be a subjective illusion which is only stochastically  supported by the quantum formalism---
in terms of Gleason's ``projective views'' on that single, value definite context.
Experiments may enquire about such value indefinite observables by ``forcing'' a measurement on
a system not prepared or encoded to be interrogated in that way.
However,  these measurements of non-existing properties,
although seemingly possessing viable outcomes
which might be interpreted as referring to some alleged hidden properties,
cannot carry any (consistent classical) content of that system alone.

To paraphrase a dictum by Peres~\cite{peres222}, unprepared contexts do not exist;
at least not in any operationally meaningful way.
If one nevertheless forces metaphysical existence on (value) indefinite, non-existing, physical entities
the price, hedged into the quantum formalism, is stochasticity.

\section*{Acknowledgements}

The author acknowledges the support by the Austrian Science Fund (FWF):
project I 4579-N and the Czech Science Foundation (GA\v CR): project 20-09869L.

Josef Tkadlec has kindly provided a program to find all two-valued states and important properties thereof,
given the set of pasted contexts of a logic. He also re-explained some nuances of previous work.
Christoph Lemell and the Nonlinear Dynamics group at the Vienna University of Technology
has provided the computational framework for the calculations performed.
There appears to be a lot of confusion---both in the community at large, as well as with some contributors
-- and I hope I could contribute towards clarification, and not inflict more damage to the subject.

The author declares no conflict of interest.







%

\begin{appendix}
\section{Supplemental Material: What Is so Special About Quantum Clicks?}

This supplement contains mostly code interpretable by Fukuda's {\em cddlib package cddlib-094h} for evaluating hull problems in quantum physical configurations. It also contains some corresponding quantum mechanical calculations.

\subsection{The cddlib package}



Fukuda's {\em cddlib package cddlib-094h} can be obtained from the package homepage~\cite{cdd-pck}. Installation on Unix-type operating systems is with {\em gcc};
the free library for arbitrary precision arithmetic {\em GMP} (currently 6.1.2)~\cite{gmplib}, must be installed first.

In its elementary form of the  {\em V-representation},  {\em cddlib}
takes in the $k$ vertices $\vert {\bf v}_1 \rangle , \ldots , \vert {\bf v}_k \rangle$ of a convex polytope in an $m$-dimensional
vector space as follows (note that all rows of vector components start with ``$1$''):

\begin{lstlisting}[backgroundcolor=\color{yellow!10},framerule=0pt,breaklines=true, frame=tb]
V-representation
begin
k  m+1  numbertype
1 v_11 ... v_1m
...............
1 v_k1 ... v_km
end
\end{lstlisting}


{\em cddlib} responds with the faces (boundaries of halfspaces), as encoded by  $n$  inequalities $ \textsf{\textbf{A}} \vert {\bf x} \rangle   \le \vert {\bf b} \rangle $
in the  {\em H-representation} as follows:

{\begin{lstlisting}[backgroundcolor=\color{yellow!10},framerule=0pt,breaklines=true, frame=tb]
H-representation
begin
n     m+1  numbertype
b    -A
end
\end{lstlisting}  }

Comments appear after an asterisk.

\subsection{Trivial examples}
\label{2017-b-teap}

\subsubsection{One observable}
\label{2017-b-ooa}

The case of a single variable has two extreme cases: false$\equiv 0$ and true$\equiv 1$,
resulting in $0\le p_1 \le 1$:

{ \begin{lstlisting}[backgroundcolor=\color{yellow!10},framerule=0pt,breaklines=true, frame=tb]

* one variable
*
V-representation
begin
2   2   integer
1   0
1   1
end

~~~~~~ cddlib response

H-representation
begin
 2 2 real
  1 -1
  0  1
end

\end{lstlisting}  }

\subsubsection{Two observables}
\label{2017-b-toa}

The case of two variables $p_1$ and $p_2$, and a joint variable $p_{12}$,
result in
\begin{eqnarray}
p_1 + p_2 - p_{12} &\le& 1,
\\
- p_1 + p_{12} &\le& 0,
\\
- p_2 + p_{12} &\le& 0,
\\
- p_{12} &\le& 0,
\label{2017-b-1-2-p-ca}
\end{eqnarray}
and thus $0  \le  p_{12}  \le  p_1 , p_2$.

{ \begin{lstlisting}[backgroundcolor=\color{yellow!10},framerule=0pt,breaklines=true, frame=tb]

* two variables: p1, p2, p12=p1*p2
*
V-representation
begin
4   4   integer
1   0   0   0
1   0   1   0
1   1   0   0
1   1   1   1
end

~~~~~~ cddlib response

H-representation
begin
 4 4 real
  1 -1 -1  1
  0  1  0 -1
  0  0  1 -1
  0  0  0  1
end

\end{lstlisting}  }

For dichotomic expectation values $\pm 1$,
{ \begin{lstlisting}[backgroundcolor=\color{yellow!10},framerule=0pt,breaklines=true, frame=tb]

* two expectation values: E1, E2, E12=E1*E2
*
V-representation
begin
4   4   integer
1   -1   -1   1
1   -1   1   -1
1   1   -1   -1
1   1   1   1
end

~~~~~~ cddlib response

H-representation
begin
 4 4 real
  1 -1 -1  1
  1  1 -1 -1
  1 -1  1 -1
  1  1  1  1
end

\end{lstlisting}  }

\subsubsection{Bounds on the (joint) probabilities and expectations of three observables}
\label{2017-b-tevoa}

{ \begin{lstlisting}[backgroundcolor=\color{yellow!10},framerule=0pt,breaklines=true, frame=tb]

* four joint expectations:
* p1, p2, p3,
* p12=p1*p2, p13=p1*p3, p23=p2*p3,
* p123=p1*p2*p3
V-representation
begin
8   8    integer
1       0    0    0    0    0    0    0
1       0    0    1    0    0    0    0
1       0    1    0    0    0    0    0
1       0    1    1    0    0    1    0
1       1    0    0    0    0    0    0
1       1    0    1    0    1    0    0
1       1    1    0    1    0    0    0
1       1    1    1    1    1    1    1
end

~~~~~~ cddlib response

H-representation
begin
 8 8 real
  1 -1 -1 -1  1  1  1 -1
  0  1  0  0 -1 -1  0  1
  0  0  1  0 -1  0 -1  1
  0  0  0  1  0 -1 -1  1
  0  0  0  0  1  0  0 -1
  0  0  0  0  0  1  0 -1
  0  0  0  0  0  0  1 -1
  0  0  0  0  0  0  0  1
end

\end{lstlisting}  }

If single observable  expectations  are set to zero by assumption (axiom) and are not-enumerated,
the table of expectation values may be redundand.

The case of three expectation value observables
$E_1$, $E_2$  and $E_3$ (which are not explicitly enumerated),
as well as all joint expectations $E_{12}$, $E_{13}$, $E_{23}$, and $E_{123}$,
result in
\begin{eqnarray}
- E_{12}- E_{13}- E_{23} &\le& 1
\\
- E_{123} &\le& 1,
\\
E_{123} &\le& 1,
\\
- E_{12}+ E_{13}+ E_{23} &\le& 1,
\\
 E_{12}- E_{13}+ E_{23} &\le& 1,
\\
 E_{12}+ E_{13}- E_{23} &\le& 1
.
\label{2017-b-1-3-e-ia}
\end{eqnarray}

{ \begin{lstlisting}[backgroundcolor=\color{yellow!10},framerule=0pt,breaklines=true, frame=tb]

* four joint expectations:
* [E1, E2, E3, not explicitly enumerated]
* E12=E1*E2, E13=E1*E3, E23=E2*E3,
* E123=E1*E2*E3
V-representation
begin
8   5    integer
1    1    1    1    1
1    1   -1   -1   -1
1   -1    1   -1   -1
1   -1   -1    1    1
1   -1   -1    1   -1
1   -1    1   -1    1
1    1   -1   -1    1
1    1    1    1   -1
end

~~~~~~ cddlib response

H-representation
begin
 6 5 real
  1  1  1  1  0
  1  0  0  0  1
  1  0  0  0 -1
  1  1 -1 -1  0
  1 -1  1 -1  0
  1 -1 -1  1  0
end

\end{lstlisting}  }

\subsection{2 observers, 2 measurement configurations per observer}
\label{2017-b-totmcpoa}
From a quantum physical standpoint the first relevant case is that of 2 observers and 2 measurement configurations per observer.

\subsubsection{Bell-Wigner-Fine case: probabilities for 2 observers, 2 measurement configurations per observer}
\label{2017-b-bwfa}

The case of four probabilities
$p_1$, $p_2$, $p_3$  and $p_4$,
as well as four joint probabilities $p_{13}$, $p_{14}$, $p_{23}$, and $p_{24}$
result in
\begin{eqnarray}
                                          -p_{14}                        &\le&  0\\
                                                            -p_{24}      &\le&  0\\
 +p_{1}                  +p_{4}  -p_{13}  -p_{14}  +p_{23}  -p_{24}      &\le&  1\\
         +p_{2}          +p_{4}  +p_{13}  -p_{14}  -p_{23}  -p_{24}      &\le&  1\\
         +p_{2}  +p_{3}          -p_{13}  +p_{14}  -p_{23}  -p_{24}      &\le&  1\\
 +p_{1}          +p_{3}          -p_{13}  -p_{14}  -p_{23}  +p_{24}      &\le&  1\\
                                 -p_{13}                                 &\le&  0\\
                                                   -p_{23}               &\le&  0\\
 -p_{1}                  -p_{4}  +p_{13}  +p_{14}  -p_{23}  +p_{24}      &\le&  0\\
         -p_{2}          -p_{4}  -p_{13}  +p_{14}  +p_{23}  +p_{24}      &\le&  0\\
         -p_{2}  -p_{3}          +p_{13}  -p_{14}  +p_{23}  +p_{24}      &\le&  0\\
 -p_{1}          -p_{3}          +p_{13}  +p_{14}  +p_{23}  -p_{24}      &\le&  0\\
 -p_{1}                                   +p_{14}                        &\le&  0\\
         -p_{2}                                             +p_{24}      &\le&  0\\
                 -p_{3}                            +p_{23}               &\le&  0\\
                 -p_{3}          +p_{13}                                 &\le&  0\\
 -p_{1}                          +p_{13}                                 &\le&  0\\
         -p_{2}                                    +p_{23}               &\le&  0\\
                         -p_{4}                             +p_{24}      &\le&  0\\
                         -p_{4}           +p_{14}                        &\le&  0\\
         +p_{2}          +p_{4}                             -p_{24}      &\le&  1\\
 +p_{1}                  +p_{4}           -p_{14}                        &\le&  1\\
         +p_{2}  +p_{3}                            -p_{23}               &\le&  1\\
 +p_{1}          +p_{3}          -p_{13}                                 &\le&  1
.
\label{2017-b-2-2-p-c}
\end{eqnarray}

{ \begin{lstlisting}[backgroundcolor=\color{yellow!10},framerule=0pt,breaklines=true, frame=tb]

* eight variables: p1, p2, p3, p4,
* p13, p14, p23, p24
*
V-representation
begin
16   9   integer
1      0    0    0    0    0    0    0    0
1      0    0    0    1    0    0    0    0
1      0    0    1    0    0    0    0    0
1      0    0    1    1    0    0    0    0
1      0    1    0    0    0    0    0    0
1      0    1    0    1    0    0    0    1
1      0    1    1    0    0    0    1    0
1      0    1    1    1    0    0    1    1
1      1    0    0    0    0    0    0    0
1      1    0    0    1    0    1    0    0
1      1    0    1    0    1    0    0    0
1      1    0    1    1    1    1    0    0
1      1    1    0    0    0    0    0    0
1      1    1    0    1    0    1    0    1
1      1    1    1    0    1    0    1    0
1      1    1    1    1    1    1    1    1
end

~~~~~~ cddlib response

H-representation
begin
 24 9 real
  0  0  0  0  0  0  1  0  0
  0  0  0  0  0  0  0  0  1
  1 -1  0  0 -1  1  1 -1  1
  1  0 -1  0 -1 -1  1  1  1
  1  0 -1 -1  0  1 -1  1  1
  1 -1  0 -1  0  1  1  1 -1
  0  0  0  0  0  1  0  0  0
  0  0  0  0  0  0  0  1  0
  0  1  0  0  1 -1 -1  1 -1
  0  0  1  0  1  1 -1 -1 -1
  0  0  1  1  0 -1  1 -1 -1
  0  1  0  1  0 -1 -1 -1  1
  0  1  0  0  0  0 -1  0  0
  0  0  1  0  0  0  0  0 -1
  0  0  0  1  0  0  0 -1  0
  0  0  0  1  0 -1  0  0  0
  0  1  0  0  0 -1  0  0  0
  0  0  1  0  0  0  0 -1  0
  0  0  0  0  1  0  0  0 -1
  0  0  0  0  1  0 -1  0  0
  1  0 -1  0 -1  0  0  0  1
  1 -1  0  0 -1  0  1  0  0
  1  0 -1 -1  0  0  0  1  0
  1 -1  0 -1  0  1  0  0  0
end

\end{lstlisting}  }

\subsubsection{Clauser-Horne-Shimony-Holt case: expectation values for 2 observers, 2 measurement configurations per observer}
\label{2017-b-chshcevta}

The case of four expectation values
$E_1$, $E_2$, $E_3$  and $E_4$ (which are not explicitly enumerated),
as well as all joint expectations $E_{13}$, $E_{14}$, $E_{23}$, and $E_{24}$
result in
\begin{eqnarray}
   + E_{13} - E_{14} - E_{23} - E_{24}       &\le&      2    \\
                              - E_{24}       &\le&      1    \\
                     - E_{23}                &\le&      1    \\
   - E_{13} + E_{14} - E_{23} - E_{24}       &\le&      2    \\
            - E_{14}                         &\le&      1    \\
   - E_{13} - E_{14} + E_{23} - E_{24}       &\le&      2    \\
   - E_{13} - E_{14} - E_{23} + E_{24}       &\le&      2    \\
   - E_{13}                                  &\le&      1    \\
   - E_{13} + E_{14} + E_{23} + E_{24}       &\le&      2    \\
                              + E_{24}       &\le&      1    \\
                     + E_{23}                &\le&      1    \\
   + E_{13} - E_{14} + E_{23} + E_{24}       &\le&      2    \\
            + E_{14}                         &\le&      1    \\
   + E_{13} + E_{14} - E_{23} + E_{24}       &\le&      2    \\
   + E_{13} + E_{14} + E_{23} - E_{24}       &\le&      2    \\
   + E_{13}                                  &\le&      1
.
\label{2017-b-2-2-e-i}
\end{eqnarray}

{ \begin{lstlisting}[backgroundcolor=\color{yellow!10},framerule=0pt,breaklines=true, frame=tb]

* four joint expectations:
* E13, E14, E23, E24
*
V-representation
begin
16   5   integer
1   1    1    1    1
1   1   -1    1   -1
1  -1    1   -1    1
1  -1   -1   -1   -1
1   1    1   -1   -1
1   1   -1   -1    1
1  -1    1    1   -1
1  -1   -1    1    1
1  -1   -1    1    1
1  -1    1    1   -1
1   1   -1   -1    1
1   1    1   -1   -1
1  -1   -1   -1   -1
1  -1    1   -1    1
1   1   -1    1   -1
1   1    1    1    1
end

~~~~~~ cddlib response

H-representation
begin
 16 5 real
  2 -1  1  1  1
  1  0  0  0  1
  1  0  0  1  0
  2  1 -1  1  1
  1  0  1  0  0
  2  1  1 -1  1
  2  1  1  1 -1
  1  1  0  0  0
  2  1 -1 -1 -1
  1  0  0  0 -1
  1  0  0 -1  0
  2 -1  1 -1 -1
  1  0 -1  0  0
  2 -1 -1  1 -1
  2 -1 -1 -1  1
  1 -1  0  0  0
end

\end{lstlisting}  }

\subsubsection{Beyond the Clauser-Horne-Shimony-Holt case: 2 observers, 3 measurement configurations per observer}
\label{2017-b-chshc1ba}

{ \begin{lstlisting}[backgroundcolor=\color{yellow!10},framerule=0pt,breaklines=true, frame=tb]

* 6  expectations:
* E1, ... , E6
* 9 joint expectations:
* E14, E15, E16, E24, E25, E26, E34, E35, E36
* 1,2,3 on one side
* 4,5,6 on other side
*
V-representation
begin
64   16    integer
1       1    1    1    1    1    1    1    1    1    1    1    1    1    1    1
1       1    1    1    1    1   -1    1    1   -1    1    1   -1    1    1   -1
1       1    1    1    1   -1    1    1   -1    1    1   -1    1    1   -1    1
1       1    1    1    1   -1   -1    1   -1   -1    1   -1   -1    1   -1   -1
1       1    1    1   -1    1    1   -1    1    1   -1    1    1   -1    1    1
1       1    1    1   -1    1   -1   -1    1   -1   -1    1   -1   -1    1   -1
1       1    1    1   -1   -1    1   -1   -1    1   -1   -1    1   -1   -1    1
1       1    1    1   -1   -1   -1   -1   -1   -1   -1   -1   -1   -1   -1   -1
1       1    1   -1    1    1    1    1    1    1    1    1    1   -1   -1   -1
1       1    1   -1    1    1   -1    1    1   -1    1    1   -1   -1   -1    1
1       1    1   -1    1   -1    1    1   -1    1    1   -1    1   -1    1   -1
1       1    1   -1    1   -1   -1    1   -1   -1    1   -1   -1   -1    1    1
1       1    1   -1   -1    1    1   -1    1    1   -1    1    1    1   -1   -1
1       1    1   -1   -1    1   -1   -1    1   -1   -1    1   -1    1   -1    1
1       1    1   -1   -1   -1    1   -1   -1    1   -1   -1    1    1    1   -1
1       1    1   -1   -1   -1   -1   -1   -1   -1   -1   -1   -1    1    1    1
1       1   -1    1    1    1    1    1    1    1   -1   -1   -1    1    1    1
1       1   -1    1    1    1   -1    1    1   -1   -1   -1    1    1    1   -1
1       1   -1    1    1   -1    1    1   -1    1   -1    1   -1    1   -1    1
1       1   -1    1    1   -1   -1    1   -1   -1   -1    1    1    1   -1   -1
1       1   -1    1   -1    1    1   -1    1    1    1   -1   -1   -1    1    1
1       1   -1    1   -1    1   -1   -1    1   -1    1   -1    1   -1    1   -1
1       1   -1    1   -1   -1    1   -1   -1    1    1    1   -1   -1   -1    1
1       1   -1    1   -1   -1   -1   -1   -1   -1    1    1    1   -1   -1   -1
1       1   -1   -1    1    1    1    1    1    1   -1   -1   -1   -1   -1   -1
1       1   -1   -1    1    1   -1    1    1   -1   -1   -1    1   -1   -1    1
1       1   -1   -1    1   -1    1    1   -1    1   -1    1   -1   -1    1   -1
1       1   -1   -1    1   -1   -1    1   -1   -1   -1    1    1   -1    1    1
1       1   -1   -1   -1    1    1   -1    1    1    1   -1   -1    1   -1   -1
1       1   -1   -1   -1    1   -1   -1    1   -1    1   -1    1    1   -1    1
1       1   -1   -1   -1   -1    1   -1   -1    1    1    1   -1    1    1   -1
1       1   -1   -1   -1   -1   -1   -1   -1   -1    1    1    1    1    1    1
1      -1    1    1    1    1    1   -1   -1   -1    1    1    1    1    1    1
1      -1    1    1    1    1   -1   -1   -1    1    1    1   -1    1    1   -1
1      -1    1    1    1   -1    1   -1    1   -1    1   -1    1    1   -1    1
1      -1    1    1    1   -1   -1   -1    1    1    1   -1   -1    1   -1   -1
1      -1    1    1   -1    1    1    1   -1   -1   -1    1    1   -1    1    1
1      -1    1    1   -1    1   -1    1   -1    1   -1    1   -1   -1    1   -1
1      -1    1    1   -1   -1    1    1    1   -1   -1   -1    1   -1   -1    1
1      -1    1    1   -1   -1   -1    1    1    1   -1   -1   -1   -1   -1   -1
1      -1    1   -1    1    1    1   -1   -1   -1    1    1    1   -1   -1   -1
1      -1    1   -1    1    1   -1   -1   -1    1    1    1   -1   -1   -1    1
1      -1    1   -1    1   -1    1   -1    1   -1    1   -1    1   -1    1   -1
1      -1    1   -1    1   -1   -1   -1    1    1    1   -1   -1   -1    1    1
1      -1    1   -1   -1    1    1    1   -1   -1   -1    1    1    1   -1   -1
1      -1    1   -1   -1    1   -1    1   -1    1   -1    1   -1    1   -1    1
1      -1    1   -1   -1   -1    1    1    1   -1   -1   -1    1    1    1   -1
1      -1    1   -1   -1   -1   -1    1    1    1   -1   -1   -1    1    1    1
1      -1   -1    1    1    1    1   -1   -1   -1   -1   -1   -1    1    1    1
1      -1   -1    1    1    1   -1   -1   -1    1   -1   -1    1    1    1   -1
1      -1   -1    1    1   -1    1   -1    1   -1   -1    1   -1    1   -1    1
1      -1   -1    1    1   -1   -1   -1    1    1   -1    1    1    1   -1   -1
1      -1   -1    1   -1    1    1    1   -1   -1    1   -1   -1   -1    1    1
1      -1   -1    1   -1    1   -1    1   -1    1    1   -1    1   -1    1   -1
1      -1   -1    1   -1   -1    1    1    1   -1    1    1   -1   -1   -1    1
1      -1   -1    1   -1   -1   -1    1    1    1    1    1    1   -1   -1   -1
1      -1   -1   -1    1    1    1   -1   -1   -1   -1   -1   -1   -1   -1   -1
1      -1   -1   -1    1    1   -1   -1   -1    1   -1   -1    1   -1   -1    1
1      -1   -1   -1    1   -1    1   -1    1   -1   -1    1   -1   -1    1   -1
1      -1   -1   -1    1   -1   -1   -1    1    1   -1    1    1   -1    1    1
1      -1   -1   -1   -1    1    1    1   -1   -1    1   -1   -1    1   -1   -1
1      -1   -1   -1   -1    1   -1    1   -1    1    1   -1    1    1   -1    1
1      -1   -1   -1   -1   -1    1    1    1   -1    1    1   -1    1    1   -1
1      -1   -1   -1   -1   -1   -1    1    1    1    1    1    1    1    1    1
end

~~~~~~ cddlib response

H-representation
begin
 684 16 real
  4  0 -1  1 -1 -1  0  1 -1  0  1  1  1 -1 -1  1
  [...]
  4  1  1  0  1  1  0  1  1  1  1  1 -1  1 -1  0
  [...]
end

\end{lstlisting}  }

\subsection{Pentagon logic}

\subsection{Probabilities but no joint probabilities}

Here is a computation which includes all probabilities but no joint probabilities:

{ \begin{lstlisting}[backgroundcolor=\color{yellow!10},framerule=0pt,breaklines=true, frame=tb]

* ten probabilities:
* p1 ... p10
*
begin
11  11  integer
1       1    0    0    1    0    1    0    1    0    0
1       1    0    0    0    1    0    0    1    0    0
1       1    0    0    1    0    0    1    0    0    0
1       0    0    1    0    0    1    0    1    0    1
1       0    0    1    0    0    0    1    0    0    1
1       0    0    1    0    0    1    0    0    1    0
1       0    1    0    0    1    0    0    1    0    1
1       0    1    0    0    1    0    0    0    1    0
1       0    1    0    1    0    0    1    0    0    1
1       0    1    0    1    0    1    0    0    1    0
1       0    1    0    1    0    1    0    1    0    1
end

~~~~~~ cddlib response

H-representation
linearity 5  12 13 14 15 16
begin
 16 11 real
  0  0  0  0  0  0  1  0  0  0  0
  0  0  0  0  0  0  0  0  1  0  0
  0 -1  0  0  1  0  0  0  1  0  0
  0  0  0  0  1  0  0  0  0  0  0
  0  1  0  0  0  0  0  0  0  0  0
  1 -1 -1  0  1  0 -1  0  0  0  0
  0  0  1  0  0  0  0  0  0  0  0
  1 -2 -1  0  1  0 -1  0  1  0  0
  0  1  1  0 -1  0  0  0  0  0  0
  0  1  1  0 -1  0  1  0 -1  0  0
  1 -1 -1  0  0  0  0  0  0  0  0
 -1  1  1  1  0  0  0  0  0  0  0
  0 -1 -1  0  1  1  0  0  0  0  0
 -1  1  1  0 -1  0  1  1  0  0  0
  0 -1 -1  0  1  0 -1  0  1  1  0
 -1  2  1  0 -1  0  1  0 -1  0  1
end

\end{lstlisting}  }

\begin{eqnarray}
                                    + p_6                                    \ge           0       \\
                                                  + p_8                      \ge           0       \\
  - p_1               + p_4                       + p_8                      \ge           0       \\
                      + p_4                                                  \ge           0       \\
  + p_1                                                                      \ge           0       \\
  - p_1 - p_2         + p_4         - p_6                                    \ge           -1      \\
        + p_2                                                                \ge           0       \\
  -2p_1 - p_2         + p_4         - p_6         + p_8                      \ge           -1      \\
  + p_1 + p_2         - p_4                                                  \ge           0       \\
  + p_1 + p_2         - p_4         + p_6         - p_8                      \ge           0       \\
  - p_1 - p_2                                                                \ge           -1      \\
  + p_1 + p_2  + p_3                                                         \ge           1       \\
  - p_1 - p_2         + p_4  + p_5                                           \ge           0       \\
  + p_1 + p_2         - p_4         + p_6  + p_7                             \ge           1       \\
  - p_1 - p_2         + p_4         - p_6         + p_8  + p_9               \ge           0       \\
   2p_1 + p_2         - p_4         + p_6         - p_8         + p_{10}     \ge           1
.
\label{2017-b-kl-p-c}
\end{eqnarray}

\subsection{Joint Expectations on all atoms}

This is a full hull computation taking all joint expectations into account:

{\footnotesize \begin{lstlisting}[backgroundcolor=\color{yellow!10},framerule=0pt,breaklines=true, frame=tb]

* 45 pair expectations:
* E12 ... E910
*
V-representation
begin
11  46  real
1 -1 -1 1 -1 1 -1 1 -1 -1 1 -1 1 -1 1 -1 1 1 -1 1 -1 1 -1 1 1 -1 1 -1 1 -1 -1 -1 1 -1 1 1 -1 1 -1 -1 -1 1 1 -1 -1 1
1 -1 -1 -1 1 -1 -1 1 -1 -1 1 1 -1 1 1 -1 1 1 1 -1 1 1 -1 1 1 -1 1 1 -1 1 1 -1 -1 1 -1 -1 1 -1 1 1 -1 1 1 -1 -1 1
1 -1 -1 1 -1 -1 1 -1 -1 -1 1 -1 1 1 -1 1 1 1 -1 1 1 -1 1 1 1 -1 -1 1 -1 -1 -1 1 -1 1 1 1 -1 1 1 1 -1 -1 -1 1 1 1
1 1 -1 1 1 -1 1 -1 1 -1 -1 1 1 -1 1 -1 1 -1 -1 -1 1 -1 1 -1 1 1 -1 1 -1 1 -1 -1 1 -1 1 -1 -1 1 -1 1 -1 1 -1 -1 1 -1
1 1 -1 1 1 1 -1 1 1 -1 -1 1 1 1 -1 1 1 -1 -1 -1 -1 1 -1 -1 1 1 1 -1 1 1 -1 1 -1 1 1 -1 -1 1 1 -1 -1 -1 1 1 -1 -1
1 1 -1 1 1 -1 1 1 -1 1 -1 1 1 -1 1 1 -1 1 -1 -1 1 -1 -1 1 -1 1 -1 1 1 -1 1 -1 1 1 -1 1 -1 -1 1 -1 1 -1 1 -1 1 -1
1 -1 1 1 -1 1 1 -1 1 -1 -1 -1 1 -1 -1 1 -1 1 1 -1 1 1 -1 1 -1 -1 1 1 -1 1 -1 -1 -1 1 -1 1 1 -1 1 -1 -1 1 -1 -1 1 -1
1 -1 1 1 -1 1 1 1 -1 1 -1 -1 1 -1 -1 -1 1 -1 1 -1 1 1 1 -1 1 -1 1 1 1 -1 1 -1 -1 -1 1 -1 1 1 -1 1 1 -1 1 -1 1 -1
1 -1 1 -1 1 1 -1 1 1 -1 -1 1 -1 -1 1 -1 -1 1 -1 1 1 -1 1 1 -1 -1 -1 1 -1 -1 1 1 -1 1 1 -1 -1 1 1 -1 -1 -1 1 1 -1 -1
1 -1 1 -1 1 -1 1 1 -1 1 -1 1 -1 1 -1 -1 1 -1 -1 1 -1 1 1 -1 1 -1 1 -1 -1 1 -1 -1 1 1 -1 1 -1 -1 1 -1 1 -1 1 -1 1 -1
1 -1 1 -1 1 -1 1 -1 1 -1 -1 1 -1 1 -1 1 -1 1 -1 1 -1 1 -1 1 -1 -1 1 -1 1 -1 1 -1 1 -1 1 -1 -1 1 -1 1 -1 1 -1 -1 1 -1
end

~~~~~~ cddlib response

H-representation
linearity 35  12 13 14 15 16 17 18 19 20 21 22 23 24 25 26 27 28 29 30 31 32 33 34 35 36 37 38 39 40 41 42 43 44 45 46
begin
 46 46 real
  1 0 -1 -1 0 0 0 0 0 0 0 0 0 0 0 0 0 0 1 0 0 0 0 0 0 0 0 0 0 0 0 0 0 0 0 0 0 0 0 0 0 0 0 0 0 0
  1 1 0 0 0 0 0 -1 0 0 0 0 0 0 0 -1 0 0 0 0 0 0 0 0 0 0 0 0 0 0 0 0 0 0 0 0 0 0 0 0 0 0 0 0 0 0
  1 0 0 -1 0 0 0 -1 0 0 0 0 0 0 0 0 0 0 0 0 0 0 0 0 0 0 0 0 1 0 0 0 0 0 0 0 0 0 0 0 0 0 0 0 0 0
 -1 -1 0 1 0 0 0 0 0 0 0 0 0 1 0 0 0 0 -1 0 1 0 0 0 0 0 0 0 0 0 0 0 0 0 0 0 0 0 0 0 0 0 0 0 0 0
  0 -1 -1 0 0 0 0 0 0 0 0 0 0 -1 0 0 0 0 0 0 -1 0 0 0 0 0 0 0 0 0 0 0 0 0 0 0 0 0 0 0 0 0 0 0 0 0
  0 0 1 1 0 -1 0 1 0 0 0 0 0 0 0 0 0 0 0 0 -1 0 0 0 0 0 0 0 -1 0 0 0 0 0 0 0 0 0 0 0 0 0 0 0 0 0
  1 1 0 0 0 1 0 0 0 0 0 0 0 1 0 0 0 0 0 0 0 0 0 0 0 0 0 0 0 0 0 0 0 0 0 0 0 0 0 0 0 0 0 0 0 0
  0 0 0 0 0 -1 0 1 0 0 0 0 0 -1 0 1 0 0 0 0 0 0 0 0 0 0 0 0 0 0 0 0 0 0 0 0 0 0 0 0 0 0 0 0 0 0
  1 0 0 0 0 0 0 0 0 0 0 0 0 -1 0 1 0 0 1 0 -1 0 0 0 0 0 0 0 -1 0 0 0 0 0 0 0 0 0 0 0 0 0 0 0 0 0
  0 0 0 -1 0 1 0 0 0 0 0 0 0 0 0 0 0 0 -1 0 1 0 0 0 0 0 0 0 0 0 0 0 0 0 0 0 0 0 0 0 0 0 0 0 0 0
  0 0 1 1 0 0 0 0 0 0 0 0 0 1 0 -1 0 0 0 0 1 0 0 0 0 0 0 0 1 0 0 0 0 0 0 0 0 0 0 0 0 0 0 0 0 0
 -1 -1 0 1 1 0 0 0 0 0 0 0 0 0 0 0 0 0 0 0 0 0 0 0 0 0 0 0 0 0 0 0 0 0 0 0 0 0 0 0 0 0 0 0 0 0
  0 0 -1 -1 0 1 1 0 0 0 0 0 0 0 0 0 0 0 0 0 0 0 0 0 0 0 0 0 0 0 0 0 0 0 0 0 0 0 0 0 0 0 0 0 0 0
 -1 -1 0 1 0 -1 0 1 1 0 0 0 0 0 0 0 0 0 0 0 0 0 0 0 0 0 0 0 0 0 0 0 0 0 0 0 0 0 0 0 0 0 0 0 0 0
  1 0 -1 -1 0 1 0 -1 0 1 0 0 0 0 0 0 0 0 0 0 0 0 0 0 0 0 0 0 0 0 0 0 0 0 0 0 0 0 0 0 0 0 0 0 0 0
  1 1 1 0 0 0 0 0 0 0 1 0 0 0 0 0 0 0 0 0 0 0 0 0 0 0 0 0 0 0 0 0 0 0 0 0 0 0 0 0 0 0 0 0 0 0
 -1 -1 0 1 0 0 0 0 0 0 0 1 0 0 0 0 0 0 0 0 0 0 0 0 0 0 0 0 0 0 0 0 0 0 0 0 0 0 0 0 0 0 0 0 0 0
  0 0 0 -1 0 0 0 0 0 0 0 0 1 0 0 0 0 0 0 0 0 0 0 0 0 0 0 0 0 0 0 0 0 0 0 0 0 0 0 0 0 0 0 0 0 0
  0 0 1 1 0 0 0 0 0 0 0 0 0 1 1 0 0 0 0 0 0 0 0 0 0 0 0 0 0 0 0 0 0 0 0 0 0 0 0 0 0 0 0 0 0 0
  0 0 0 -1 0 0 0 0 0 0 0 0 0 -1 0 1 1 0 0 0 0 0 0 0 0 0 0 0 0 0 0 0 0 0 0 0 0 0 0 0 0 0 0 0 0 0
  0 1 1 1 0 0 0 0 0 0 0 0 0 1 0 -1 0 1 0 0 0 0 0 0 0 0 0 0 0 0 0 0 0 0 0 0 0 0 0 0 0 0 0 0 0 0
  1 1 0 0 0 0 0 0 0 0 0 0 0 0 0 0 0 0 1 1 0 0 0 0 0 0 0 0 0 0 0 0 0 0 0 0 0 0 0 0 0 0 0 0 0 0
 -1 0 0 0 0 0 0 0 0 0 0 0 0 0 0 0 0 0 -1 0 1 1 0 0 0 0 0 0 0 0 0 0 0 0 0 0 0 0 0 0 0 0 0 0 0 0
  1 0 -1 -1 0 0 0 0 0 0 0 0 0 -1 0 1 0 0 1 0 -1 0 1 0 0 0 0 0 0 0 0 0 0 0 0 0 0 0 0 0 0 0 0 0 0 0
  0 1 1 1 0 0 0 0 0 0 0 0 0 1 0 -1 0 0 0 0 0 0 0 1 0 0 0 0 0 0 0 0 0 0 0 0 0 0 0 0 0 0 0 0 0 0
  0 0 0 -1 0 0 0 0 0 0 0 0 0 -1 0 1 0 0 0 0 0 0 0 0 1 0 0 0 0 0 0 0 0 0 0 0 0 0 0 0 0 0 0 0 0 0
  0 -1 0 0 0 0 0 0 0 0 0 0 0 0 0 0 0 0 0 0 0 0 0 0 0 1 0 0 0 0 0 0 0 0 0 0 0 0 0 0 0 0 0 0 0 0
 -1 0 0 0 0 0 0 0 0 0 0 0 0 0 0 0 0 0 -1 0 1 0 0 0 0 0 1 0 0 0 0 0 0 0 0 0 0 0 0 0 0 0 0 0 0 0
  0 0 0 0 0 0 0 0 0 0 0 0 0 0 0 0 0 0 0 0 -1 0 0 0 0 0 0 1 0 0 0 0 0 0 0 0 0 0 0 0 0 0 0 0 0 0
 -1 -1 0 0 0 0 0 0 0 0 0 0 0 0 0 0 0 0 -1 0 1 0 0 0 0 0 0 0 1 1 0 0 0 0 0 0 0 0 0 0 0 0 0 0 0 0
  0 0 0 1 0 0 0 0 0 0 0 0 0 0 0 0 0 0 0 0 -1 0 0 0 0 0 0 0 -1 0 1 0 0 0 0 0 0 0 0 0 0 0 0 0 0 0
  1 0 0 0 0 -1 0 0 0 0 0 0 0 -1 0 0 0 0 1 0 -1 0 0 0 0 0 0 0 0 0 0 1 0 0 0 0 0 0 0 0 0 0 0 0 0 0
  0 0 0 0 0 1 0 0 0 0 0 0 0 1 0 0 0 0 0 0 1 0 0 0 0 0 0 0 0 0 0 0 1 0 0 0 0 0 0 0 0 0 0 0 0 0
  0 0 0 0 0 0 0 -1 0 0 0 0 0 0 0 -1 0 0 0 0 0 0 0 0 0 0 0 0 1 0 0 0 0 1 0 0 0 0 0 0 0 0 0 0 0 0
  0 0 0 0 0 -1 0 1 0 0 0 0 0 -1 0 1 0 0 1 0 -1 0 0 0 0 0 0 0 -1 0 0 0 0 0 1 0 0 0 0 0 0 0 0 0 0 0
  1 1 0 -1 0 1 0 -1 0 0 0 0 0 1 0 -1 0 0 0 0 1 0 0 0 0 0 0 0 1 0 0 0 0 0 0 1 0 0 0 0 0 0 0 0 0 0
  0 0 0 0 0 0 0 0 0 0 0 0 0 0 0 0 0 0 -1 0 0 0 0 0 0 0 0 0 0 0 0 0 0 0 0 0 1 0 0 0 0 0 0 0 0 0
  0 0 0 0 0 -1 0 1 0 0 0 0 0 -1 0 1 0 0 1 0 -1 0 0 0 0 0 0 0 -1 0 0 0 0 0 0 0 0 1 0 0 0 0 0 0 0 0
  0 0 0 0 0 0 0 -1 0 0 0 0 0 0 0 -1 0 0 0 0 0 0 0 0 0 0 0 0 1 0 0 0 0 0 0 0 0 0 1 0 0 0 0 0 0 0
  0 0 0 0 0 0 0 1 0 0 0 0 0 -1 0 1 0 0 0 0 -1 0 0 0 0 0 0 0 -1 0 0 0 0 0 0 0 0 0 0 1 0 0 0 0 0 0
  1 0 -1 -1 0 1 0 -1 0 0 0 0 0 0 0 0 0 0 0 0 0 0 0 0 0 0 0 0 0 0 0 0 0 0 0 0 0 0 0 0 1 0 0 0 0 0
 -1 0 1 1 0 0 0 1 0 0 0 0 0 1 0 0 0 0 -1 0 1 0 0 0 0 0 0 0 0 0 0 0 0 0 0 0 0 0 0 0 0 1 0 0 0 0
  0 0 0 0 0 0 0 -1 0 0 0 0 0 0 0 0 0 0 0 0 0 0 0 0 0 0 0 0 0 0 0 0 0 0 0 0 0 0 0 0 0 0 1 0 0 0
  1 0 0 0 0 -1 0 0 0 0 0 0 0 -1 0 0 0 0 1 0 -1 0 0 0 0 0 0 0 0 0 0 0 0 0 0 0 0 0 0 0 0 0 0 1 0 0
  0 0 -1 -1 0 1 0 0 0 0 0 0 0 0 0 0 0 0 0 0 0 0 0 0 0 0 0 0 0 0 0 0 0 0 0 0 0 0 0 0 0 0 0 0 1 0
  1 1 1 0 0 0 0 0 0 0 0 0 0 0 0 0 0 0 0 0 0 0 0 0 0 0 0 0 0 0 0 0 0 0 0 0 0 0 0 0 0 0 0 0 0 1
end

\end{lstlisting}  }

\begin{eqnarray}
 E_{13} + E_{14} - E_{34} &\le&  1,                                                                          \\
-E_{12} + E_{18} + E_{28} &\le&  1,                                                                          \\
 E_{14} + E_{18} - E_{48} &\le&  1,                                                                          \\
 E_{12} - E_{14} - E_{26} + E_{34} - E_{36} &\le&  -1,                                                       \\
 E_{12} + E_{13} + E_{26} + E_{36} &\le&  0,                                                                 \\
-E_{13} - E_{14} + E_{16} - E_{18} + E_{36} + E_{48} &\le&    0,                                             \\
-E_{12} - E_{16} - E_{26} &\le&  1,                                                                          \\
E_{16} - E_{18} + E_{26} - E_{28} &\le&  0,                                                                  \\
 E_{26} - E_{28} - E_{34} + E_{36} + E_{48} &\le&  1,                                                        \\
 E_{14} - E_{16} + E_{34} - E_{36} &\le&  0,                                                                 \\
-E_{13} - E_{14} - E_{26} + E_{28} - E_{36} - E_{48} &\le&  0,                                               \\
 E_{12} - E_{14} - E_{15} &\le&  -1,                                                                         \\
 E_{13} + E_{14} - E_{16} - E_{17} &\le&  0,                                                                 \\
 E_{12} - E_{14} + E_{16} - E_{18} - E_{19} &\le&  -1,                                                       \\
-E_{1,10} + E_{13} + E_{14} - E_{16} + E_{18} &\le&    1,                                                    \\
-E_{12} - E_{13} - E_{23} &\le&  1,                                                                          \\
 E_{12} - E_{14} - E_{24} &\le&  -1,                                                                         \\
 E_{14} - E_{25} &\le&  0,                                                                                   \\
-E_{13} - E_{14} - E_{26} - E_{27} &\le&  0,                                                                 \\
 E_{14} + E_{26} - E_{28} - E_{29} &\le&  0,                                                                 \\
-E_{12} - E_{13} - E_{14} - E_{2,10} - E_{26} + E_{28} &\le&    0,                                           \\
-E_{12} - E_{34} - E_{35} &\le&  1,                                                                          \\
 E_{34} - E_{36} - E_{37} &\le&  -1,                                                                         \\
 E_{13} + E_{14} + E_{26} - E_{28} - E_{34} + E_{36} - E_{38} &\le&   1,                                     \\
-E_{12} - E_{13} - E_{14} - E_{26} + E_{28} - E_{39} &\le&  0,                                               \\
 E_{14} + E_{26} - E_{28} - E_{3,10} &\le&  0,                                                               \\
 E_{12} - E_{45} &\le&  0,                                                                                   \\
 E_{34} - E_{36} - E_{46} &\le&  -1,                                                                         \\
 E_{36} - E_{47} &\le&  0,                                                                                   \\
 E_{12} + E_{34} - E_{36} - E_{48} - E_{49} &\le&  -1,                                                       \\
-E_{14} + E_{36} - E_{4,10} + E_{48} &\le&  0,                                                               \\
 E_{16} + E_{26} - E_{34} + E_{36} - E_{56} &\le&  1,                                                        \\
-E_{16} - E_{26} - E_{36} - E_{57} &\le&  0,                                                                 \\
 E_{18} + E_{28} - E_{48} - E_{58} &\le&  0,                                                                 \\
 E_{16} - E_{18} + E_{26} - E_{28} - E_{34} + E_{36} + E_{48} - E_{59} &\le&   0,                            \\
-E_{12} + E_{14} - E_{16} + E_{18} - E_{26} + E_{28} - E_{36} - E_{48} - E_{5,10} &\le&  1,                  \\
 E_{34} - E_{67} &\le&  0,                                                                                   \\
 E_{16} - E_{18} + E_{26} - E_{28} - E_{34} + E_{36} + E_{48} - E_{68} &\le&  0,                             \\
 E_{18} + E_{28} - E_{48} - E_{69} &\le&  0,                                                                 \\
-E_{18} + E_{26} - E_{28} + E_{36} + E_{48} - E_{6,10} &\le&  0,                                             \\
 E_{13} + E_{14} - E_{16} + E_{18} - E_{78} &\le&  1,                                                        \\
-E_{13} - E_{14} - E_{18} - E_{26} + E_{34} - E_{36} - E_{79} &\le&  -1,                                     \\
 E_{18} - E_{7,10} &\le&  0,                                                                                 \\
 E_{16} + E_{26} - E_{34} + E_{36} - E_{89} &\le&  1,                                                        \\
 E_{13} + E_{14} - E_{16} - E_{8,10} &\le&  0,                                                               \\
-E_{12} - E_{13} - E_{9,10} &\le&  1
.
\label{2017-b-kl-e-c}
\end{eqnarray}

\section{House/pentagon/pentagram gadget}

\subsubsection{Bub-Stairs inequality}

If one considers only the five probabilities on the intertwining atoms,
then the following Bub-Stairs inequality  $p_1+p_3+p_5+p_7+p_9  \le  2$, among others,
results:

{ \begin{lstlisting}[backgroundcolor=\color{yellow!10},framerule=0pt,breaklines=true, frame=tb]

* five probabilities on intertwining contexts
* p1, p3, p5, p7, p9
*
V-representation
begin
11  6  integer
1       1    0    0    0    0
1       1    0    1    0    0
1       1    0    0    1    0
1       0    1    0    0    0
1       0    1    0    1    0
1       0    1    0    0    1
1       0    0    1    0    0
1       0    0    1    0    1
1       0    0    0    1    0
1       0    0    0    0    1
1       0    0    0    0    0
end

~~~~~~ cddlib response

H-representation
begin
 11 6 real
  0  0  0  1  0  0
  1  0  0  0 -1 -1
  0  1  0  0  0  0
  1  0 -1 -1  0  0
  2 -1 -1 -1 -1 -1
  1 -1 -1  0  0  0
  0  0  0  0  1  0
  1 -1  0  0  0 -1
  1  0  0 -1 -1  0
  0  0  1  0  0  0
  0  0  0  0  0  1
end

\end{lstlisting}  }

One could also consider probabilities on the non-intertwining atoms yielding; in particular,
$p_2+p_4+p_6+p_8+p_{10} \ge 1$.

{ \begin{lstlisting}[backgroundcolor=\color{yellow!10},framerule=0pt,breaklines=true, frame=tb]

* five probabilities
* on non-intertwining atoms
* p2, p4, p6, p8, p10
*
V-representation
begin
11  6  integer
1       0    1    1    1    0
1       0    0    0    1    0
1       0    1    0    0    0
1       0    0    1    1    1
1       0    0    0    0    1
1       0    0    1    0    0
1       1    0    0    1    1
1       1    0    0    0    0
1       1    1    0    0    1
1       1    1    1    0    0
1       1    1    1    1    1
end

~~~~~~ cddlib response

H-representation
begin
 11 6 real
  0  0  0  0  1  0
  0  0  0  0  0  1
  0  0  1  0  0  0
 -1  1  1  1  1  1
  0  1  0  0  0  0
  0  0  0  1  0  0
  1  1 -1  1 -1 -1
  1 -1  1 -1 -1  1
  1  1 -1 -1  1 -1
  1 -1  1 -1  1 -1
  1 -1 -1  1 -1  1
end

\end{lstlisting}  }

\subsubsection{Klyachko-Can-Biniciogolu-Shumovsky inequalities}
\label{2017-b-kcbsia}

The following hull computation is limited to adjacent pair expectations;
it yields the Klyachko-Can-Biniciogolu-Shumovsky inequality
$  E_{13}  + E_{35}  + E_{57}  + E_{79}  + E_{91}   \ge 3   $:

{ \begin{lstlisting}[backgroundcolor=\color{yellow!10},framerule=0pt,breaklines=true, frame=tb]

* five joint Expectations:
* E13 E35 E57 E79 E91
*
V-representation
begin
11  6  real
1      -1    1    1    1   -1
1      -1   -1   -1    1   -1
1      -1    1   -1   -1   -1
1      -1   -1    1    1    1
1      -1   -1   -1   -1    1
1      -1   -1    1   -1   -1
1       1   -1   -1    1    1
1       1   -1   -1   -1   -1
1       1    1   -1   -1    1
1       1    1    1   -1   -1
1       1    1    1    1    1
end

~~~~~~ cddlib response

H-representation
begin
 11 6 real
  1  0  0  0  1  0
  1  0  0  0  0  1
  1  0  1  0  0  0
  3  1  1  1  1  1
  1  1  0  0  0  0
  1  0  0  1  0  0
  1  1 -1  1 -1 -1
  1 -1  1 -1 -1  1
  1  1 -1 -1  1 -1
  1 -1  1 -1  1 -1
  1 -1 -1  1 -1  1
end

\end{lstlisting}  }

\begin{eqnarray}
                                 - E_{79}             &\le& 1       \\
                                           - E_{91}   &\le& 1       \\
             - E_{35}                                 &\le& 1       \\
   - E_{13}  - E_{35}  - E_{57}  - E_{79}  - E_{91}   &\le& 3       \\
   - E_{13}                                           &\le& 1       \\
                       - E_{57}                       &\le& 1       \\
   - E_{13}  + E_{35}  - E_{57}  + E_{79}  + E_{91}   &\le& 1       \\
   + E_{13}  - E_{35}  + E_{57}  + E_{79}  - E_{91}   &\le& 1       \\
   - E_{13}  + E_{35}  + E_{57}  - E_{79}  + E_{91}   &\le& 1       \\
   + E_{13}  - E_{35}  + E_{57}  - E_{79}  + E_{91}   &\le& 1       \\
   + E_{13}  + E_{35}  - E_{57}  + E_{79}  - E_{91}   &\le& 1
.
\label{2017-b-kl-e-i}
\end{eqnarray}

\subsection{Two intertwined pentagon logics forming a Specker K\"afer (bug) or cat's cradle logic}

\subsubsection{Probabilities on the Specker bug logic}

A {\em Mathematica}~\cite{Mathematica11.1} code to reduce probabilities on the Specker bug logic:

{ \begin{lstlisting}[backgroundcolor=\color{yellow!10},framerule=0pt,breaklines=true, frame=tb]

Reduce[
p1 + p2 + p3 == 1
&& p3 + p4 + p5 == 1
&& p5 + p6 + p7 == 1
&&   p7 + p8 + p9 == 1
&& p9 + p10 + p11 == 1
&& p11 + p12 + p1 == 1
&&  p4 + p10 + p13 == 1,
{p3, p11, p5, p9, p4, p10}, Reals]

~~~~~~ Mathematica response

p1 == 3/2 - p12/2 - p13/2 - p2/2 - p6/2 - p7 - p8/2 &&
 p3 == -(1/2) + p12/2 + p13/2 - p2/2 + p6/2 + p7 + p8/2 &&
 p11 == -(1/2) - p12/2 + p13/2 + p2/2 + p6/2 + p7 + p8/2 &&
 p5 == 1 - p6 - p7 && p9 == 1 - p7 - p8 &&
 p4 == 1/2 - p12/2 - p13/2 + p2/2 + p6/2 - p8/2 &&
 p10 == 1/2 + p12/2 - p13/2 - p2/2 - p6/2 + p8/2

\end{lstlisting}  }

Computation of all the two-valued states thereon:

{ \begin{lstlisting}[backgroundcolor=\color{yellow!10},framerule=0pt,breaklines=true, frame=tb]

Reduce[p1 + p2 + p3 == 1 && p3 + p4 + p5 == 1 && p5 + p6 + p7 == 1 &&
  p7 + p8 + p9 == 1 && p9 + p10 + p11 == 1 && p11 + p12 + p1 == 1 &&
  p4 + p10 + p13 == 1 && p1^2 == p1 && p2^2 == p2 && p3^2 == p3 &&
  p4^2 == p4 && p5^2 == p5 && p6^2 == p6 && p7^2 == p7 && p8^2 == p8 &&
   p9^2 == p9 && p10^2 == p10 && p11^2 == p11 && p12^2 == p12 &&
  p13^2 == p13]

~~~~~~ Mathematica response

(p9 == 0 && p8 == 0 && p7 == 1 && p6 == 0 && p5 == 0 && p4 == 0 &&
   p3 == 1 && p2 == 0 && p13 == 0 && p12 == 1 && p11 == 0 &&
   p10 == 1 && p1 == 0) || (p9 == 0 && p8 == 0 && p7 == 1 && p6 == 0 &&
    p5 == 0 && p4 == 0 && p3 == 1 && p2 == 0 && p13 == 1 && p12 == 0 &&
    p11 == 1 && p10 == 0 && p1 == 0) || (p9 == 0 && p8 == 0 &&
   p7 == 1 && p6 == 0 && p5 == 0 && p4 == 1 && p3 == 0 && p2 == 1 &&
   p13 == 0 && p12 == 0 && p11 == 1 && p10 == 0 &&
   p1 == 0) || (p9 == 0 && p8 == 1 && p7 == 0 && p6 == 0 && p5 == 1 &&
    p4 == 0 && p3 == 0 && p2 == 0 && p13 == 0 && p12 == 0 &&
   p11 == 0 && p10 == 1 && p1 == 1) || (p9 == 0 && p8 == 1 &&
   p7 == 0 && p6 == 0 && p5 == 1 && p4 == 0 && p3 == 0 && p2 == 1 &&
   p13 == 0 && p12 == 1 && p11 == 0 && p10 == 1 &&
   p1 == 0) || (p9 == 0 && p8 == 1 && p7 == 0 && p6 == 0 && p5 == 1 &&
    p4 == 0 && p3 == 0 && p2 == 1 && p13 == 1 && p12 == 0 &&
   p11 == 1 && p10 == 0 && p1 == 0) || (p9 == 0 && p8 == 1 &&
   p7 == 0 && p6 == 1 && p5 == 0 && p4 == 0 && p3 == 1 && p2 == 0 &&
   p13 == 0 && p12 == 1 && p11 == 0 && p10 == 1 &&
   p1 == 0) || (p9 == 0 && p8 == 1 && p7 == 0 && p6 == 1 && p5 == 0 &&
    p4 == 0 && p3 == 1 && p2 == 0 && p13 == 1 && p12 == 0 &&
   p11 == 1 && p10 == 0 && p1 == 0) || (p9 == 0 && p8 == 1 &&
   p7 == 0 && p6 == 1 && p5 == 0 && p4 == 1 && p3 == 0 && p2 == 1 &&
   p13 == 0 && p12 == 0 && p11 == 1 && p10 == 0 &&
   p1 == 0) || (p9 == 1 && p8 == 0 && p7 == 0 && p6 == 0 && p5 == 1 &&
    p4 == 0 && p3 == 0 && p2 == 0 && p13 == 1 && p12 == 0 &&
   p11 == 0 && p10 == 0 && p1 == 1) || (p9 == 1 && p8 == 0 &&
   p7 == 0 && p6 == 0 && p5 == 1 && p4 == 0 && p3 == 0 && p2 == 1 &&
   p13 == 1 && p12 == 1 && p11 == 0 && p10 == 0 &&
   p1 == 0) || (p9 == 1 && p8 == 0 && p7 == 0 && p6 == 1 && p5 == 0 &&
    p4 == 0 && p3 == 1 && p2 == 0 && p13 == 1 && p12 == 1 &&
   p11 == 0 && p10 == 0 && p1 == 0) || (p9 == 1 && p8 == 0 &&
   p7 == 0 && p6 == 1 && p5 == 0 && p4 == 1 && p3 == 0 && p2 == 0 &&
   p13 == 0 && p12 == 0 && p11 == 0 && p10 == 0 &&
   p1 == 1) || (p9 == 1 && p8 == 0 && p7 == 0 && p6 == 1 && p5 == 0 &&
    p4 == 1 && p3 == 0 && p2 == 1 && p13 == 0 && p12 == 1 &&
   p11 == 0 && p10 == 0 && p1 == 0)

\end{lstlisting}  }

\subsubsection{Hull calculation for the probabilities on the Specker bug logic}

{ \begin{lstlisting}[backgroundcolor=\color{yellow!10},framerule=0pt,breaklines=true, frame=tb]

* 13 probabilities on atoms a1...a13:
* p1 ... p13
*
V-representation
begin
14  14  real
1 1 0 0 0 1 0 0 0 1 0 0 0 1
1 1 0 0 1 0 1 0 0 1 0 0 0 0
1 1 0 0 0 1 0 0 1 0 1 0 0 0
1 0 1 0 0 1 0 0 0 1 0 0 1 1
1 0 1 0 0 1 0 0 1 0 0 1 0 1
1 0 1 0 1 0 1 0 0 1 0 0 1 0
1 0 1 0 1 0 0 1 0 0 0 1 0 0
1 0 1 0 1 0 1 0 1 0 0 1 0 0
1 0 1 0 0 1 0 0 1 0 1 0 1 0
1 0 0 1 0 0 0 1 0 0 0 1 0 1
1 0 0 1 0 0 1 0 1 0 0 1 0 1
1 0 0 1 0 0 1 0 0 1 0 0 1 1
1 0 0 1 0 0 0 1 0 0 1 0 1 0
1 0 0 1 0 0 1 0 1 0 1 0 1 0
end

~~~~~~ cddlib response


H-representation
linearity 7  17 18 19 20 21 22 23
begin
 23 14 real
  0  0  0  0  1  0  0  0  0  0  0  0  0  0
  0  0  0  0  0  0  1  0  0  0  0  0  0  0
  0  1  1  0 -1  0  1  0 -1  0  0  0  0  0
  0  1  0  0  0  0  0  0  0  0  0  0  0  0
  0  1  1  0 -1  0  0  0  0  0  0  0  0  0
  0  1  2  0 -2  0  1  0 -1  0  0  0  0  0
  0  0  1  0 -1  0  1  0  0  0  0  0  0  0
  0  0  1  0  0  0  0  0  0  0  0  0  0  0
  0  0  0  0  0  0  0  0  0  0  1  0  0  0
  0  0  0  0  0  0  0  0  1  0  0  0  0  0
  0  0  1  0 -1  0  1  0 -1  0  1  0  0  0
  1  0  0  0 -1  0  0  0  0  0 -1  0  0  0
  1 -1 -1  0  1  0 -1  0  1  0 -1  0  0  0
  1 -1 -1  0  0  0  0  0  1  0 -1  0  0  0
  1 -1 -1  0  0  0  0  0  0  0  0  0  0  0
  1 -1 -1  0  1  0 -1  0  0  0  0  0  0  0
 -1  1  1  1  0  0  0  0  0  0  0  0  0  0
  0 -1 -1  0  1  1  0  0  0  0  0  0  0  0
 -1  1  1  0 -1  0  1  1  0  0  0  0  0  0
  0 -1 -1  0  1  0 -1  0  1  1  0  0  0  0
 -1  1  1  0 -1  0  1  0 -1  0  1  1  0  0
  0  0 -1  0  1  0 -1  0  1  0 -1  0  1  0
 -1  0  0  0  1  0  0  0  0  0  1  0  0  1
end

\end{lstlisting}  }

The resulting face inequalities are
\begin{eqnarray}
                     - p_4                                                                                 \le     0 , \\
                                   - p_6                                                                   \le     0 , \\
- p_1  - p_2         + p_4         - p_6         + p_8                                                     \le     0 , \\
- p_1                                                                                                      \le     0 , \\
- p_1  - p_2         + p_4                                                                                 \le     0 , \\
- p_1  -2p_2         +2p_4         - p_6         + p_8                                                     \le     0 , \\
       - p_2         + p_4         - p_6                                                                   \le     0 , \\
       - p_2                                                                                               \le     0 , \\
                                                               - p_{10}                                    \le     0 , \\
                                                 - p_8                                                     \le     0 , \\
       - p_2         + p_4         - p_6         + p_8         - p_{10}                                    \le     0 , \\
                     + p_4                                     + p_{10}                                    \le    +1 , \\
+ p_1  + p_2         - p_4         + p_6         - p_8         + p_{10}                                    \le    +1 , \\
+ p_1  + p_2                                     - p_8         + p_{10}                                    \le    +1 , \\
+ p_1  + p_2                                                                                               \le    +1 , \\
+ p_1  + p_2         - p_4         + p_6                                                                   \le    +1 , \\
- p_1  - p_2  - p_3                                                                                        \le    -1 , \\
+ p_1  + p_2         - p_4  - p_5                                                                          \le     0 , \\
- p_1  - p_2         + p_4         - p_6  - p_7                                                            \le    -1 ,  \\
+ p_1  + p_2         - p_4         + p_6         - p_8  - p_9                                              \le     0 , \\
- p_1  - p_2         + p_4         - p_6         + p_8         - p_{10}  - p_{11}                          \le    -1 ,  \\
       + p_2         - p_4         + p_6         - p_8         + p_{10}            - p_{12}                \le     0 , \\
                     - p_4                                     - p_{10}                      - p_{13}      \le    -1
.
\label{2017-b-sb-p-c}
\end{eqnarray}

\subsubsection{Hull calculation for the expectations on the Specker bug logic}

{ \begin{lstlisting}[backgroundcolor=\color{yellow!10},framerule=0pt,breaklines=true, frame=tb]

* (13 expectations on atoms a1...a13:
* E1 ... E13 not enumerated)
* 6 joint expectations  E1*E3, E3*E5, ..., E11*E1
*
V-representation
begin
14  7  integer
1      -1   -1   -1   -1   -1   -1
1      -1    1    1   -1   -1   -1
1      -1   -1   -1    1    1   -1
1       1   -1   -1   -1   -1    1
1       1   -1   -1    1   -1   -1
1       1    1    1   -1   -1    1
1       1    1   -1   -1   -1   -1
1       1    1    1    1   -1   -1
1       1   -1   -1    1    1    1
1      -1   -1   -1   -1   -1   -1
1      -1   -1    1    1   -1   -1
1      -1   -1    1   -1   -1    1
1      -1   -1   -1   -1    1    1
1      -1   -1    1    1    1    1
end

~~~~~~ cddlib response


H-representation
linearity 1  18
begin
 18 7 real
  1  0  0  0  1  0  0
  1 -1  0  0  1 -1  0
  1 -1  1 -1  1 -1  0
  1  0  0 -1  1 -1  0
  1  0  1  0  0  0  0
  1  1  0  0  0  0  0
  1  1 -1  1  0  0  0
  1  0  0  1  0  0  0
  1  1 -1  0 -1  0  0
  1  0  0  0 -1  0  0
  1  0 -1  1 -1  0  0
  1  1 -1  1 -1  1  0
  1  0  0 -1  0  0  0
  1 -1  1 -1  0  0  0
  1 -1  0  0  0  0  0
  1  0  0  0  0  1  0
  0  0 -1  0  0 -1  0
  0 -1  1 -1  1 -1  1
end

\end{lstlisting}  }

\subsubsection{Extended Specker bug logic}

Here is the {\em Mathematica}~\cite{Mathematica11.1} code to reduce probabilities on the extended (by two contexts) Specker bug logics:

{ \begin{lstlisting}[backgroundcolor=\color{yellow!10},framerule=0pt,breaklines=true, frame=tb]

Reduce[
p1 + p2 + p3 == 1
&& p3 + p4 + p5 == 1
&& p5 + p6 + p7 == 1
&&   p7 + p8 + p9 == 1
&& p9 + p10 + p11 == 1
&& p11 + p12 + p1 == 1
&&  p4 + p10 + p13 == 1
&& p1 + pc + q7 ==1
&& p7 + pc + q1 ==1,
{p3, p11, p5, p9, p4, p10, q3, q11, q5, q9, q4, q10, p13, q13, pc}]

~~~~~~ Mathematica response

p1 == p7 + q1 - q7 && p3 == 1 - p2 - p7 - q1 + q7 &&
 p11 == 1 - p12 - p7 - q1 + q7 && p5 == 1 - p6 - p7 &&
 p9 == 1 - p7 - p8 && p4 == -1 + p2 + p6 + 2 p7 + q1 - q7 &&
 p10 == -1 + p12 + 2 p7 + p8 + q1 - q7 &&
 p13 == 3 - p12 - p2 - p6 - 4 p7 - p8 - 2 q1 + 2 q7 &&
 pc == 1 - p7 - q1

\end{lstlisting}  }

Computation of all the 112 two-valued states thereon:

{ \begin{lstlisting}[backgroundcolor=\color{yellow!10},framerule=0pt,breaklines=true, frame=tb]

Reduce[p1 + p2 + p3 == 1 && p3 + p4 + p5 == 1 && p5 + p6 + p7 == 1 &&
  p7 + p8 + p9 == 1 && p9 + p10 + p11 == 1 && p11 + p12 + p1 == 1 &&
  p4 + p10 + p13 == 1 && p1^2 == p1 && p2^2 == p2 && p3^2 == p3 &&
  p4^2 == p4 && p5^2 == p5 && p6^2 == p6 && p7^2 == p7 && p8^2 == p8 &&
   p9^2 == p9 && p10^2 == p10 && p11^2 == p11 && p12^2 == p12 &&
  p13^2 == p13 &&  q1^2 == q1 &&  q7^2 == q7 &&  pc^2 == pc]

~~~~~~ Mathematica response

q7 == 0 && q1 == 0 && pc == 0 && p9 == 0 && p8 == 0 && p7 == 1 &&
   p6 == 0 && p5 == 0 && p4 == 0 && p3 == 1 && p2 == 0 && p13 == 0 &&
   p12 == 1 && p11 == 0 && p10 == 1 && p1 == 0) || (q7 == 0 &&
   q1 == 0 && pc == 0 && p9 == 0 && p8 == 0 && p7 == 1 && p6 == 0 &&
   p5 == 0 && p4 == 0 && p3 == 1 && p2 == 0 && p13 == 1 && p12 == 0 &&
    p11 == 1 && p10 == 0 && p1 == 0) ||
    [...]
  || (q7 == 1 && q1 == 1 && pc == 1 && p9 == 1 && p8 == 0 &&
    p7 == 0 && p6 == 1 && p5 == 0 && p4 == 1 && p3 == 0 && p2 == 1 &&
   p13 == 0 && p12 == 1 && p11 == 0 && p10 == 0 && p1 == 0)

\end{lstlisting}  }

\subsection{Two intertwined Specker bug logics}
\label{2017-b-bugscombinoa}

Here is the {\em Mathematica}~\cite{Mathematica11.1} code to reduce probabilities on two intertwined Specker bug logics:

{ \begin{lstlisting}[backgroundcolor=\color{yellow!10},framerule=0pt,breaklines=true, frame=tb]

Reduce[
p1 + p2 + p3 == 1
&& p3 + p4 + p5 == 1
&& p5 + p6 + p7 == 1
&&   p7 + p8 + p9 == 1
&& p9 + p10 + p11 == 1
&& p11 + p12 + p1 == 1
&&  p4 + p10 + p13 == 1
&& q1 + q2 + q3 == 1
&& q3 + q4 + q5 == 1
&& q5 + q6 + q7 == 1
&&   q7 + q8 + q9 == 1
&& q9 + q10 + q11 == 1
&& q11 + q12 + q1 == 1
&&  q4 + q10 + q13 == 1
&& p1 + pc + q7 ==1
&& p7 + pc + q1 ==1,
{p3, p11, p5, p9, p4, p10, q3, q11, q5, q9, q4, q10, p13, q13, pc}]

~~~~~~ Mathematica response

p1 == p7 + q1 - q7 && p3 == 1 - p2 - p7 - q1 + q7 &&
 p11 == 1 - p12 - p7 - q1 + q7 && p5 == 1 - p6 - p7 &&
 p9 == 1 - p7 - p8 && p4 == -1 + p2 + p6 + 2 p7 + q1 - q7 &&
 p10 == -1 + p12 + 2 p7 + p8 + q1 - q7 && q3 == 1 - q1 - q2 &&
 q11 == 1 - q1 - q12 && q5 == 1 - q6 - q7 && q9 == 1 - q7 - q8 &&
 q4 == -1 + q1 + q2 + q6 + q7 && q10 == -1 + q1 + q12 + q7 + q8 &&
 p13 == 3 - p12 - p2 - p6 - 4 p7 - p8 - 2 q1 + 2 q7 &&
 q13 == 3 - 2 q1 - q12 - q2 - q6 - 2 q7 - q8 && pc == 1 - p7 - q1

\end{lstlisting}  }

%
%
%
%
%

\subsubsection{Hull calculation for the contexual inequalities corresponding to the Cabello, Estebaranz and Garc{\'{i}}a-Alcaine logic}

{ \begin{lstlisting}[backgroundcolor=\color{yellow!10},framerule=0pt,breaklines=true, frame=tb]

* (13 expectations on atoms A1...A18:
*  not enumerated)
*  9 4th order expectations  A1A2A3A4 A4A5A6A7 ... A2A9A11A18
*
V-representation
begin
262144  10   real
1   1    1    1    1    1    1    1    1    1
1   1    1    1    1    1   -1   -1    1    1
1   1    1    1    1    1   -1    1    1   -1
[[...]]
1   1    1    1    1    1   -1    1    1   -1
1   1    1    1    1    1   -1   -1    1    1
1   1    1    1    1    1    1    1    1    1
end

~~~~~~ cddlib response

H-representation
begin
 274 10 real
  1  0  0  0  0  0  0  0  0  1
  1  0  0  0  0  0  0  0  1  0
  7 -1 -1 -1 -1 -1 -1  1  1  1
  7 -1 -1 -1 -1 -1  1 -1  1  1
  7 -1 -1 -1 -1  1 -1 -1  1  1
  7 -1 -1 -1  1 -1 -1 -1  1  1
  7 -1 -1  1 -1 -1 -1 -1  1  1
  7 -1  1 -1 -1 -1 -1 -1  1  1
  7  1 -1 -1 -1 -1 -1 -1  1  1
  1  0  0  0  0  0  0  1  0  0
  7 -1 -1 -1 -1 -1  1  1 -1  1
  7 -1 -1 -1 -1  1 -1  1 -1  1
  7 -1 -1 -1  1 -1 -1  1 -1  1
  7 -1 -1  1 -1 -1 -1  1 -1  1
  7 -1  1 -1 -1 -1 -1  1 -1  1
  7  1 -1 -1 -1 -1 -1  1 -1  1
  7 -1 -1 -1 -1 -1  1  1  1 -1
  7 -1 -1 -1 -1  1 -1  1  1 -1
  7 -1 -1 -1  1 -1 -1  1  1 -1
  7 -1 -1  1 -1 -1 -1  1  1 -1
  7 -1  1 -1 -1 -1 -1  1  1 -1
  7  1 -1 -1 -1 -1 -1  1  1 -1
  1  0  0  0  0  0  1  0  0  0
  7 -1 -1 -1 -1  1  1 -1 -1  1
  7 -1 -1 -1  1 -1  1 -1 -1  1
  7 -1 -1  1 -1 -1  1 -1 -1  1
  7 -1  1 -1 -1 -1  1 -1 -1  1
  7  1 -1 -1 -1 -1  1 -1 -1  1
  7 -1 -1 -1 -1  1  1 -1  1 -1
  7 -1 -1 -1  1 -1  1 -1  1 -1
  7 -1 -1  1 -1 -1  1 -1  1 -1
  7 -1  1 -1 -1 -1  1 -1  1 -1
  7  1 -1 -1 -1 -1  1 -1  1 -1
  7 -1 -1 -1 -1  1  1  1 -1 -1
  7 -1 -1 -1  1 -1  1  1 -1 -1
  7 -1 -1  1 -1 -1  1  1 -1 -1
  7 -1  1 -1 -1 -1  1  1 -1 -1
  7  1 -1 -1 -1 -1  1  1 -1 -1
  7 -1 -1 -1 -1  1  1  1  1  1
  7 -1 -1 -1  1 -1  1  1  1  1
  7 -1 -1  1 -1 -1  1  1  1  1
  7 -1  1 -1 -1 -1  1  1  1  1
  7  1 -1 -1 -1 -1  1  1  1  1
  1  0  0  0  0  1  0  0  0  0
  7 -1 -1 -1  1  1 -1 -1 -1  1
  7 -1 -1  1 -1  1 -1 -1 -1  1
  7 -1  1 -1 -1  1 -1 -1 -1  1
  7  1 -1 -1 -1  1 -1 -1 -1  1
  7 -1 -1 -1  1  1 -1 -1  1 -1
  7 -1 -1  1 -1  1 -1 -1  1 -1
  7 -1  1 -1 -1  1 -1 -1  1 -1
  7  1 -1 -1 -1  1 -1 -1  1 -1
  7 -1 -1 -1  1  1 -1  1 -1 -1
  7 -1 -1  1 -1  1 -1  1 -1 -1
  7 -1  1 -1 -1  1 -1  1 -1 -1
  7  1 -1 -1 -1  1 -1  1 -1 -1
  7 -1 -1 -1  1  1 -1  1  1  1
  7 -1 -1  1 -1  1 -1  1  1  1
  7 -1  1 -1 -1  1 -1  1  1  1
  7  1 -1 -1 -1  1 -1  1  1  1
  7 -1 -1 -1  1  1  1 -1 -1 -1
  7 -1 -1  1 -1  1  1 -1 -1 -1
  7 -1  1 -1 -1  1  1 -1 -1 -1
  7  1 -1 -1 -1  1  1 -1 -1 -1
  7 -1 -1 -1  1  1  1 -1  1  1
  7 -1 -1  1 -1  1  1 -1  1  1
  7 -1  1 -1 -1  1  1 -1  1  1
  7  1 -1 -1 -1  1  1 -1  1  1
  7 -1 -1 -1  1  1  1  1 -1  1
  7 -1 -1  1 -1  1  1  1 -1  1
  7 -1  1 -1 -1  1  1  1 -1  1
  7  1 -1 -1 -1  1  1  1 -1  1
  7 -1 -1 -1  1  1  1  1  1 -1
  7 -1 -1  1 -1  1  1  1  1 -1
  7 -1  1 -1 -1  1  1  1  1 -1
  7  1 -1 -1 -1  1  1  1  1 -1
  1  0  0  0  1  0  0  0  0  0
  7 -1 -1  1  1 -1 -1 -1 -1  1
  7 -1  1 -1  1 -1 -1 -1 -1  1
  7  1 -1 -1  1 -1 -1 -1 -1  1
  7 -1 -1  1  1 -1 -1 -1  1 -1
  7 -1  1 -1  1 -1 -1 -1  1 -1
  7  1 -1 -1  1 -1 -1 -1  1 -1
  7 -1 -1  1  1 -1 -1  1 -1 -1
  7 -1  1 -1  1 -1 -1  1 -1 -1
  7  1 -1 -1  1 -1 -1  1 -1 -1
  7 -1 -1  1  1 -1 -1  1  1  1
  7 -1  1 -1  1 -1 -1  1  1  1
  7  1 -1 -1  1 -1 -1  1  1  1
  7 -1 -1  1  1 -1  1 -1 -1 -1
  7 -1  1 -1  1 -1  1 -1 -1 -1
  7  1 -1 -1  1 -1  1 -1 -1 -1
  7 -1 -1  1  1 -1  1 -1  1  1
  7 -1  1 -1  1 -1  1 -1  1  1
  7  1 -1 -1  1 -1  1 -1  1  1
  7 -1 -1  1  1 -1  1  1 -1  1
  7 -1  1 -1  1 -1  1  1 -1  1
  7  1 -1 -1  1 -1  1  1 -1  1
  7 -1 -1  1  1 -1  1  1  1 -1
  7 -1  1 -1  1 -1  1  1  1 -1
  7  1 -1 -1  1 -1  1  1  1 -1
  7 -1 -1  1  1  1 -1 -1 -1 -1
  7 -1  1 -1  1  1 -1 -1 -1 -1
  7  1 -1 -1  1  1 -1 -1 -1 -1
  7 -1 -1  1  1  1 -1 -1  1  1
  7 -1  1 -1  1  1 -1 -1  1  1
  7  1 -1 -1  1  1 -1 -1  1  1
  7 -1 -1  1  1  1 -1  1 -1  1
  7 -1  1 -1  1  1 -1  1 -1  1
  7  1 -1 -1  1  1 -1  1 -1  1
  7 -1 -1  1  1  1 -1  1  1 -1
  7 -1  1 -1  1  1 -1  1  1 -1
  7  1 -1 -1  1  1 -1  1  1 -1
  7 -1 -1  1  1  1  1 -1 -1  1
  7 -1  1 -1  1  1  1 -1 -1  1
  7  1 -1 -1  1  1  1 -1 -1  1
  7 -1 -1  1  1  1  1 -1  1 -1
  7 -1  1 -1  1  1  1 -1  1 -1
  7  1 -1 -1  1  1  1 -1  1 -1
  7 -1 -1  1  1  1  1  1 -1 -1
  7 -1  1 -1  1  1  1  1 -1 -1
  7  1 -1 -1  1  1  1  1 -1 -1
  7 -1 -1  1  1  1  1  1  1  1
  7 -1  1 -1  1  1  1  1  1  1
  7  1 -1 -1  1  1  1  1  1  1
  1  0  0  1  0  0  0  0  0  0
  7 -1  1  1 -1 -1 -1 -1 -1  1
  7  1 -1  1 -1 -1 -1 -1 -1  1
  7 -1  1  1 -1 -1 -1 -1  1 -1
  7  1 -1  1 -1 -1 -1 -1  1 -1
  7 -1  1  1 -1 -1 -1  1 -1 -1
  7  1 -1  1 -1 -1 -1  1 -1 -1
  7 -1  1  1 -1 -1 -1  1  1  1
  7  1 -1  1 -1 -1 -1  1  1  1
  7 -1  1  1 -1 -1  1 -1 -1 -1
  7  1 -1  1 -1 -1  1 -1 -1 -1
  7 -1  1  1 -1 -1  1 -1  1  1
  7  1 -1  1 -1 -1  1 -1  1  1
  7 -1  1  1 -1 -1  1  1 -1  1
  7  1 -1  1 -1 -1  1  1 -1  1
  7 -1  1  1 -1 -1  1  1  1 -1
  7  1 -1  1 -1 -1  1  1  1 -1
  7 -1  1  1 -1  1 -1 -1 -1 -1
  7  1 -1  1 -1  1 -1 -1 -1 -1
  7 -1  1  1 -1  1 -1 -1  1  1
  7  1 -1  1 -1  1 -1 -1  1  1
  7 -1  1  1 -1  1 -1  1 -1  1
  7  1 -1  1 -1  1 -1  1 -1  1
  7 -1  1  1 -1  1 -1  1  1 -1
  7  1 -1  1 -1  1 -1  1  1 -1
  7 -1  1  1 -1  1  1 -1 -1  1
  7  1 -1  1 -1  1  1 -1 -1  1
  7 -1  1  1 -1  1  1 -1  1 -1
  7  1 -1  1 -1  1  1 -1  1 -1
  7 -1  1  1 -1  1  1  1 -1 -1
  7  1 -1  1 -1  1  1  1 -1 -1
  7 -1  1  1 -1  1  1  1  1  1
  7  1 -1  1 -1  1  1  1  1  1
  7 -1  1  1  1 -1 -1 -1 -1 -1
  7  1 -1  1  1 -1 -1 -1 -1 -1
  7 -1  1  1  1 -1 -1 -1  1  1
  7  1 -1  1  1 -1 -1 -1  1  1
  7 -1  1  1  1 -1 -1  1 -1  1
  7  1 -1  1  1 -1 -1  1 -1  1
  7 -1  1  1  1 -1 -1  1  1 -1
  7  1 -1  1  1 -1 -1  1  1 -1
  7 -1  1  1  1 -1  1 -1 -1  1
  7  1 -1  1  1 -1  1 -1 -1  1
  7 -1  1  1  1 -1  1 -1  1 -1
  7  1 -1  1  1 -1  1 -1  1 -1
  7 -1  1  1  1 -1  1  1 -1 -1
  7  1 -1  1  1 -1  1  1 -1 -1
  7 -1  1  1  1 -1  1  1  1  1
  7  1 -1  1  1 -1  1  1  1  1
  7 -1  1  1  1  1 -1 -1 -1  1
  7  1 -1  1  1  1 -1 -1 -1  1
  7 -1  1  1  1  1 -1 -1  1 -1
  7  1 -1  1  1  1 -1 -1  1 -1
  7 -1  1  1  1  1 -1  1 -1 -1
  7  1 -1  1  1  1 -1  1 -1 -1
  7 -1  1  1  1  1 -1  1  1  1
  7  1 -1  1  1  1 -1  1  1  1
  7 -1  1  1  1  1  1 -1 -1 -1
  7  1 -1  1  1  1  1 -1 -1 -1
  7 -1  1  1  1  1  1 -1  1  1
  7  1 -1  1  1  1  1 -1  1  1
  7 -1  1  1  1  1  1  1 -1  1
  7  1 -1  1  1  1  1  1 -1  1
  7 -1  1  1  1  1  1  1  1 -1
  7  1 -1  1  1  1  1  1  1 -1
  1  0  1  0  0  0  0  0  0  0
  7  1  1 -1 -1 -1 -1 -1 -1  1
  7  1  1 -1 -1 -1 -1 -1  1 -1
  7  1  1 -1 -1 -1 -1  1 -1 -1
  7  1  1 -1 -1 -1 -1  1  1  1
  7  1  1 -1 -1 -1  1 -1 -1 -1
  7  1  1 -1 -1 -1  1 -1  1  1
  7  1  1 -1 -1 -1  1  1 -1  1
  7  1  1 -1 -1 -1  1  1  1 -1
  7  1  1 -1 -1  1 -1 -1 -1 -1
  7  1  1 -1 -1  1 -1 -1  1  1
  7  1  1 -1 -1  1 -1  1 -1  1
  7  1  1 -1 -1  1 -1  1  1 -1
  7  1  1 -1 -1  1  1 -1 -1  1
  7  1  1 -1 -1  1  1 -1  1 -1
  7  1  1 -1 -1  1  1  1 -1 -1
  7  1  1 -1 -1  1  1  1  1  1
  7  1  1 -1  1 -1 -1 -1 -1 -1
  7  1  1 -1  1 -1 -1 -1  1  1
  7  1  1 -1  1 -1 -1  1 -1  1
  7  1  1 -1  1 -1 -1  1  1 -1
  7  1  1 -1  1 -1  1 -1 -1  1
  7  1  1 -1  1 -1  1 -1  1 -1
  7  1  1 -1  1 -1  1  1 -1 -1
  7  1  1 -1  1 -1  1  1  1  1
  7  1  1 -1  1  1 -1 -1 -1  1
  7  1  1 -1  1  1 -1 -1  1 -1
  7  1  1 -1  1  1 -1  1 -1 -1
  7  1  1 -1  1  1 -1  1  1  1
  7  1  1 -1  1  1  1 -1 -1 -1
  7  1  1 -1  1  1  1 -1  1  1
  7  1  1 -1  1  1  1  1 -1  1
  7  1  1 -1  1  1  1  1  1 -1
  7  1  1  1 -1 -1 -1 -1 -1 -1
  7  1  1  1 -1 -1 -1 -1  1  1
  7  1  1  1 -1 -1 -1  1 -1  1
  7  1  1  1 -1 -1 -1  1  1 -1
  7  1  1  1 -1 -1  1 -1 -1  1
  7  1  1  1 -1 -1  1 -1  1 -1
  7  1  1  1 -1 -1  1  1 -1 -1
  7  1  1  1 -1 -1  1  1  1  1
  7  1  1  1 -1  1 -1 -1 -1  1
  7  1  1  1 -1  1 -1 -1  1 -1
  7  1  1  1 -1  1 -1  1 -1 -1
  7  1  1  1 -1  1 -1  1  1  1
  7  1  1  1 -1  1  1 -1 -1 -1
  7  1  1  1 -1  1  1 -1  1  1
  7  1  1  1 -1  1  1  1 -1  1
  7  1  1  1 -1  1  1  1  1 -1
  7  1  1  1  1 -1 -1 -1 -1  1
  7  1  1  1  1 -1 -1 -1  1 -1
  7  1  1  1  1 -1 -1  1 -1 -1
  7  1  1  1  1 -1 -1  1  1  1
  7  1  1  1  1 -1  1 -1 -1 -1
  7  1  1  1  1 -1  1 -1  1  1
  7  1  1  1  1 -1  1  1 -1  1
  7  1  1  1  1 -1  1  1  1 -1
  7  1  1  1  1  1 -1 -1 -1 -1
  7  1  1  1  1  1 -1 -1  1  1
  7  1  1  1  1  1 -1  1 -1  1
  7  1  1  1  1  1 -1  1  1 -1
  7  1  1  1  1  1  1 -1 -1  1
  7  1  1  1  1  1  1 -1  1 -1
  7  1  1  1  1  1  1  1 -1 -1
  7  1  1  1  1  1  1  1  1  1
  1  1  0  0  0  0  0  0  0  0
  7  1 -1 -1 -1 -1 -1 -1 -1 -1
  7 -1  1 -1 -1 -1 -1 -1 -1 -1
  7 -1 -1  1 -1 -1 -1 -1 -1 -1
  7 -1 -1 -1  1 -1 -1 -1 -1 -1
  7 -1 -1 -1 -1  1 -1 -1 -1 -1
  7 -1 -1 -1 -1 -1  1 -1 -1 -1
  7 -1 -1 -1 -1 -1 -1  1 -1 -1
  7 -1 -1 -1 -1 -1 -1 -1  1 -1
  7 -1 -1 -1 -1 -1 -1 -1 -1  1
  1  0  0  0  0  0  0  0  0 -1
  1  0  0  0  0  0  0  0 -1  0
  1  0  0  0  0  0  0 -1  0  0
  1  0  0  0  0  0 -1  0  0  0
  1  0  0  0  0 -1  0  0  0  0
  1  0  0  0 -1  0  0  0  0  0
  1  0  0 -1  0  0  0  0  0  0
  1  0 -1  0  0  0  0  0  0  0
  1 -1  0  0  0  0  0  0  0  0
end

~~~~~~ cddlib reverse vertex computation


V-representation
begin
 256 10 real
  1 -1 -1 -1 -1 -1 -1  1  1  1
  1 -1 -1 -1 -1 -1  1 -1  1  1
  1 -1 -1 -1 -1  1 -1 -1  1  1
  1 -1 -1 -1  1 -1 -1 -1  1  1
  1 -1 -1  1 -1 -1 -1 -1  1  1
  1 -1  1 -1 -1 -1 -1 -1  1  1
  1  1 -1 -1 -1 -1 -1 -1  1  1
  1  1 -1 -1 -1 -1 -1  1  1 -1
  1 -1  1 -1 -1 -1 -1  1  1 -1
  1 -1 -1  1 -1 -1 -1  1  1 -1
  1 -1 -1 -1  1 -1 -1  1  1 -1
  1 -1 -1 -1 -1  1 -1  1  1 -1
  1 -1 -1 -1 -1 -1  1  1  1 -1
  1  1 -1 -1 -1 -1  1 -1  1 -1
  1 -1  1 -1 -1 -1  1 -1  1 -1
  1 -1 -1  1 -1 -1  1 -1  1 -1
  1 -1 -1 -1  1 -1  1 -1  1 -1
  1 -1 -1 -1 -1  1  1 -1  1 -1
  1  1 -1 -1 -1 -1  1  1  1  1
  1 -1  1 -1 -1 -1  1  1  1  1
  1 -1 -1  1 -1 -1  1  1  1  1
  1 -1 -1 -1  1 -1  1  1  1  1
  1 -1 -1 -1 -1  1  1  1  1  1
  1  1 -1 -1 -1  1 -1 -1  1 -1
  1 -1  1 -1 -1  1 -1 -1  1 -1
  1 -1 -1  1 -1  1 -1 -1  1 -1
  1 -1 -1 -1  1  1 -1 -1  1 -1
  1  1 -1 -1 -1  1 -1  1  1  1
  1 -1  1 -1 -1  1 -1  1  1  1
  1 -1 -1  1 -1  1 -1  1  1  1
  1 -1 -1 -1  1  1 -1  1  1  1
  1  1 -1 -1 -1  1  1 -1  1  1
  1 -1  1 -1 -1  1  1 -1  1  1
  1 -1 -1  1 -1  1  1 -1  1  1
  1 -1 -1 -1  1  1  1 -1  1  1
  1  1 -1 -1 -1  1  1  1  1 -1
  1 -1  1 -1 -1  1  1  1  1 -1
  1 -1 -1  1 -1  1  1  1  1 -1
  1 -1 -1 -1  1  1  1  1  1 -1
  1  1 -1 -1  1 -1 -1 -1  1 -1
  1 -1  1 -1  1 -1 -1 -1  1 -1
  1 -1 -1  1  1 -1 -1 -1  1 -1
  1  1 -1 -1  1 -1 -1  1  1  1
  1 -1  1 -1  1 -1 -1  1  1  1
  1 -1 -1  1  1 -1 -1  1  1  1
  1  1 -1 -1  1 -1  1 -1  1  1
  1 -1  1 -1  1 -1  1 -1  1  1
  1 -1 -1  1  1 -1  1 -1  1  1
  1  1 -1 -1  1 -1  1  1  1 -1
  1 -1  1 -1  1 -1  1  1  1 -1
  1 -1 -1  1  1 -1  1  1  1 -1
  1  1 -1 -1  1  1 -1 -1  1  1
  1 -1  1 -1  1  1 -1 -1  1  1
  1 -1 -1  1  1  1 -1 -1  1  1
  1  1 -1 -1  1  1 -1  1  1 -1
  1 -1  1 -1  1  1 -1  1  1 -1
  1 -1 -1  1  1  1 -1  1  1 -1
  1  1 -1 -1  1  1  1 -1  1 -1
  1 -1  1 -1  1  1  1 -1  1 -1
  1 -1 -1  1  1  1  1 -1  1 -1
  1  1 -1 -1  1  1  1  1  1  1
  1 -1  1 -1  1  1  1  1  1  1
  1 -1 -1  1  1  1  1  1  1  1
  1  1 -1  1 -1 -1 -1 -1  1 -1
  1 -1  1  1 -1 -1 -1 -1  1 -1
  1  1 -1  1 -1 -1 -1  1  1  1
  1 -1  1  1 -1 -1 -1  1  1  1
  1  1 -1  1 -1 -1  1 -1  1  1
  1 -1  1  1 -1 -1  1 -1  1  1
  1  1 -1  1 -1 -1  1  1  1 -1
  1 -1  1  1 -1 -1  1  1  1 -1
  1  1 -1  1 -1  1 -1 -1  1  1
  1 -1  1  1 -1  1 -1 -1  1  1
  1  1 -1  1 -1  1 -1  1  1 -1
  1 -1  1  1 -1  1 -1  1  1 -1
  1  1 -1  1 -1  1  1 -1  1 -1
  1 -1  1  1 -1  1  1 -1  1 -1
  1  1 -1  1 -1  1  1  1  1  1
  1 -1  1  1 -1  1  1  1  1  1
  1  1 -1  1  1 -1 -1 -1  1  1
  1 -1  1  1  1 -1 -1 -1  1  1
  1  1 -1  1  1 -1 -1  1  1 -1
  1 -1  1  1  1 -1 -1  1  1 -1
  1  1 -1  1  1 -1  1 -1  1 -1
  1 -1  1  1  1 -1  1 -1  1 -1
  1  1 -1  1  1 -1  1  1  1  1
  1 -1  1  1  1 -1  1  1  1  1
  1  1 -1  1  1  1 -1 -1  1 -1
  1 -1  1  1  1  1 -1 -1  1 -1
  1  1 -1  1  1  1 -1  1  1  1
  1 -1  1  1  1  1 -1  1  1  1
  1  1 -1  1  1  1  1 -1  1  1
  1 -1  1  1  1  1  1 -1  1  1
  1  1 -1  1  1  1  1  1  1 -1
  1 -1  1  1  1  1  1  1  1 -1
  1  1  1 -1 -1 -1 -1 -1  1 -1
  1  1  1 -1 -1 -1 -1  1  1  1
  1  1  1 -1 -1 -1  1 -1  1  1
  1  1  1 -1 -1 -1  1  1  1 -1
  1  1  1 -1 -1  1 -1 -1  1  1
  1  1  1 -1 -1  1 -1  1  1 -1
  1  1  1 -1 -1  1  1 -1  1 -1
  1  1  1 -1 -1  1  1  1  1  1
  1  1  1 -1  1 -1 -1 -1  1  1
  1  1  1 -1  1 -1 -1  1  1 -1
  1  1  1 -1  1 -1  1 -1  1 -1
  1  1  1 -1  1 -1  1  1  1  1
  1  1  1 -1  1  1 -1 -1  1 -1
  1  1  1 -1  1  1 -1  1  1  1
  1  1  1 -1  1  1  1 -1  1  1
  1  1  1 -1  1  1  1  1  1 -1
  1  1  1  1 -1 -1 -1 -1  1  1
  1  1  1  1 -1 -1 -1  1  1 -1
  1  1  1  1 -1 -1  1 -1  1 -1
  1  1  1  1 -1 -1  1  1  1  1
  1  1  1  1 -1  1 -1 -1  1 -1
  1  1  1  1 -1  1 -1  1  1  1
  1  1  1  1 -1  1  1 -1  1  1
  1  1  1  1 -1  1  1  1  1 -1
  1  1  1  1  1 -1 -1 -1  1 -1
  1  1  1  1  1 -1 -1  1  1  1
  1  1  1  1  1 -1  1 -1  1  1
  1  1  1  1  1 -1  1  1  1 -1
  1  1  1  1  1  1 -1 -1  1  1
  1  1  1  1  1  1 -1  1  1 -1
  1  1  1  1  1  1  1 -1  1 -1
  1  1  1  1  1  1  1  1  1  1
  1  1  1  1  1  1  1  1 -1 -1
  1  1  1  1  1  1  1 -1 -1  1
  1  1  1  1  1  1 -1  1 -1  1
  1  1  1  1  1  1 -1 -1 -1 -1
  1  1  1  1  1 -1  1  1 -1  1
  1  1  1  1  1 -1  1 -1 -1 -1
  1  1  1  1  1 -1 -1  1 -1 -1
  1  1  1  1  1 -1 -1 -1 -1  1
  1  1  1  1 -1  1  1  1 -1  1
  1  1  1  1 -1  1  1 -1 -1 -1
  1  1  1  1 -1  1 -1  1 -1 -1
  1  1  1  1 -1  1 -1 -1 -1  1
  1  1  1  1 -1 -1  1  1 -1 -1
  1  1  1  1 -1 -1  1 -1 -1  1
  1  1  1  1 -1 -1 -1  1 -1  1
  1  1  1  1 -1 -1 -1 -1 -1 -1
  1  1  1 -1  1  1  1  1 -1  1
  1  1  1 -1  1  1  1 -1 -1 -1
  1  1  1 -1  1  1 -1  1 -1 -1
  1  1  1 -1  1  1 -1 -1 -1  1
  1  1  1 -1  1 -1  1  1 -1 -1
  1  1  1 -1  1 -1  1 -1 -1  1
  1  1  1 -1  1 -1 -1  1 -1  1
  1  1  1 -1  1 -1 -1 -1 -1 -1
  1  1  1 -1 -1  1  1  1 -1 -1
  1  1  1 -1 -1  1  1 -1 -1  1
  1  1  1 -1 -1  1 -1  1 -1  1
  1  1  1 -1 -1  1 -1 -1 -1 -1
  1  1  1 -1 -1 -1  1  1 -1  1
  1  1  1 -1 -1 -1  1 -1 -1 -1
  1  1  1 -1 -1 -1 -1  1 -1 -1
  1  1  1 -1 -1 -1 -1 -1 -1  1
  1 -1  1  1  1  1  1  1 -1  1
  1  1 -1  1  1  1  1  1 -1  1
  1 -1  1  1  1  1  1 -1 -1 -1
  1  1 -1  1  1  1  1 -1 -1 -1
  1 -1  1  1  1  1 -1  1 -1 -1
  1  1 -1  1  1  1 -1  1 -1 -1
  1 -1  1  1  1  1 -1 -1 -1  1
  1  1 -1  1  1  1 -1 -1 -1  1
  1 -1  1  1  1 -1  1  1 -1 -1
  1  1 -1  1  1 -1  1  1 -1 -1
  1 -1  1  1  1 -1  1 -1 -1  1
  1  1 -1  1  1 -1  1 -1 -1  1
  1 -1  1  1  1 -1 -1  1 -1  1
  1  1 -1  1  1 -1 -1  1 -1  1
  1 -1  1  1  1 -1 -1 -1 -1 -1
  1  1 -1  1  1 -1 -1 -1 -1 -1
  1 -1  1  1 -1  1  1  1 -1 -1
  1  1 -1  1 -1  1  1  1 -1 -1
  1 -1  1  1 -1  1  1 -1 -1  1
  1  1 -1  1 -1  1  1 -1 -1  1
  1 -1  1  1 -1  1 -1  1 -1  1
  1  1 -1  1 -1  1 -1  1 -1  1
  1 -1  1  1 -1  1 -1 -1 -1 -1
  1  1 -1  1 -1  1 -1 -1 -1 -1
  1 -1  1  1 -1 -1  1  1 -1  1
  1  1 -1  1 -1 -1  1  1 -1  1
  1 -1  1  1 -1 -1  1 -1 -1 -1
  1  1 -1  1 -1 -1  1 -1 -1 -1
  1 -1  1  1 -1 -1 -1  1 -1 -1
  1  1 -1  1 -1 -1 -1  1 -1 -1
  1 -1  1  1 -1 -1 -1 -1 -1  1
  1  1 -1  1 -1 -1 -1 -1 -1  1
  1 -1 -1  1  1  1  1  1 -1 -1
  1 -1  1 -1  1  1  1  1 -1 -1
  1  1 -1 -1  1  1  1  1 -1 -1
  1 -1 -1  1  1  1  1 -1 -1  1
  1 -1  1 -1  1  1  1 -1 -1  1
  1  1 -1 -1  1  1  1 -1 -1  1
  1 -1 -1  1  1  1 -1  1 -1  1
  1 -1  1 -1  1  1 -1  1 -1  1
  1  1 -1 -1  1  1 -1  1 -1  1
  1 -1 -1  1  1  1 -1 -1 -1 -1
  1 -1  1 -1  1  1 -1 -1 -1 -1
  1  1 -1 -1  1  1 -1 -1 -1 -1
  1 -1 -1  1  1 -1  1  1 -1  1
  1 -1  1 -1  1 -1  1  1 -1  1
  1  1 -1 -1  1 -1  1  1 -1  1
  1 -1 -1  1  1 -1  1 -1 -1 -1
  1 -1  1 -1  1 -1  1 -1 -1 -1
  1  1 -1 -1  1 -1  1 -1 -1 -1
  1 -1 -1  1  1 -1 -1  1 -1 -1
  1 -1  1 -1  1 -1 -1  1 -1 -1
  1  1 -1 -1  1 -1 -1  1 -1 -1
  1 -1 -1  1  1 -1 -1 -1 -1  1
  1 -1  1 -1  1 -1 -1 -1 -1  1
  1  1 -1 -1  1 -1 -1 -1 -1  1
  1 -1 -1 -1  1  1  1  1 -1  1
  1 -1 -1  1 -1  1  1  1 -1  1
  1 -1  1 -1 -1  1  1  1 -1  1
  1  1 -1 -1 -1  1  1  1 -1  1
  1 -1 -1 -1  1  1  1 -1 -1 -1
  1 -1 -1  1 -1  1  1 -1 -1 -1
  1 -1  1 -1 -1  1  1 -1 -1 -1
  1  1 -1 -1 -1  1  1 -1 -1 -1
  1 -1 -1 -1  1  1 -1  1 -1 -1
  1 -1 -1  1 -1  1 -1  1 -1 -1
  1 -1  1 -1 -1  1 -1  1 -1 -1
  1  1 -1 -1 -1  1 -1  1 -1 -1
  1 -1 -1 -1  1  1 -1 -1 -1  1
  1 -1 -1  1 -1  1 -1 -1 -1  1
  1 -1  1 -1 -1  1 -1 -1 -1  1
  1  1 -1 -1 -1  1 -1 -1 -1  1
  1 -1 -1 -1 -1  1  1  1 -1 -1
  1 -1 -1 -1  1 -1  1  1 -1 -1
  1 -1 -1  1 -1 -1  1  1 -1 -1
  1 -1  1 -1 -1 -1  1  1 -1 -1
  1  1 -1 -1 -1 -1  1  1 -1 -1
  1 -1 -1 -1 -1  1  1 -1 -1  1
  1 -1 -1 -1  1 -1  1 -1 -1  1
  1 -1 -1  1 -1 -1  1 -1 -1  1
  1 -1  1 -1 -1 -1  1 -1 -1  1
  1  1 -1 -1 -1 -1  1 -1 -1  1
  1 -1 -1 -1 -1 -1  1  1 -1  1
  1 -1 -1 -1 -1  1 -1  1 -1  1
  1 -1 -1 -1  1 -1 -1  1 -1  1
  1 -1 -1  1 -1 -1 -1  1 -1  1
  1 -1  1 -1 -1 -1 -1  1 -1  1
  1  1 -1 -1 -1 -1 -1  1 -1  1
  1 -1 -1 -1 -1 -1 -1  1 -1 -1
  1 -1 -1 -1 -1 -1  1 -1 -1 -1
  1 -1 -1 -1 -1  1 -1 -1 -1 -1
  1 -1 -1 -1  1 -1 -1 -1 -1 -1
  1 -1 -1  1 -1 -1 -1 -1 -1 -1
  1 -1  1 -1 -1 -1 -1 -1 -1 -1
  1  1 -1 -1 -1 -1 -1 -1 -1 -1
  1 -1 -1 -1 -1 -1 -1 -1  1 -1
  1 -1 -1 -1 -1 -1 -1 -1 -1  1
end

\end{lstlisting}  }

\subsubsection{Hull calculation for the contexual inequalities corresponding to the pentagon logic}

{ \begin{lstlisting}[backgroundcolor=\color{yellow!10},framerule=0pt,breaklines=true, frame=tb]

* (10 expectations on atoms A1...A10:
*  not enumerated)
*  5 3th order expectations  A1A2A3 A3A4A5 ... A9A10A1
*  obtained through reverse Hull computation
V-representation
begin
 32 6 real
  1  1 -1 -1 -1 -1
  1  1 -1 -1 -1  1
  1  1 -1 -1  1 -1
  1  1 -1 -1  1  1
  1  1 -1  1 -1 -1
  1  1 -1  1 -1  1
  1  1 -1  1  1 -1
  1  1 -1  1  1  1
  1  1  1  1 -1 -1
  1  1  1  1 -1  1
  1  1  1  1  1  1
  1  1  1  1  1 -1
  1  1  1 -1  1  1
  1  1  1 -1  1 -1
  1  1  1 -1 -1  1
  1  1  1 -1 -1 -1
  1 -1  1  1  1  1
  1 -1  1  1  1 -1
  1 -1  1  1 -1  1
  1 -1  1  1 -1 -1
  1 -1  1 -1  1  1
  1 -1  1 -1  1 -1
  1 -1  1 -1 -1  1
  1 -1  1 -1 -1 -1
  1 -1 -1  1  1  1
  1 -1 -1  1  1 -1
  1 -1 -1  1 -1  1
  1 -1 -1  1 -1 -1
  1 -1 -1 -1  1  1
  1 -1 -1 -1  1 -1
  1 -1 -1 -1 -1  1
  1 -1 -1 -1 -1 -1
end

~~~~~~ cddlib response

H-representation
begin
 10 6 real
  1  0  0  0  0  1
  1  0  0  0  1  0
  1  0  0  1  0  0
  1  0  1  0  0  0
  1  1  0  0  0  0
  1  0  0  0  0 -1
  1  0  0  0 -1  0
  1  0  0 -1  0  0
  1  0 -1  0  0  0
  1 -1  0  0  0  0
end

\end{lstlisting}  }

\subsubsection{Hull calculation for the contexual inequalities corresponding to Specker bug logics}

{ \begin{lstlisting}[backgroundcolor=\color{yellow!10},framerule=0pt,breaklines=true, frame=tb]

* (13 expectations on atoms A1...A13:
*  not enumerated)
*  7 3th order expectations  A1A2A3 A3A4A5 ... A11A12A1 A4A13A10
*  obtained through reverse Hull computation
V-representation
begin
 128 8 real
  1  1 -1 -1 -1 -1 -1 -1
  1  1 -1 -1 -1 -1 -1  1
  1  1 -1 -1 -1 -1  1 -1
  1  1 -1 -1 -1 -1  1  1
  1  1 -1 -1 -1  1 -1 -1
  1  1 -1 -1 -1  1 -1  1
  1  1 -1 -1 -1  1  1 -1
  1  1 -1 -1 -1  1  1  1
  1  1 -1 -1  1 -1 -1 -1
  1  1 -1 -1  1 -1 -1  1
  1  1 -1 -1  1 -1  1 -1
  1  1 -1 -1  1 -1  1  1
  1  1 -1 -1  1  1 -1 -1
  1  1 -1 -1  1  1 -1  1
  1  1 -1 -1  1  1  1 -1
  1  1 -1 -1  1  1  1  1
  1  1 -1  1 -1 -1 -1 -1
  1  1 -1  1 -1 -1 -1  1
  1  1 -1  1 -1 -1  1 -1
  1  1 -1  1 -1 -1  1  1
  1  1 -1  1 -1  1 -1 -1
  1  1 -1  1 -1  1 -1  1
  1  1 -1  1 -1  1  1 -1
  1  1 -1  1 -1  1  1  1
  1  1 -1  1  1 -1 -1 -1
  1  1 -1  1  1 -1 -1  1
  1  1 -1  1  1 -1  1 -1
  1  1 -1  1  1 -1  1  1
  1  1 -1  1  1  1 -1 -1
  1  1 -1  1  1  1 -1  1
  1  1 -1  1  1  1  1 -1
  1  1 -1  1  1  1  1  1
  1  1  1  1 -1 -1 -1 -1
  1  1  1  1 -1 -1 -1  1
  1  1  1  1 -1 -1  1 -1
  1  1  1  1 -1 -1  1  1
  1  1  1  1 -1  1 -1 -1
  1  1  1  1 -1  1 -1  1
  1  1  1  1 -1  1  1 -1
  1  1  1  1 -1  1  1  1
  1  1  1  1  1  1 -1 -1
  1  1  1  1  1  1 -1  1
  1  1  1  1  1  1  1  1
  1  1  1  1  1  1  1 -1
  1  1  1  1  1 -1  1  1
  1  1  1  1  1 -1  1 -1
  1  1  1  1  1 -1 -1  1
  1  1  1  1  1 -1 -1 -1
  1  1  1 -1  1  1  1  1
  1  1  1 -1  1  1  1 -1
  1  1  1 -1  1  1 -1  1
  1  1  1 -1  1  1 -1 -1
  1  1  1 -1  1 -1  1  1
  1  1  1 -1  1 -1  1 -1
  1  1  1 -1  1 -1 -1  1
  1  1  1 -1  1 -1 -1 -1
  1  1  1 -1 -1  1  1  1
  1  1  1 -1 -1  1  1 -1
  1  1  1 -1 -1  1 -1  1
  1  1  1 -1 -1  1 -1 -1
  1  1  1 -1 -1 -1  1  1
  1  1  1 -1 -1 -1  1 -1
  1  1  1 -1 -1 -1 -1  1
  1  1  1 -1 -1 -1 -1 -1
  1 -1  1  1  1  1  1  1
  1 -1  1  1  1  1  1 -1
  1 -1  1  1  1  1 -1  1
  1 -1  1  1  1  1 -1 -1
  1 -1  1  1  1 -1  1  1
  1 -1  1  1  1 -1  1 -1
  1 -1  1  1  1 -1 -1  1
  1 -1  1  1  1 -1 -1 -1
  1 -1  1  1 -1  1  1  1
  1 -1  1  1 -1  1  1 -1
  1 -1  1  1 -1  1 -1  1
  1 -1  1  1 -1  1 -1 -1
  1 -1  1  1 -1 -1  1  1
  1 -1  1  1 -1 -1  1 -1
  1 -1  1  1 -1 -1 -1  1
  1 -1  1  1 -1 -1 -1 -1
  1 -1  1 -1  1  1  1  1
  1 -1  1 -1  1  1  1 -1
  1 -1  1 -1  1  1 -1  1
  1 -1  1 -1  1  1 -1 -1
  1 -1  1 -1  1 -1  1  1
  1 -1  1 -1  1 -1  1 -1
  1 -1  1 -1  1 -1 -1  1
  1 -1  1 -1  1 -1 -1 -1
  1 -1  1 -1 -1  1  1  1
  1 -1  1 -1 -1  1  1 -1
  1 -1  1 -1 -1  1 -1  1
  1 -1  1 -1 -1  1 -1 -1
  1 -1  1 -1 -1 -1  1  1
  1 -1  1 -1 -1 -1  1 -1
  1 -1  1 -1 -1 -1 -1  1
  1 -1  1 -1 -1 -1 -1 -1
  1 -1 -1  1  1  1  1  1
  1 -1 -1  1  1  1  1 -1
  1 -1 -1  1  1  1 -1  1
  1 -1 -1  1  1  1 -1 -1
  1 -1 -1  1  1 -1  1  1
  1 -1 -1  1  1 -1  1 -1
  1 -1 -1  1  1 -1 -1  1
  1 -1 -1  1  1 -1 -1 -1
  1 -1 -1  1 -1  1  1  1
  1 -1 -1  1 -1  1  1 -1
  1 -1 -1  1 -1  1 -1  1
  1 -1 -1  1 -1  1 -1 -1
  1 -1 -1  1 -1 -1  1  1
  1 -1 -1  1 -1 -1  1 -1
  1 -1 -1  1 -1 -1 -1  1
  1 -1 -1  1 -1 -1 -1 -1
  1 -1 -1 -1  1  1  1  1
  1 -1 -1 -1  1  1  1 -1
  1 -1 -1 -1  1  1 -1  1
  1 -1 -1 -1  1  1 -1 -1
  1 -1 -1 -1  1 -1  1  1
  1 -1 -1 -1  1 -1  1 -1
  1 -1 -1 -1  1 -1 -1  1
  1 -1 -1 -1  1 -1 -1 -1
  1 -1 -1 -1 -1  1  1  1
  1 -1 -1 -1 -1  1  1 -1
  1 -1 -1 -1 -1  1 -1  1
  1 -1 -1 -1 -1  1 -1 -1
  1 -1 -1 -1 -1 -1  1  1
  1 -1 -1 -1 -1 -1  1 -1
  1 -1 -1 -1 -1 -1 -1  1
  1 -1 -1 -1 -1 -1 -1 -1
end

~~~~~~ cddlib response

H-representation
begin
 14 8 real
  1  0  0  0  0  0  0  1
  1  0  0  0  0  0  1  0
  1  0  0  0  0  1  0  0
  1  0  0  0  1  0  0  0
  1  0  0  1  0  0  0  0
  1  0  1  0  0  0  0  0
  1  1  0  0  0  0  0  0
  1  0  0  0  0  0  0 -1
  1  0  0  0  0  0 -1  0
  1  0  0  0  0 -1  0  0
  1  0  0  0 -1  0  0  0
  1  0  0 -1  0  0  0  0
  1  0 -1  0  0  0  0  0
  1 -1  0  0  0  0  0  0
end

\end{lstlisting}  }

\subsubsection{Min-max calculation for the quantum bounds of two-two-state particles}

{ \begin{lstlisting}[backgroundcolor=\color{yellow!10},framerule=0pt,breaklines=true, frame=tb]


(* ~~~~~~~~~~~~~~~~~~~~~~~~~~~~~~~~~~~~~~~~~~~~~~~~~~~~~~~~~~~~~~~~~~~~~~~ *)
(* ~~~~~~~~~~~~~~~~~~Start Mathematica Code~~~~~~~~~~~~~~~~~~~~~~~~~~~~~~~ *)
(* ~~~~~~~~~~~~~~~~~~~~~~~~~~~~~~~~~~~~~~~~~~~~~~~~~~~~~~~~~~~~~~~~~~~~~~~ *)

(* old stuff

<<Algebra`ReIm`

Normalize[z_]:= z/Sqrt[z.Conjugate[z]];    *)

(*Definition of "my" Tensor Product*)
(*a,b are nxn and mxm-matrices*)

MyTensorProduct[a_, b_] :=
  Table[
   a[[Ceiling[s/Length[b]], Ceiling[t/Length[b]]]]*
    b[[s - Floor[(s - 1)/Length[b]]*Length[b],
      t - Floor[(t - 1)/Length[b]]*Length[b]]], {s, 1,
    Length[a]*Length[b]}, {t, 1, Length[a]*Length[b]}];


(*Definition of the Tensor Product between two vectors*)

TensorProductVec[x_, y_] :=
  Flatten[Table[
    x[[i]] y[[j]], {i, 1, Length[x]}, {j, 1, Length[y]}]];


(*Definition of the Dyadic Product*)

DyadicProductVec[x_] :=
  Table[x[[i]] Conjugate[x[[j]]], {i, 1, Length[x]}, {j, 1,
    Length[x]}];

(*Definition of the sigma matrices*)


vecsig[r_, tt_, p_] :=
 r*{{Cos[tt], Sin[tt] Exp[-I p]}, {Sin[tt] Exp[I p], -Cos[tt]}}

(*Definition of some vectors*)

BellBasis = (1/Sqrt[2]) {{1, 0, 0, 1}, {0, 1, 1, 0}, {0, 1, -1,
     0}, {1, 0, 0, -1}};

Basis = {{1, 0, 0, 0}, {0, 1, 0, 0}, {0, 0, 1, 0}, {0, 0, 0, 1}};



(*~~~~~~~~~~~~~~~~~~~~~~~~~  2  PARTICLES ~~~~~~~~~~~~~~~~~~~~~~~~~~~~~~~~~~~~~~~*)

(*~~~~~~~~~~~~~~~~~~~~~~~~~  2  State System ~~~~~~~~~~~~~~~~~~~~~~~~~~~~~~~~~~~~~~~

% ~~~~~~~~~~~~~~~   2 x 2
% ~~~~~~~~~~~~~~~   2 x 2
% ~~~~~~~~~~~~~~~   2 x 2
% ~~~~~~~~~~~~~~~   2 x 2
% ~~~~~~~~~~~~~~~   2 x 2
% ~~~~~~~~~~~~~~~   2 x 2

*)


(*Definition of singlet state*)
vp = {1,0};
vm = {0,1};
psi2s = (1/Sqrt[2])*(TensorProductVec[vp, vm] -
    TensorProductVec[vm, vp])

DyadicProductVec[psi2s]

(*Definition of operators*)

(* Definition of one-particle operator *)

M2X = (1/2) {{0, 1}, {1, 0}};
M2Y = (1/2) {{0, -I}, {I, 0}};
M2Z = (1/2) {{1, 0}, {0, -1}};


Eigenvectors[M2X]
Eigenvectors[M2Y]
Eigenvectors[M2Z]

S2[t_, p_] := FullSimplify[M2X *Sin[t] Cos[p] + M2Y *Sin[t] Sin[p] + M2Z *Cos[t]]

FullSimplify[S2[\[Theta], \[Phi]]] // MatrixForm

FullSimplify[ComplexExpand[S2[Pi/2, 0]]] // MatrixForm
FullSimplify[ComplexExpand[S2[Pi/2, Pi/2]]] // MatrixForm
FullSimplify[ComplexExpand[S2[0, 0]]] // MatrixForm

Assuming[{0 <= \[Theta] <= Pi, 0 <= \[Phi] <= 2 Pi}, FullSimplify[Eigensystem[S2[\[Theta], \[Phi]]], {Element[\[Theta], Reals],
  Element[\[Phi], Reals]}]]



FullSimplify[
 Normalize[
  Eigenvectors[S2[\[Theta], \[Phi]]][[1]]], {Element[\[Theta], Reals],
   Element[\[Phi], Reals]}]

ES2M[\[Theta]_,\[Phi]_] := {-E^(-I \[Phi]) Tan[\[Theta]/2], 1}*Cos[\[Theta]/2]*E^(I \[Phi]/2)
ES2P[\[Theta]_,\[Phi]_] := {E^(-I \[Phi]) Cot[\[Theta]/2], 1}*Sin[\[Theta]/2]*E^(I \[Phi]/2)

FullSimplify[ES2M[\[Theta],\[Phi]] .Conjugate[ES2M [\[Theta],\[Phi]]], {Element[\[Theta], Reals],
  Element[\[Phi], Reals]}]
FullSimplify[ES2P[\[Theta],\[Phi]] .Conjugate[ES2P [\[Theta],\[Phi]]], {Element[\[Theta], Reals],
  Element[\[Phi], Reals]}]
FullSimplify[ES2P[\[Theta],\[Phi]] .Conjugate[ES2M[\[Theta],\[Phi]]], {Element[\[Theta], Reals],
  Element[\[Phi], Reals]}]


ProjectorES2M[\[Theta]_,\[Phi]_] := FullSimplify[DyadicProductVec[ES2M[\[Theta],\[Phi]]], {Element[\[Theta], Reals],
  Element[\[Phi], Reals]}]
ProjectorES2P[\[Theta]_,\[Phi]_] := FullSimplify[DyadicProductVec[ES2P[\[Theta],\[Phi]]], {Element[\[Theta], Reals],
  Element[\[Phi], Reals]}]

 ProjectorES2M[\[Theta],\[Phi]] //MatrixForm
 ProjectorES2P[\[Theta],\[Phi]] //MatrixForm


(* verification of spectral form *)

FullSimplify[(-1/2)ProjectorES2M[\[Theta],\[Phi]] + (1/2)ProjectorES2P[\[Theta],\[Phi]], {Element[\[Theta], Reals],
  Element[\[Phi], Reals]}]


SingleParticleSpinOneHalfeObservable[x_, p_] :=   FullSimplify[(1/2) (IdentityMatrix[2] + vecsig[1, x, p])] ;

SingleParticleSpinOneHalfeObservable[\[Theta], \[Phi]] // MatrixForm

Eigensystem[FullSimplify[SingleParticleSpinOneHalfeObservable[x, p]]]


(*Definition of single operators for occurrence of spin up*)

SingleParticleProjector2first[x_, p_, pm_] :=   MyTensorProduct[1/2 (IdentityMatrix[2] + pm*vecsig[1, x, p]),  IdentityMatrix[2]]

SingleParticleProjector2second[x_, p_, pm_] :=  MyTensorProduct[IdentityMatrix[2], 1/2 (IdentityMatrix[2] + pm*vecsig[1, x, p])]



(*Definition of two-particle joint operator for occurrence of spin up \
and down*)

JointProjector2[x1_, x2_, p1_, p2_, pm1_, pm2_] :=  MyTensorProduct[1/2 (IdentityMatrix[2] + pm1*vecsig[1, x1, p1]),  1/2 (IdentityMatrix[2] + pm2*vecsig[1, x2, p2])]


(*Definition of probabilities*)


(*Probability of concurrence of two equal events for two-particle \
probability in singlet Bell state for occurrence of spin up*)

JointProb2s[x1_, x2_, p1_, p2_, pm1_, pm2_] :=
 FullSimplify[
  Tr[DyadicProductVec[psi2s].JointProjector2[x1, x2, p1, p2, pm1,
     pm2]]]

JointProb2s[x1, x2, p1, p2, pm1, pm2]

JointProb2s[x1, x2, p1, p2, pm1, pm2] // TeXForm

(*sum of joint probabilities add up to one*)

FullSimplify[
 Sum[JointProb2s[x1, x2, p1, p2, pm1, pm2], {pm1, -1, 1, 2}, {pm2, -1,
    1, 2}]]

(*Probability of concurrence of two equal events*)

P2Es[x1_, x2_, p1_, p2_] =
  FullSimplify[
   Sum[UnitStep[pm1*pm2]*
     JointProb2s[x1, x2, p1, p2, pm1, pm2], {pm1, -1, 1, 2}, {pm2, -1,
      1, 2}]];

P2Es[x1, x2, p1, p2]

(*Probability of concurrence of two non-equal events*)

P2NEs[x1_, x2_, p1_, p2_] =
  FullSimplify[
   Sum[UnitStep[-pm1*pm2]*
     JointProb2s[x1, x2, p1, p2, pm1, pm2], {pm1, -1, 1, 2}, {pm2, -1,
      1, 2}]];

P2NEs[x1, x2, p1, p2]

(*Expectation function*)

Expectation2s[x1_, x2_, p1_, p2_] =
 FullSimplify[P2Es[x1, x2, p1, p2] - P2NEs[x1, x2, p1, p2]]


(* ~~~~~~~~~~~~~~~~~~~~~~~ Min-Max calculation of the quantum correlation function ~~~~~~~~~~~~~~~~~~~~~~~ *)

JointExpectation2[t1_, t2_, p1_, p2_] :=  MyTensorProduct[ 2 * S2[t1, p1] , 2 * S2[t2, p2] ]


FullSimplify[
 Eigensystem[
 JointExpectation2[t1 , t2 , p1 , p2 ]  ]]    // MatrixForm

FullSimplify[
 Eigensystem[
 DyadicProductVec[psi2s]. JointExpectation2[t1, t2, p1 , p2 ] . DyadicProductVec[psi2s]  ]]   // MatrixForm

FullSimplify[
 Eigensystem[
 JointExpectation2[Pi/2 , Pi/2 , p1 , p2 ]  ]]        // MatrixForm

FullSimplify[
  Eigensystem[
   DyadicProductVec[psi2s].JointExpectation2[Pi/2, Pi/2, p1,  p2 ].DyadicProductVec[psi2s]]] // MatrixForm


psi2mp = (1/Sqrt[2])*(TensorProductVec[vp, vm] +
    TensorProductVec[vm, vp])

psi2mm = (1/Sqrt[2])*(TensorProductVec[vp, vp] -
    TensorProductVec[vm, vm])

psi2pp = (1/Sqrt[2])*(TensorProductVec[vp, vp] +
    TensorProductVec[vm, vm])


FullSimplify[
  Eigensystem[
   DyadicProductVec[psi2mp].JointExpectation2[Pi/2, Pi/2, p1,
     p2].DyadicProductVec[psi2mp]]] // MatrixForm

FullSimplify[
  Eigensystem[
   DyadicProductVec[psi2mm].JointExpectation2[Pi/2, Pi/2, p1,
     p2].DyadicProductVec[psi2mm]]] // MatrixForm

FullSimplify[
  Eigensystem[
   DyadicProductVec[psi2pp].JointExpectation2[Pi/2, Pi/2, p1,
     p2].DyadicProductVec[psi2pp]]] // MatrixForm

(* ~~~~~~~~~~~~~~~~~~~~~~~ Min-Max calculation of the Tsirelson bound ~~~~~~~~~~~~~~~~~~~~~~~ *)

JointProjector2Red[ p1_, p2_, pm1_, pm2_] :=  JointProjector2[ Pi/2 , Pi/2 , p1, p2, pm1, pm2]

FullSimplify[ JointProjector2Red[ p1 , p2 , pm1 , pm2 ]]

(* ~~~~~~~~~~~~~~~~~~~~~~~~~ plausibility check *)

JointProb2sRed[p1_, p2_, pm1_, pm2_] :=
 FullSimplify[
  Tr[DyadicProductVec[psi2s].JointProjector2Red[p1, p2, pm1, pm2]]]

JointProb2sRed[p1, p2, pm1, pm2]

FullSimplify[
 JointProb2sRed[p1, p2, 1, 1] + JointProb2sRed[p1, p2, -1, -1] -
  JointProb2sRed[p1, p2, -1, 1] - JointProb2sRed[p1, p2, 1, -1]]

(* ~~~~~~~~~~~~~~~~~~~~~~~~~ end plausibility check *)


TwoParticleExpectationsRed[ p1_, p2_] := JointProjector2Red[ p1, p2, 1, 1]  + JointProjector2Red[ p1, p2, -1, -1] -
                                         JointProjector2Red[ p1, p2, -1, 1] - JointProjector2Red[ p1, p2, 1, -1]


(* ~~~~~~~~~~~~~~~~~~~~~~~~~ plausibility check *)

FullSimplify[ Tr[DyadicProductVec[psi2s].TwoParticleExpectationsRed[A1, B1]] ]

(* ~~~~~~~~~~~~~~~~~~~~~~~~~ end plausibility check *)

TwoParticleExpectationsRed[A1, B1] // MatrixForm
TwoParticleExpectationsRed[A1, B1] // TeXForm

Eigenvalues[
 ComplexExpand[
  TwoParticleExpectationsRed[A1, B1] +
   TwoParticleExpectationsRed[A2, B1] +
   TwoParticleExpectationsRed[A1, B2] -
   TwoParticleExpectationsRed[A2, B2] ]]

FullSimplify[
 Eigenvalues[
  ComplexExpand[
   TwoParticleExpectationsRed[A1, B1] +
    TwoParticleExpectationsRed[A2, B1] +
    TwoParticleExpectationsRed[A1, B2] -
    TwoParticleExpectationsRed[A2, B2] ]]]


FullSimplify[
   TwoParticleExpectationsRed[A1, B1] +
    TwoParticleExpectationsRed[A2, B1] +
    TwoParticleExpectationsRed[A1, B2] -
    TwoParticleExpectationsRed[A2, B2] ]

(*  observables along psi_singlet *)


Eigenvalues[
 ComplexExpand[
  DyadicProductVec[
    psi2s].(TwoParticleExpectationsRed[A1, B1] +
     TwoParticleExpectationsRed[A2, B1] +
     TwoParticleExpectationsRed[A1, B2] -
     TwoParticleExpectationsRed[A2, B2]).DyadicProductVec[psi2s]]]

FullSimplify[
 TrigExpand[
  Eigenvalues[
   ComplexExpand[
    DyadicProductVec[
      psi2s].(TwoParticleExpectationsRed[0, Pi/4] +
       TwoParticleExpectationsRed[Pi/2, Pi/4] +
       TwoParticleExpectationsRed[0, -Pi/4] -
       TwoParticleExpectationsRed[Pi/2, -Pi/4]).DyadicProductVec[
      psi2s]]]]]

(*  observables along psi_+ *)


Eigenvalues[
 ComplexExpand[
  DyadicProductVec[
    psi2mp].(TwoParticleExpectationsRed[A1, B1] +
     TwoParticleExpectationsRed[A2, B1] +
     TwoParticleExpectationsRed[A1, B2] -
     TwoParticleExpectationsRed[A2, B2]).DyadicProductVec[psi2mp]]]

FullSimplify[
 TrigExpand[
  Eigenvalues[
   ComplexExpand[
    DyadicProductVec[
      psi2mp].(TwoParticleExpectationsRed[0, Pi/4] +
       TwoParticleExpectationsRed[Pi/2, Pi/4] +
       TwoParticleExpectationsRed[0, -Pi/4] -
       TwoParticleExpectationsRed[Pi/2, -Pi/4]).DyadicProductVec[
      psi2mp]]]]]

(*** observables along phi_+ ***)


Eigenvalues[
 ComplexExpand[
  DyadicProductVec[
    psi2mm].(TwoParticleExpectationsRed[A1, B1] +
     TwoParticleExpectationsRed[A2, B1] +
     TwoParticleExpectationsRed[A1, B2] -
     TwoParticleExpectationsRed[A2, B2]).DyadicProductVec[psi2mm]]]

FullSimplify[
 TrigExpand[
  Eigenvalues[
   ComplexExpand[
    DyadicProductVec[
      psi2mm].(TwoParticleExpectationsRed[0, -Pi/4] +
       TwoParticleExpectationsRed[Pi/2, -Pi/4] +
       TwoParticleExpectationsRed[0, Pi/4] -
       TwoParticleExpectationsRed[Pi/2, Pi/4]).DyadicProductVec[
      psi2mm]]]]]


(*** observables along phi_+ ***)


Eigenvalues[
 ComplexExpand[
  DyadicProductVec[
    psi2pp].(TwoParticleExpectationsRed[A1, B1] +
     TwoParticleExpectationsRed[A2, B1] +
     TwoParticleExpectationsRed[A1, B2] -
     TwoParticleExpectationsRed[A2, B2]).DyadicProductVec[psi2pp]]]

FullSimplify[
 TrigExpand[
  Eigenvalues[
   ComplexExpand[
    DyadicProductVec[
      psi2pp].(TwoParticleExpectationsRed[0, -Pi/4] +
       TwoParticleExpectationsRed[Pi/2, -Pi/4] +
       TwoParticleExpectationsRed[0, Pi/4] -
       TwoParticleExpectationsRed[Pi/2, Pi/4]).DyadicProductVec[
      psi2pp]]]]]

\end{lstlisting}  }

\subsubsection{Min-max calculation for the quantum bounds of two three-state particles}

{ \begin{lstlisting}[backgroundcolor=\color{yellow!10},framerule=0pt,breaklines=true, frame=tb]

(* ~~~~~~~~~~~~~~~~~~~~~~~~~~~~~~~~~~~~~~~~~~~~~~~~~~~~~~~~~~~~~~~~~~~~~~~ *)
(* ~~~~~~~~~~~~~~~~~~Start Mathematica Code~~~~~~~~~~~~~~~~~~~~~~~~~~~~~~~ *)
(* ~~~~~~~~~~~~~~~~~~~~~~~~~~~~~~~~~~~~~~~~~~~~~~~~~~~~~~~~~~~~~~~~~~~~~~~ *)

(* old stuff

<<Algebra`ReIm`

Normalize[z_]:= z/Sqrt[z.Conjugate[z]];    *)

(*Definition of "my" Tensor Product*)
(*a,b are nxn and mxm-matrices*)

MyTensorProduct[a_, b_] :=
  Table[
   a[[Ceiling[s/Length[b]], Ceiling[t/Length[b]]]]*
    b[[s - Floor[(s - 1)/Length[b]]*Length[b],
      t - Floor[(t - 1)/Length[b]]*Length[b]]], {s, 1,
    Length[a]*Length[b]}, {t, 1, Length[a]*Length[b]}];


(*Definition of the Tensor Product between two vectors*)

TensorProductVec[x_, y_] :=
  Flatten[Table[
    x[[i]] y[[j]], {i, 1, Length[x]}, {j, 1, Length[y]}]];


(*Definition of the Dyadic Product*)

DyadicProductVec[x_] :=
  Table[x[[i]] Conjugate[x[[j]]], {i, 1, Length[x]}, {j, 1,
    Length[x]}];

(*Definition of the sigma matrices*)


vecsig[r_, tt_, p_] :=
 r*{{Cos[tt], Sin[tt] Exp[-I p]}, {Sin[tt] Exp[I p], -Cos[tt]}}

(*Definition of some vectors*)

BellBasis = (1/Sqrt[2]) {{1, 0, 0, 1}, {0, 1, 1, 0}, {0, 1, -1,
     0}, {1, 0, 0, -1}};

Basis = {{1, 0, 0, 0}, {0, 1, 0, 0}, {0, 0, 1, 0}, {0, 0, 0, 1}};


(*~~~~~~~~~~~~~~~~~~~~~~~~~  3  State System ~~~~~~~~~~~~~~~~~~~~~~~~~~~~~~~~~~~~~~~

% ~~~~~~~~~~~~~~~   2 x 3
% ~~~~~~~~~~~~~~~   2 x 3
% ~~~~~~~~~~~~~~~   2 x 3
% ~~~~~~~~~~~~~~~   2 x 3
% ~~~~~~~~~~~~~~~   2 x 3
% ~~~~~~~~~~~~~~~   2 x 3
% ~~~~~~~~~~~~~~~   2 x 3
% ~~~~~~~~~~~~~~~   2 x 3

*)



(*Definition of operators*)

(* Definition of one-particle operator *)

M3X = (1/Sqrt[2]) {{0, 1, 0}, {1, 0, 1},{0, 1, 0}};
M3Y = (1/Sqrt[2]) {{0, -I, 0}, {I, 0, -I}, {0, I, 0}};
M3Z =  {{1, 0, 0}, {0, 0, 0},{0, 0, -1}};


Eigenvectors[M3X]
Eigenvectors[M3Y]
Eigenvectors[M3Z]

S3[t_, p_] := M3X *Sin[t] Cos[p] + M3Y *Sin[t] Sin[p] + M3Z *Cos[t]

FullSimplify[S3[\[Theta], \[Phi]]] // MatrixForm

FullSimplify[ComplexExpand[S3[Pi/2, 0]]] // MatrixForm
FullSimplify[ComplexExpand[S3[Pi/2, Pi/2]]] // MatrixForm
FullSimplify[ComplexExpand[S3[0, 0]]] // MatrixForm

Assuming[{0 <= \[Theta] <= Pi, 0 <= \[Phi] <= 2 Pi}, FullSimplify[Eigensystem[S3[\[Theta], \[Phi]]], {Element[\[Theta], Reals],
  Element[\[Phi], Reals]}]]



FullSimplify[ComplexExpand[
 Normalize[
  Eigenvectors[S3[\[Theta], \[Phi]]][[1]]], {Element[\[Theta], Reals],
   Element[\[Phi], Reals]}]]

ES3M[\[Theta]_,\[Phi]_] := FullSimplify[ ComplexExpand[
 Normalize[
  Eigenvectors[S3[\[Theta], \[Phi]]][[1]]]*E^(I \[Phi])  , {Element[\[Theta], Reals], Element[\[Phi], Reals]}]]

ES3M[\[Theta],\[Phi]]


ES3P[\[Theta]_,\[Phi]_] := FullSimplify[ComplexExpand[
 Normalize[
  Eigenvectors[S3[\[Theta], \[Phi]]][[2]]]*E^(I \[Phi])  , {Element[\[Theta], Reals], Element[\[Phi], Reals]}]]

ES3P[\[Theta],\[Phi]]

ES30[\[Theta]_,\[Phi]_] := FullSimplify[ComplexExpand[
 Normalize[
  Eigenvectors[S3[\[Theta], \[Phi]]][[3]]]*E^(I \[Phi])  , {Element[\[Theta], Reals], Element[\[Phi], Reals]}]]

ES30[\[Theta],\[Phi]]

FullSimplify[ES3M[\[Theta],\[Phi]] .Conjugate[ES3M [\[Theta],\[Phi]]], {Element[\[Theta], Reals],
  Element[\[Phi], Reals]}]
FullSimplify[ES3P[\[Theta],\[Phi]] .Conjugate[ES3P [\[Theta],\[Phi]]], {Element[\[Theta], Reals],
  Element[\[Phi], Reals]}]
FullSimplify[ES30[\[Theta],\[Phi]] .Conjugate[ES30 [\[Theta],\[Phi]]], {Element[\[Theta], Reals],
  Element[\[Phi], Reals]}]
FullSimplify[ES3P[\[Theta],\[Phi]] .Conjugate[ES3M[\[Theta],\[Phi]]], {Element[\[Theta], Reals],
  Element[\[Phi], Reals]}]
FullSimplify[ES3P[\[Theta],\[Phi]] .Conjugate[ES30[\[Theta],\[Phi]]], {Element[\[Theta], Reals],
  Element[\[Phi], Reals]}]
FullSimplify[ES30[\[Theta],\[Phi]] .Conjugate[ES3M[\[Theta],\[Phi]]], {Element[\[Theta], Reals],
  Element[\[Phi], Reals]}]


ProjectorES30[\[Theta]_,\[Phi]_] := FullSimplify[ComplexExpand[DyadicProductVec[ES30[\[Theta],\[Phi]]], {Element[\[Theta], Reals],
  Element[\[Phi], Reals]}]]
ProjectorES3M[\[Theta]_,\[Phi]_] := FullSimplify[ComplexExpand[DyadicProductVec[ES3M[\[Theta],\[Phi]]], {Element[\[Theta], Reals],
  Element[\[Phi], Reals]}]]
ProjectorES3P[\[Theta]_,\[Phi]_] := FullSimplify[ComplexExpand[DyadicProductVec[ES3P[\[Theta],\[Phi]]], {Element[\[Theta], Reals],
  Element[\[Phi], Reals]}]]

 ProjectorES30[\[Theta],\[Phi]] //MatrixForm
 ProjectorES3M[\[Theta],\[Phi]] //MatrixForm
 ProjectorES3P[\[Theta],\[Phi]] //MatrixForm

ProjectorES30[\[Theta], \[Phi]] // MatrixForm // TeXForm
ProjectorES3M[\[Theta], \[Phi]] // MatrixForm // TeXForm
ProjectorES3P[\[Theta], \[Phi]] // MatrixForm // TeXForm

(* verification of spectral form *)

FullSimplify[0 * ProjectorES30[\[Theta],\[Phi]] +  (-1) * ProjectorES3M[\[Theta],\[Phi]] + (+1) * ProjectorES3P[\[Theta],\[Phi]], {Element[\[Theta], Reals],
  Element[\[Phi], Reals]}] //MatrixForm


(*  ~~~~~~~~~~~~~~~~~~~ general operator ~~~~~~~~~~~~~~~~~~~~~~~  *)

Operator3GEN[\[Theta]_,\[Phi]_] := FullSimplify[LM * ProjectorES3M[\[Theta],\[Phi]] + L0 * ProjectorES30[\[Theta],\[Phi]] + LP * ProjectorES3P[\[Theta],\[Phi]], {Element[\[Theta], Reals], Element[\[Phi], Reals]}];

Operator3GEN[\[Theta],\[Phi]]

JointProjector3GEN[x1_, x2_, p1_, p2_] :=  MyTensorProduct[Operator3GEN[x1,p1],Operator3GEN[x2,p2]];

v3p = {1,0,0};
v30 = {0,1,0};
v3m = {0,0,1};

psi3s = (1/Sqrt[3])*(-TensorProductVec[v30, v30] + TensorProductVec[v3m, v3p] + TensorProductVec[v3p, v3m])


Expectation3sGEN[x1_, x2_, p1_, p2_] := FullSimplify[ Tr[DyadicProductVec[psi3s].JointProjector3GEN[x1, x2, p1, p2]]];

Expectation3sGEN[x1, x2, p1, p2]


Ex3[LM_,L0_,LP_,x1_,x2_,p1_,p2_]:=FullSimplify[1/192 (24 L0^2 + 40 L0 (LM + LP) + 22 (LM + LP)^2 -
   32 (LM - LP)^2 Cos[x1] Cos[x2] +
   2 (-2 L0 + LM + LP)^2 Cos[
     2 x2] ((3 + Cos[2 (p1 - p2)]) Cos[2 x1] + 2 Sin[p1 - p2]^2) +
   2 (-2 L0 + LM + LP)^2 (Cos[2 (p1 - p2)] +
      2 Cos[2 x1] Sin[p1 - p2]^2) -
   32 (LM - LP)^2 Cos[p1 - p2] Sin[x1] Sin[x2] +
   8 (-2 L0 + LM + LP)^2 Cos[p1 - p2] Sin[2 x1] Sin[2 x2])];

Ex3[-1,0,1,x1,x2,p1,p2]



(* ~~~~~~~~~~~ natural spin observables ~~~~~~~~~~~~~~~~~~~~~ *)



JointProjector3NAT[x1_, x2_, p1_, p2_] :=  MyTensorProduct[S3[x1,p1],S3[x2,p2]];

Expectation3sNAT[x1_, x2_, p1_, p2_] := FullSimplify[ Tr[DyadicProductVec[psi3s].JointProjector3NAT[x1, x2, p1, p2]]];

Expectation3sNAT[x1, x2, p1, p2]



(* ~~~~~~~~~~~ Kochen-Specker observables ~~~~~~~~~~~~~~~~~~~~~ *)


(*
S3[t_, p_] := M3X *Sin[t] Cos[p] + M3Y *Sin[t] Sin[p] + M3Z *Cos[t]

MM3X[ \[Alpha]_ ] := FullSimplify[S3[Pi/2, \[Alpha]]];
MM3Y[ \[Alpha]_ ] := FullSimplify[S3[Pi/2, \[Alpha]+Pi/2]];
MM3Z[ \[Alpha]_ ] := FullSimplify[S3[0, 0]];

SKS[ \[Alpha]_ ] := FullSimplify[ MM3X[\[Alpha]].MM3X[\[Alpha]] + MM3Y[\[Alpha]].MM3Y[\[Alpha]] + MM3Z[\[Alpha]].MM3Z[\[Alpha]] ];

FullSimplify[SKS[ \[Alpha] ]] // MatrixForm

FullSimplify[ComplexExpand[SKS[ 0]]] // MatrixForm
FullSimplify[ComplexExpand[SKS[ Pi/2]]] // MatrixForm

Assuming[{0 <= \[Theta] <= Pi, 0 <= \[Phi] <= 2 Pi}, FullSimplify[Eigensystem[SKS[ \[Alpha] ]], {Element[\[Alpha], Reals]}]]

*)

Ex3[1, 0, 1, \[Theta]1, \[Theta]2, \[CurlyPhi]1, \[CurlyPhi]2]

Ex3[0, 1, 0, \[Theta]1, \[Theta]2, \[CurlyPhi]1, \[CurlyPhi]2]

Ex3[1, 0, 1, Pi/2, Pi/2, \[CurlyPhi]1, \[CurlyPhi]2]

Ex3[0, 1, 0, Pi/2, Pi/2, \[CurlyPhi]1, \[CurlyPhi]2]

Ex3[1, 0, 1, \[Theta]1, \[Theta]2, 0, 0]

Ex3[0, 1, 0, \[Theta]1, \[Theta]2, 0, 0]




(* min-max computation *)

(* define dichotomic operator based on spin-1  expectation value , take \[Phi] = Pi/2 *)

(* old, invalid parameterization
A[ \[Theta]1_ , \[Theta]2_  ] :=   MyTensorProduct[ S3[\[Theta]1, Pi/2] ,  S3[\[Theta]2, Pi/2] ]

(* Form the Klyachko-Can-Biniciogolu-Shumovsky operator  *)

T[\[Theta]1_, \[Theta]3_, \[Theta]5_, \[Theta]7_, \[Theta]9_] :=
 A[\[Theta]1,\[Theta]3] + A[\[Theta]3,\[Theta]5] +
  A[\[Theta]5,\[Theta]7] + A[\[Theta]7,\[Theta]9] +
  A[\[Theta]9,\[Theta]1]


FullSimplify[
 Eigenvalues[
  FullSimplify[
  T[\[Theta]1, \[Theta]3, \[Theta]5, \[Theta]7, \[Theta]9]]]]

FullSimplify[
 Eigenvalues[
  T[2 Pi/5 , 4 Pi/5, 6 Pi/5, 8 Pi/5, 2 Pi]]]

 *)

A[ \[Theta]1_ , \[Theta]2_ ,\[CurlyPhi]1_, \[CurlyPhi]2_  ] :=   MyTensorProduct[ S3[\[Theta]1, \[CurlyPhi]1] ,  S3[\[Theta]2, \[CurlyPhi]2] ]

(* Form the Klyachko-Can-Biniciogolu-Shumovsky operator  *)

T[\[Theta]1_, \[Theta]3_, \[Theta]5_, \[Theta]7_, \[Theta]9_,\[CurlyPhi]1_, \[CurlyPhi]3_,\[CurlyPhi]5_, \[CurlyPhi]7_,\[CurlyPhi]9_] :=
 A[\[Theta]1,\[Theta]3, \[CurlyPhi]1 ,\[CurlyPhi]3] + A[\[Theta]3,\[Theta]5,\[CurlyPhi]3,\[CurlyPhi]5] +
  A[\[Theta]5,\[Theta]7,\[CurlyPhi]5,\[CurlyPhi]7] + A[\[Theta]7,\[Theta]9,\[CurlyPhi]7,\[CurlyPhi]9] +
  A[\[Theta]9,\[Theta]1,\[CurlyPhi]9,\[CurlyPhi]1]


A1   =  CoordinateTransformData[ "Cartesian" -> "Spherical", "Mapping", {1,0,0  }] ;
A2   =  CoordinateTransformData[ "Cartesian" -> "Spherical", "Mapping", {0,1,0  }] ;
A3   =  (* CoordinateTransformData[ "Cartesian" -> "Spherical", "Mapping", {0,0,1  }] *)  {1,0,Pi/2} ;
A4   =  CoordinateTransformData[ "Cartesian" -> "Spherical", "Mapping", {1,-1,0 }] ;
A5   =  CoordinateTransformData[ "Cartesian" -> "Spherical", "Mapping", {1,1,0  }] ;
A6   =  CoordinateTransformData[ "Cartesian" -> "Spherical", "Mapping", {1,-1,2 }] ;
A7   =  CoordinateTransformData[ "Cartesian" -> "Spherical", "Mapping", {-1,1,1 }] ;
A8   =  CoordinateTransformData[ "Cartesian" -> "Spherical", "Mapping", {2,1,1  }] ;
A9   =  CoordinateTransformData[ "Cartesian" -> "Spherical", "Mapping", {0,1,-1 }] ;
A10  =  CoordinateTransformData[ "Cartesian" -> "Spherical", "Mapping", {0,1,1  }] ;



FullSimplify[
 Eigenvalues[
  FullSimplify[
  T[ A1[[2]],  A3[[2]],  A5[[2]],  A7[[2]],  A9[[2]] , A1[[3]],  A3[[3]],  A5[[3]],  A7[[3]],  A9[[3]]]]]]



{A1,
A2 ,
A3 ,
A4 ,
A5 ,
A6 ,
A7 ,
A8 ,
A9 ,
A10} //TexForm

\end{lstlisting}  }

\subsubsection{Min-max calculation for two four-state particles}

{ \begin{lstlisting}[backgroundcolor=\color{yellow!10},framerule=0pt,breaklines=true, frame=tb]

(* ~~~~~~~~~~~~~~~~~~~~~~~~~~~~~~~~~~~~~~~~~~~~~~~~~~~~~~~~~~~~~~~~~~~~~~~ *)
(* ~~~~~~~~~~~~~~~~~~Start Mathematica Code~~~~~~~~~~~~~~~~~~~~~~~~~~~~~~~ *)
(* ~~~~~~~~~~~~~~~~~~~~~~~~~~~~~~~~~~~~~~~~~~~~~~~~~~~~~~~~~~~~~~~~~~~~~~~ *)

(* old stuff

<<Algebra`ReIm`

Normalize[z_]:= z/Sqrt[z.Conjugate[z]];    *)

(*Definition of "my" Tensor Product*)
(*a,b are nxn and mxm-matrices*)

MyTensorProduct[a_, b_] :=
  Table[
   a[[Ceiling[s/Length[b]], Ceiling[t/Length[b]]]]*
    b[[s - Floor[(s - 1)/Length[b]]*Length[b],
      t - Floor[(t - 1)/Length[b]]*Length[b]]], {s, 1,
    Length[a]*Length[b]}, {t, 1, Length[a]*Length[b]}];


(*Definition of the Tensor Product between two vectors*)

TensorProductVec[x_, y_] :=
  Flatten[Table[
    x[[i]] y[[j]], {i, 1, Length[x]}, {j, 1, Length[y]}]];


(*Definition of the Dyadic Product*)

DyadicProductVec[x_] :=
  Table[x[[i]] Conjugate[x[[j]]], {i, 1, Length[x]}, {j, 1,
    Length[x]}];

(*Definition of the sigma matrices*)


vecsig[r_, tt_, p_] :=
 r*{{Cos[tt], Sin[tt] Exp[-I p]}, {Sin[tt] Exp[I p], -Cos[tt]}}

(*Definition of some vectors*)

BellBasis = (1/Sqrt[2]) {{1, 0, 0, 1}, {0, 1, 1, 0}, {0, 1, -1,
     0}, {1, 0, 0, -1}};

Basis = {{1, 0, 0, 0}, {0, 1, 0, 0}, {0, 0, 1, 0}, {0, 0, 0, 1}};


(*~~~~~~~~~~~~~~~~~~~~~~~~~  4  State System ~~~~~~~~~~~~~~~~~~~~~~~~~~~~~~~~~~~~~~~

% ~~~~~~~~~~~~~~~   2 x 4
% ~~~~~~~~~~~~~~~   2 x 4
% ~~~~~~~~~~~~~~~   2 x 4
% ~~~~~~~~~~~~~~~   2 x 4
% ~~~~~~~~~~~~~~~   2 x 4
% ~~~~~~~~~~~~~~~   2 x 4
% ~~~~~~~~~~~~~~~   2 x 4
% ~~~~~~~~~~~~~~~   2 x 4

*)





(*Definition of operators*)

(* Definition of one-particle operator *)

M4X = (1/2) {{0,Sqrt[3],0,0 },{Sqrt[3],0,2,0 },{0,2,0,Sqrt[3] },{0,0,Sqrt[3],0 }};
M4Y = (1/2) {{0,-Sqrt[3]I,0,0 },{Sqrt[3]I,0,-2I,0},{0,2I,0,-Sqrt[3]I},{0,0,Sqrt[3]I,0  } };
M4Z = (1/2) {{3,0,0,0 },{0,1,0,0 },{0,0,-1,0},{0,0,0,-3}};

Eigenvectors[M4X]
Eigenvectors[M4Y]
Eigenvectors[M4Z]

S4[t_, p_] :=  FullSimplify[M4X *Sin[t] Cos[p] + M4Y *Sin[t] Sin[p] + M4Z *Cos[t]];



(*  ~~~~~~~~~~~~~~~~~~~ general operator ~~~~~~~~~~~~~~~~~~~~~~~  *)

LM32 =-3/2;
LM12 =-1/2;
LP32 =3/2;
LP12 =1/2;

ES4M32[\[Theta]_, \[Phi]_] :=   FullSimplify[   Assuming[{0 < \[Theta] < Pi, 0 <= \[Phi] <= 2 Pi},    Normalize[     Eigenvectors[S4[\[Theta], \[Phi]]][[1]]]], {Element[\[Theta],     Reals], Element[\[Phi], Reals]}];
ES4P32[\[Theta]_, \[Phi]_] :=  FullSimplify[   Assuming[{0 < \[Theta] < Pi, 0 <= \[Phi] <= 2 Pi},    Normalize[     Eigenvectors[S4[\[Theta], \[Phi]]][[2]]]], {Element[\[Theta],     Reals], Element[\[Phi], Reals]}];
ES4M12[\[Theta]_, \[Phi]_] :=  FullSimplify[   Assuming[{0 < \[Theta] < Pi, 0 <= \[Phi] <= 2 Pi},    Normalize[     Eigenvectors[S4[\[Theta], \[Phi]]][[3]]]], {Element[\[Theta],     Reals], Element[\[Phi], Reals]}];
ES4P12[\[Theta]_, \[Phi]_] :=  FullSimplify[   Assuming[{0 < \[Theta] < Pi, 0 <= \[Phi] <= 2 Pi},    Normalize[     Eigenvectors[S4[\[Theta], \[Phi]]][[4]]]], {Element[\[Theta],     Reals], Element[\[Phi], Reals]}];



JointProjector4GEN[x1_, x2_, p1_, p2_] :=  TensorProduct[S4[x1,p1],S4[x2,p2]];

v4P32 = ES4P32[0,0]
v4P12 = ES4P12[0,0]
v4M12 = ES4M12[0,0]
v4M32 = ES4M32[0,0]



psi4s = (1/2)*(TensorProductVec[v4P32, v4M32]-TensorProductVec[v4M32, v4P32] - TensorProductVec[v4P12 , v4M12 ] + TensorProductVec[v4M12 , v4P12 ])


Expectation4sGEN[x1_, x2_, p1_, p2_] := Tr[DyadicProductVec[psi4s].JointProjector4GEN[x1, x2, p1, p2]];

FullSimplify[Expectation4sGEN[x1, x2, p1, p2]]


(* ~~~~~~~~ general case ~~~~~~~~~ *)

EPPMM1[L4M32_ , L4M12_ , L4P12_ , L4P32_ ,  \[Theta]_, \[Phi]_] :=   Assuming[{0 < \[Theta] < Pi, 0 <= \[Phi] <= 2 Pi}, FullSimplify[
L4M32 * Assuming[{0 < \[Theta] < Pi, 0 <= \[Phi] <= 2 Pi},
 FullSimplify[
  DyadicProductVec[
   ES4M32[\[Theta], \[Phi]]], {Element[\[Theta], Reals],
   Element[\[Phi], Reals]}] ]   + L4M12 *    Assuming[{0 < \[Theta] < Pi, 0 <= \[Phi] <= 2 Pi},
 FullSimplify[
  DyadicProductVec[
   ES4M12[\[Theta], \[Phi]]], {Element[\[Theta], Reals],
   Element[\[Phi], Reals]}] ]+
L4P32 * Assuming[{0 < \[Theta] < Pi, 0 <= \[Phi] <= 2 Pi},
 FullSimplify[
  DyadicProductVec[
   ES4P32[\[Theta], \[Phi]]], {Element[\[Theta], Reals],
   Element[\[Phi], Reals]}] ]+
L4P12 * Assuming[{0 < \[Theta] < Pi, 0 <= \[Phi] <= 2 Pi},
 FullSimplify[
  DyadicProductVec[
   ES4P12[\[Theta], \[Phi]]], {Element[\[Theta], Reals],
   Element[\[Phi], Reals]}] ]
]]


EPPMM1[-1,-1,1,1,\[Theta], \[Phi]] //MatrixForm

JointProjector4PPMM1[L4M32_ , L4M12_ , L4P12_ , L4P32_ , x1_, x2_, p1_, p2_] :=  Assuming[{0 < \[Theta] < Pi, 0 <= \[Phi] <= 2 Pi},
 FullSimplify[TensorProduct[EPPMM1[L4M32 , L4M12 , L4P12 , L4P32 , x1,p1],EPPMM1[L4M32 , L4M12 , L4P12 , L4P32 ,x2,p2]], {Element[\[Theta], Reals],
   Element[\[Phi], Reals]}] ];

Expectation4PPMM1[L4M32_ , L4M12_ , L4P12_ , L4P32_ , x1_, x2_, p1_, p2_] := Tr[DyadicProductVec[psi4s].JointProjector4PPMM1[L4M32 , L4M12 , L4P12 , L4P32 ,x1, x2, p1, p2]];

FullSimplify[Expectation4PPMM1[-1,-1,1,1,x1, x2, p1, p2]]

Emmpp[x1_ ]= FullSimplify[Expectation4PPMM1[-1, -1, 1, 1, x1, 0, 0, 0]];
Emppm[x1_ ]= FullSimplify[Expectation4PPMM1[-1, 1, 1, -1, x1, 0, 0, 0]];
Empmp[x1_ ]= FullSimplify[Expectation4PPMM1[-1, 1, -1, 1, x1, 0, 0, 0]];


(*********** minmax calculation  *************)

v12   = Normalize [ { 1,0,0,0     } ]  ;
v18   = Normalize [ { 0,1,0,0     } ]  ;
v17   = Normalize [ { 0,0,1,1     } ] ;
v16   = Normalize [ { 0,0,1,-1   } ]  ;
v67   = Normalize [ { 1,-1,0,0   } ]  ;
v69   = Normalize [ { 1,1,-1,-1  } ]  ;
v56   = Normalize [ { 1,1,1,1    } ]  ;
v59   = Normalize [ { 1,-1,1,-1  } ]  ;
v58   = Normalize [ { 1,0,-1,0   } ]  ;
v45   = Normalize [ { 0,1,0,-1   } ]  ;
v48   = Normalize [ { 1,0,1,0    } ]  ;
v47   = Normalize [ { 1,1,-1,1   } ]  ;
v34   = Normalize [ { -1,1,1,1   } ]  ;
v37   = Normalize [ { 1,1,1,-1   } ]  ;
v39   = Normalize [ { 1,0,0,1    } ]  ;
v23   = Normalize [ { 0,1,-1,0   } ]  ;
v29   = Normalize [ { 0,1,1,0    } ]  ;
v28   = Normalize [ { 0,0,0,1     } ]  ;

A12   = 2 * DyadicProductVec[ v12 ] -  IdentityMatrix[4];
A18   = 2 * DyadicProductVec[ v18 ] -  IdentityMatrix[4];
A17   = 2 * DyadicProductVec[ v17 ] -  IdentityMatrix[4];
A16   = 2 * DyadicProductVec[ v16 ] -  IdentityMatrix[4];
A67   = 2 * DyadicProductVec[ v67 ] -  IdentityMatrix[4];
A69   = 2 * DyadicProductVec[ v69 ] -  IdentityMatrix[4];
A56   = 2 * DyadicProductVec[ v56 ] -  IdentityMatrix[4];
A59   = 2 * DyadicProductVec[ v59 ] -  IdentityMatrix[4];
A58   = 2 * DyadicProductVec[ v58 ] -  IdentityMatrix[4];
A45   = 2 * DyadicProductVec[ v45 ] -  IdentityMatrix[4];
A48   = 2 * DyadicProductVec[ v48 ] -  IdentityMatrix[4];
A47   = 2 * DyadicProductVec[ v47 ] -  IdentityMatrix[4];
A34   = 2 * DyadicProductVec[ v34 ] -  IdentityMatrix[4];
A37   = 2 * DyadicProductVec[ v37 ] -  IdentityMatrix[4];
A39   = 2 * DyadicProductVec[ v39 ] -  IdentityMatrix[4];
A23   = 2 * DyadicProductVec[ v23 ] -  IdentityMatrix[4];
A29   = 2 * DyadicProductVec[ v29 ] -  IdentityMatrix[4];
A28   = 2 * DyadicProductVec[ v28 ] -  IdentityMatrix[4];


T=- MyTensorProduct[ A12, MyTensorProduct[ A16, MyTensorProduct[ A17,  A18]]]-
    MyTensorProduct[ A34, MyTensorProduct[ A45, MyTensorProduct[ A47,  A48]]]-
    MyTensorProduct[ A17, MyTensorProduct[ A37, MyTensorProduct[ A47,  A67]]]-
    MyTensorProduct[ A12, MyTensorProduct[ A23, MyTensorProduct[ A28,  A29]]]-
    MyTensorProduct[ A45, MyTensorProduct[ A56, MyTensorProduct[ A58,  A59]]]-
    MyTensorProduct[ A18, MyTensorProduct[ A28, MyTensorProduct[ A48,  A58]]]-
    MyTensorProduct[ A23, MyTensorProduct[ A34, MyTensorProduct[ A37,  A39]]]-
    MyTensorProduct[ A16, MyTensorProduct[ A56, MyTensorProduct[ A67,  A69]]]-
    MyTensorProduct[ A29, MyTensorProduct[ A39, MyTensorProduct[ A59,  A69]]];


Sort[N[ Eigenvalues[FullSimplify[T]] ]]

~~~~~~~~~~~~~~~ Mathematica responds with

-6.94177, -6.67604, -6.33701, -6.28615, -6.23127, -6.16054, -6.03163, \
-5.96035, -5.93383, -5.84682, -5.73132, -5.69364, -5.56816, -5.51187, \
-5.41033, -5.37887, -5.30655, -5.19379, -5.16625, -5.14571, -5.10303, \
-5.05058, -4.94995, -4.88683, -4.81198, -4.76875, -4.64477, -4.59783, \
-4.51564, -4.46342, -4.44793, -4.36655, -4.33535, -4.26487, -4.24242, \
-4.18346, -4.11958, -4.05858, -4.00766, -3.94818, -3.91915, -3.86835, \
-3.83409, -3.77134, -3.7264, -3.68635, -3.63589, -3.59371, -3.54261, \
-3.48718, -3.47436, -3.4259, -3.35916, -3.35162, -3.29849, -3.24756, \
-3.23809, -3.18265, -3.14344, -3.09402, -3.07889, -3.03559, -3.02288, \
-2.98647, -2.88163, -2.84532, -2.80141, -2.76377, -2.72709, -2.67779, \
-2.65641, -2.64092, -2.5736, -2.53695, -2.48594, -2.46943, -2.42826, \
-2.40909, -2.3199, -2.27146, -2.26781, -2.23017, -2.19853, -2.14537, \
-2.1276, -2.1156, -2.08393, -2.02886, -2.01068, -1.95272, -1.90585, \
-1.8751, -1.81924, -1.80788, -1.77317, -1.71073, -1.67061, -1.61881, \
-1.58689, -1.56025, -1.52167, -1.47029, -1.43804, -1.41839, -1.39628, \
-1.33188, -1.2978, -1.26275, -1.24332, -1.17988, -1.16121, -1.12508, \
-1.06344, -1.04392, -0.981618, -0.9452, -0.93099, -0.902773, \
-0.866424, -0.847618, -0.797269, -0.749678, -0.718776, -0.667079, \
-0.655403, -0.621519, -0.563475, -0.535886, -0.505914, -0.488961, \
-0.477695, -0.438752, -0.413149, -0.385094, -0.329761, -0.313382, \
-0.267465, -0.251247, -0.186771, -0.162663, -0.135313, -0.115949, \
-0.0388241, -0.0285473, 0.0336107, 0.0472502, 0.0664514, 0.0818923, \
0.137393, 0.170784, 0.18296, 0.254586, 0.311604, 0.337846, 0.347853, \
0.351775, 0.395505, 0.422414, 0.481815, 0.515078, 0.57488, 0.600515, \
0.655748, 0.703362, 0.727865, 0.763394, 0.782482, 0.81889, 0.844406, \
0.888659, 0.920904, 1.00356, 1.02312, 1.03976, 1.08469, 1.1021, \
1.11609, 1.14654, 1.20192, 1.22992, 1.28624, 1.29287, 1.32196, \
1.36147, 1.43187, 1.52158, 1.5859, 1.61094, 1.62377, 1.66645, \
1.68222, 1.77266, 1.8082, 1.86793, 1.92219, 1.94603, 1.98741, \
2.04197, 2.06058, 2.12728, 2.16917, 2.20299, 2.20934, 2.2568, \
2.34362, 2.38008, 2.38999, 2.44382, 2.47456, 2.49679, 2.57822, \
2.62572, 2.63375, 2.67809, 2.73929, 2.81403, 2.82569, 2.87209, \
2.94084, 2.94773, 2.99356, 3.03768, 3.0484, 3.09975, 3.2194, 3.26743, \
3.2782, 3.30107, 3.41633, 3.43565, 3.49832, 3.62058, 3.6639, 3.7087, \
3.78394, 3.83644, 3.94999, 3.98744, 4.01948, 4.12536, 4.33452, \
4.37928, 4.42565, 4.47313, 4.53695, 4.71925, 4.84841, 4.90328, \
4.95742, 5.0169, 5.17123, 5.28471, 5.39555, 5.68376, 5.78503, 6.023}

\end{lstlisting}  }

\end{appendix}
\end{document}